\documentclass[acmsmall]{acmart}

\AtBeginDocument{%
  }

\usepackage{makecell}
\usepackage{graphicx}
\usepackage{subcaption}
\usepackage{tcolorbox} %
\usepackage{xspace}
\usepackage{float}
\usepackage{stfloats}
\usepackage{enumitem}
\usepackage{multirow}

\newcommand{\framework}{\textsc{KG4RecEval}\xspace}

\newcommand{\commentout}[1]{}
\newcommand{\s}[0]{\mathcal{S}}
\newcommand{\g}[0]{\mathcal{G}}
\newcommand{\m}[0]{\mathcal{M}}

\newcommand{\dw}[1]{}
\newcommand{\zhn}[1]

\newcommand{\youcheng}[1]{}
\newcommand{\append}[1]{}
\newcommand{\zhntodo}[1]{}
\newcommand{\revision}[1]{#1}

\begin{document}


\title{KG4RecEval: Does Knowledge Graph Really Matter for Recommender Systems?}

\author{Haonan Zhang}
\email{haonanzhang@zju.edu.cn}
\affiliation{
  \institution{Zhejiang University}
  \country{China}
}

\author{Dongxia Wang}
\email{dxwang@zju.edu.cn}
\authornote{Corresponding author.}
\affiliation{
  \institution{Zhejiang University}
  \country{China}
}

\author{Zhu Sun}
\email{sunzhuntu@gmail.com}
\affiliation{
  \institution{Singapore University of Technology and Design}
  \country{Singapore}
}

\author{Yanhui Li}
\email{3190102121@zju.edu.cn}
\affiliation{
  \institution{Zhejiang University}
  \country{China}
}

\author{Youcheng Sun}
\email{youcheng.sun@manchester.ac.uk}
\affiliation{
  \institution{The University of Manchester}
  \country{United Kingdom}
}

\author{Huizhi Liang}
\email{huizhi.liang@newcastle.ac.uk}
\affiliation{
  \institution{Newcastle University}
  \country{United Kingdom}
}

\author{Wenhai Wang}
\email{zdzzlab@zju.edu.cn}
\affiliation{
  \institution{Zhejiang University}
  \country{China}
}
\renewcommand{\shortauthors}{Haonan Zhang et al.}


\begin{abstract}
  Recommender systems (RSs) are designed to provide personalized recommendations to users.
  Recently, knowledge graphs (KGs) have been widely introduced in RSs to
  improve recommendation accuracy. 
  In this study, however, we demonstrate that RSs do not necessarily perform worse even if the KG is downgraded to the user-item interaction graph only  (or removed).
  We propose an evaluation framework \framework to systematically evaluate how much a KG contributes to the recommendation accuracy of a KG-based RS, using our defined metric KGER (\emph{KG utilization efficiency in recommendation}).
  We consider the scenarios where knowledge in a KG gets completely removed, randomly distorted and decreased, and also where recommendations are for cold-start users.
  Our extensive experiments on four commonly used datasets and a number of state-of-the-art KG-based RSs reveal that: to remove, randomly distort or decrease knowledge does not necessarily decrease recommendation accuracy, even for cold-start users. 
  These findings inspire us to rethink how to better utilize knowledge from existing KGs, whereby we discuss and provide insights into what characteristics of datasets and KG-based RSs may help improve KG utilization efficiency.
  The code and supplementary material of this paper are available at: \href{https://github.com/HotBento/KG4RecEval}{https://github.com/HotBento/KG4RecEval}.
  
\end{abstract}


\setcopyright{acmlicensed}
\acmJournal{TOIS}
\acmYear{2025} \acmVolume{1} \acmNumber{1} \acmArticle{1} \acmMonth{1}\acmDOI{10.1145/3713071}




\maketitle

\section{Introduction}
The rapid development of recommender systems (RSs) has enabled users to obtain their preferred content from a vast amount of information. 
RSs have made significant contributions to improving user experience in domains such as movie~\cite{wang2020setrank,bobadilla2010new}, web~\cite{castellano2011newer,ochi2010predictors}, and music~\cite{hu2018leveraging} recommendation. 
By exploring user historical interactions, RSs aim to predict user preference and make recommendations as accurately as possible. 

Recently, knowledge graph-based recommender systems (KG-based RSs) have garnered considerable attention due to their potential to leverage additional knowledge or facts to improve recommendation accuracy. 
For instance, in an e-commerce RS, a user who has purchased \emph{Cola} may be recommended \emph{Sprite} since they are both linked to the entity \emph{soda} in a KG via the relation \emph{belongs\_to}. 
Moreover, introducing KGs into RSs has been proposed as a solution to cold-start issues~\cite{wang2018ripplenet, wang2019KGCN, wang2019KGNNLS}, arising where new users or items have limited interaction data. 
Despite the widespread perception that KGs can improve recommendations with additional facts, few studies have critically examined how much role KGs actually play in KG-based RSs.
Intuitively, if the knowledge provided by a KG is false, less, or even completely removed, then the recommendation accuracy should be negatively influenced.
Specifically, we are interested in exploring the research questions:


\begin{itemize}
    \item \textbf{RQ1:} What if there is no KG?
    \item \textbf{RQ2:} What if the knowledge is false?
    \item \textbf{RQ3:} What if the knowledge decreases?
    \item \textbf{RQ4:} What if the KG is for the cold-start users?
    \item \textbf{RQ5:} Authenticity and amount, which matters more?
\end{itemize}

With RQ1,2,3, we explore whether the recommendation accuracy of a KG-based RS decreases when the knowledge gets removed, randomly distorted, or decreased by a ratio, respectively. 
With RQ4, we evaluate still the former questions (RQ2 and RQ3) not for normal users but for cold-start users, for whom KGs are supposed to contribute the main part of their information.
With RQ5, we aim to figure out which of the two factors, authenticity and amount of knowledge, influences the recommendation accuracy more.

To answer these questions, we design an evaluation framework named \framework, to investigate the role of the KG in a KG-based RS.
In \framework, we propose an evaluation metric KGER (\emph{KG utilization efficiency in recommendation}) to measure how much a KG contributes to improving recommendation accuracy (with a larger positive value denoting more contribution and a negative value denoting negative effect on accuracy).
Then, we design a series of experiments to investigate the various scenarios in the RQs, including no knowledge (Section \ref{noknowledge}), false knowledge (Section \ref{random}), decreasing knowledge (Section \ref{decrease}), and cold-start (Section \ref{coldstart}) experiments. 
Diverse types of state-of-the-art (SOTA) KG-based RSs and real-world datasets are considered. 
Recommendation accuracy is assessed by various metrics such as MRR~\cite{voorhees1999trec}, hit ratio, NDCG~\cite{jarvelin2002cumulated}, precision, and recall.
We obtain multiple counter-intuitive results.
For RQ1, we found that the recommendation accuracy of a KG-based RS does not necessarily decrease when the KG is downgraded to the user-item interaction graph only (or removed).
For RQ2,3,4, we found that random distortion/decrease in knowledge of a KG does not necessarily decrease recommendation accuracy, which is also the case even for cold-start users. 
Across the datasets, models and experimental settings, MRR values mostly fluctuate, with the highest values sometimes obtained not with the original KG, but with a distorted/decreased one. 
For example, the KGER values are often negative in the results for RQ1, indicating that the removal of KGs does not necessarily decrease recommendation accuracy.  
And whether KGER takes positive or negative values is highly dependent on which dataset and RS model are used.
For RQ5, we found that while for normal users, knowledge authenticity has a slightly stronger influence on KG utilization efficiency of KG-based RSs, for cold-start users, knowledge amount has a slightly stronger influence.


Overall, our contributions can be summarised as follows:
\begin{itemize}[leftmargin=*]
    \item To the best of our knowledge, this is the first work to systematically evaluate how much a KG contributes to the recommendation accuracy of a KG-based RS. 
    \item We proposed an evaluation framework \framework to explore the role of KGs in KG-based RSs from multiple perspectives, using our proposed KG utilization efficiency metric KGER. \framework and KGER can be generalized to evaluate the contribution of other types of side information to decision accuracy of a system.
    \item With extensive experiments on multiple datasets and SOTA KG-based RSs, we found that neither the decrease of \emph{authenticity} nor that of \emph{amount} of knowledge necessarily degrades the recommendation accuracy, regardless of whether it is for normal or cold-start users, which challenges the common perception that KGs empower RSs with additional knowledge.
    \item We also found how different characteristics of datasets and RSs influence the effect of KGs on recommendation accuracy, which allows us to provide insights on how to use KGs more efficiently in RSs for the future investigation/study.
\end{itemize}

\section{Background}

This section briefly introduces the related concepts, including RSs, knowledge graphs (KGs), and KG-based RSs, and also reviews the related works.

\paragraph{Recommender Systems} A RS aims to recommend items or content to a user that he/she would be interested in, usually by predicting his/her preference based on the historical (user-item) interactions~\cite{adomavicius2005toward}.
Regarding how user preference is learned, there are several types of RSs, such as content-based RSs~\cite{javed2021review,lops2019trends}, collaborative filtering (CF)-based RSs~\cite{su2009survey},
and hybrid RSs~\cite{burke2002hybrid,nassar2020novel}.
Hybrid RSs receive more attention due to their better performance on spare user-item interactions, for which KG-based RSs are a typical type. 
Regarding the methods of recommendation employed, there are memory-based methods~\cite{deshpande2004item}, latent factorization-based methods~\cite{rendle2012bpr,koren2008factorization} and deep learning (DL)-based methods~\cite{gao2020deep}, evolving towards improving the quality and efficiency of recommendation.
Roy et al.~\cite{roy2022systematic} have provided a detailed summary of the development in the field of RSs in recent years. 
In this paper, we focus on RSs based on KGs.

\begin{figure*}
    \centering
    \includegraphics[width=0.95\linewidth]{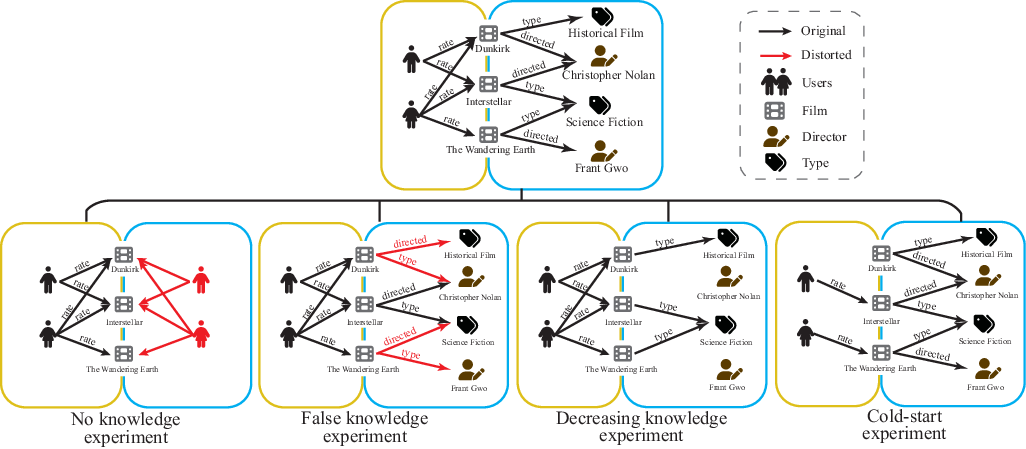}
    \caption{An illustration of our evaluation design. For ease of presentation, we take the KG in ML-1M as an example. All the five figures consist of two boxes, with the left (yellow) one denoting user-item interactions, and the right (blue) one denoting the KG about movies. The KG at the top denotes the original one, while those in no/false/decreasing knowledge experiments denote different settings. For example, for the false setting, the director of \emph{Dunkirk} is considered to be \emph{Historical Film}, which is clearly false. And for the decreasing setting, the fact that \emph{The Wandering Earth} is directed by \emph{Frant Gwo} is deleted. For the cold-start setting, only users with a small number (e.g., 1) of interactions are considered.}
    \label{fig:structure of experiment}
\end{figure*}

\paragraph{Knowledge Graphs (KGs)}
A knowledge graph (KG) is used to represent the entities and their relations in the real world~\cite{ehrlinger2016towards}.
For example, the KG of MovieLens~\cite{harper2015movielens} consists of a variety of facts related to movies, such as the producers, actors and production companies of movies, as shown in Figure~\ref{fig:structure of experiment}. 
A KG is a directed graph $\mathcal{G}=(V,E)$, where $V$ and $E$ denote the sets of nodes (entities) and edges (facts), respectively.
Each fact takes the form of <$e_h, r, e_t$>, where $e_h$ and $e_t \in V$ denote the head and tail entities, respectively, and $r\in R$ denotes the relation between them, e.g., <\emph{Interstellar, belongs\_to, Science Fiction}>. 

\paragraph{KG-based RSs} 
KG-based RSs utilize KGs as additional information (besides historical user-item interactions) to enhance recommendation performance~\cite{yang2022knowledge,wang2021learning}.
They are of interest especially for issues of data sparsity and cold-start scenarios~\cite{wang2021KGIN,wang2018ripplenet} in RSs.

In a KG-based RS, a link function $g_{link}: \mathcal{I}\rightarrow E$ transforms the items into entities in the KG\footnote{Note that existing KGs are mostly for items and their associated features but not for users.}.  
KG-based RSs aggregate the features of neighboring entities to obtain knowledge-aware representations of user and item features. 
They can be classified into
1) embedding-based methods such as CFKG~\cite{ai2018CFKG}, KTUP~\cite{cao2019KTUP}, HAKG~\cite{du2022hakg}, LKGR~\cite{chen2022modeling} and KGRec~\cite{yang2023knowledge}, which aim to learn low-dimensional representations of users, entities, and
relations in the KG; 
2) path-based methods such as MCRec~\cite{hu2018leveraging}, TNE~\cite{liang2022tne}, HeteRec~\cite{yu2013HeteRec} and RuleRec~\cite{ma2019RuleRec}, which utilize the connected paths between entities in the KG to generate recommendations; 
3) propagation-based methods such as RippleNet~\cite{wang2018ripplenet}, KGCN~\cite{wang2019KGCN} and DSKReG~\cite{wang2021dskreg}, which propagate user-specified information between entities in the KG to improve recommendation performance. 
Recently, Pugazhenthi et al. utilize KG to enhance conversational RSs, which is shown to significantly improve performance~\cite{pugazhenthi2022improving}.
There are also works surveying the existing KG-based RSs, such as~\cite{guo2020survey} and~\cite{gao2020deep}, which systematically summarize the development, classification, and methods of KG-based RSs.

\paragraph{Evaluating the Impacts of KGs} 
Several studies attempt to investigate the impact of KGs in RSs, which are relevant to our work.
For instance, Lv et al.~\cite{lv2021we} and Faber et al.~\cite{faber2021should} study the roles played by different elements of graphs and look for ways to make better use of the graph structure.
Zhang et al. modify the relations in the graph based on a fact plausibility scoring function to influence downstream tasks, e.g., link prediction and recommendation~\cite{zhang2019data}. 
Wu et al. use Q-learning to identify the most influential facts that affect the recommendation results of target items~\cite{wu2022poisoning}. 
Wang et al.~\cite{wang2021dskreg}, Yang et al.~\cite{yang2023knowledge} and Wang et al.~\cite{wang2023knowledge} study and distinguish the role played by different relations in KGs for recommendation as a way to improve the recommendation performance of KG-based RSs.
The above works aim to manipulate recommendations by analyzing which part of a KG makes a difference in recommendation, from the perspective of attacking. In contrast, our work has a different goal, namely to systematically uncover how much a KG contributes to recommendation performance, for a variety of KG-based RSs across a wide range of application domains (e.g., movie, music, and book).

There are also works evaluating or interpreting the influence of KGs in other domains.
For instance, Rossi et al.~\cite{rossi2022explaining} propose the Kelpie framework to explain the results of KG-based link prediction systems. 
Ettorre et al.~\cite{ettorre2022stunning} propose Stunning Doodle framework to enable visual analysis and comprehension of KG embeddings.  
\section{Evaluation Framework Design}
\label{sec:design}
Often, when it comes to showing that a KG-based RS performs better than other types of RS models without KGs, as in the existing works~\cite{ai2018CFKG,wang2018ripplenet}, it is difficult to tell whether and how much the KG deserves the credit.
Here, we design an evaluation framework, named \framework, to investigate the role of KGs in improving recommendation accuracy of KG-based RSs.

We first need a metric to measure how much a KG contributes (or is exploited) to improving the recommendation accuracy of a KG-based RS.
Intuitively, if the knowledge in a KG is less or less useful, then the recommendation accuracy of the KG-based RS should decrease. 
For example, given an amount of knowledge decrease of a KG, if the recommendation accuracy of a KG-based RS decreases more than the other, then it means that it exploits more (or ``wastes'' less) that knowledge. To capture this, we define KGER:
\begin{definition}\label{Definition:kger}
\emph{KG utilization efficiency in recommendation (KGER)}:
\[
    KGER(\s, \g, \m, \Delta)=\frac{\m(\s, \g)-\m(\s, (\g {-} \Delta))}{\Delta \cdot \m(\s, \g) }
\]
\end{definition}

KGER metric measures how efficiently a KG-based RS exploits the KG to improve its recommendation accuracy.
It has four parameters: 1) $\s$ denotes the recommendation model used, 2) $\g$ denotes the KG, 3) $\m$ denotes the metric used to evaluate the recommendation performance (e.g., MRR) and has $\s, \g$ as parameters\footnote{Unless otherwise specified, we use MRR as the metric to calculate KGER in this paper.}, and 4) $\Delta$ denotes the amount of decrease in authentic knowledge (e.g., the ratio of facts decreased).
A positive (negative) KGER value denotes positive (negative) effect of a KG on recommendation accuracy, with the absolute value $|KGER|$ denoting the amount of the KG's influence. 
Given $\Delta$, a larger positive KGER value means a KG contributes more to recommendation accuracy of the KG-based RS.
\revision{We also introduce a new metric, KGUS (Knowledge Graph Utilization Score), in Appendix~\ref{appendix:KGUS} that is similar to KGER but does not involve dividing by the ratio. This new metric is designed to provide an alternative perspective where we want to directly assess the relationship between KGs and recommendation performance without introducing additional scaling factors.
We find that using either KGER or KGUS as the metric leads to the same conclusions in our study.}

We then employ KGER to quantitatively evaluate the role of KG in recommendation.
Intuitively, if we reduce the amount of authentic knowledge in a KG, then the accuracy of the KG-based RS should decrease.
Below, we propose three types of ways to implement this, i.e., to completely remove the original knowledge (RQ1), to make knowledge randomly false (RQ2) and to make knowledge less (RQ3).
The design of the exploration of these RQs serves to address our main concern about the role of KG in recommendation.

\subsection{What if there is no KG? (RQ1)}
\label{sec:setup_RQ1}
A straightforward idea to understand whether a KG plays a role in the recommendation is to remove the KG from the KG-based RS and compare the performance before and after.
For most KG-based RSs, it is difficult to directly remove the KG as it is highly coupled with the recommendation process.
To address this issue, we propose:

\paragraph{KG removal} The first method forms the user-item interaction data (which serves as a training input of a RS besides the KG of a dataset) as a graph and uses it to replace the original KG, in which way the interactions become the only training input of a RS without additional modifications to its structure.
For each user-item interaction \textit{<u, i>} in the training dataset, we introduce the facts \textit{<u, interact, i>} and \textit{<i, interact\_tans, u>}\footnote{We create facts in both directions considering the needs of some KG-based RSs to build paths longer than one-hop, e.g., RippleNet.} in the graph initialized as empty. 
For instance, if user ``Mike'' interacts with item ``Interstellar'', we generate facts \textit{<Mike, interact, Interstellar>} and \textit{<Interstellar, interact\_tans, Mike>}.
KGs generated in this way are named ``\textbf{Interaction KG}''.
The second method retains only self-to-self facts (or self-connected edges) while removing all other facts of a KG. 
For each item $i$, we introduce a self-to-self fact \textit{<i, self\_to\_self, i>} in the graph which is initialized as empty.
For example, for the item ``Interstellar'', only the fact \textit{<Interstellar, self\_to\_self, Interstellar>} is retained.
KGs generated in this way are named ``\textbf{Self KG}''.
Both methods ensure that user-item interactions are the only input for KG-based RSs.

Note that while both ``Interaction KG'' and ``Self KG'' employ no more information than user-item interactions as the input to a RS, they are not strictly the same.
For instance, they may result in different number of entities and facts of a graph. We will compare their results in the experiments.
The study of RQ1 would basically provide us a binary answer about whether the existence of KG makes a difference in recommendation accuracy. For a more fine-grained study, we further introduce RQ2 and RQ3.

\subsection{What if the knowledge is false? (RQ2)}
\label{sec:setup_RQ2}

If we randomly distort the knowledge in a KG, then the false knowledge generated is unreal and should not be useful to a RS\footnote{Note that by random distortion, 
we attempt not to confine the direction in which knowledge is false, which is different from works on attacks or robustness in RSs, where false knowledge is carefully crafted~\cite{ovaisi2022rgrecsys}.}.
To put it in another way, with more knowledge randomly distorted, recommendations should be less accurate.
For example, we can switch the type of the movies ``Interstellar'' (``Science Fiction'') and ``Godfather'' (``Crime Drama''), and users preferring either the category may receive inaccurate recommendations.
Moreover, we can also change a ``type'' relation to a ``directed'' relation, making the affected facts meaningless, as shown in Figure~\ref{fig:structure of experiment}. Formally, we propose:

\paragraph{KG random distortion} 
For an arbitrarily chosen fact <$e_h, r, e_t$>, we distort it by replacing its three entries, namely randomly picking entities in the node set $V$ to replace $e_h, e_t$, and at the same time randomly picking a relation in the relation set $R$ to replace $r$.
We randomly distort $0{-}100\%$ facts: $0\%$ means that it is the original KG, and $100\%$ means that all facts in the KG are altered.
We define the ratio of distorted facts as \emph{distortion degree}, used as $\Delta$ in KGER.

\subsection{What if the knowledge decreases? (RQ3)}
\label{sec:setup_RQ3}
Another way to make a KG less useful is to reduce the amount of its knowledge, e.g., deleting some facts. Formally, we propose:

\paragraph{KG random decrease} For decreasing facts, we randomly select $0\%{-}100\%$ edges and delete them, with their head and tail entities remained in a KG. $0\%$ denotes the original KG. 
For instance, consider the fact \textit{<Mike, interact, Interstellar>}, we delete the edge in the graph but have the nodes ``Mike'' and ``Interstellar'' remain. For decreasing entities, we randomly select $0\%{-}100\%$ entities (nodes) to delete, with all the facts (edges) they are connected to also deleted. For decreasing relations, we randomly select $0\%{-}100\%$ relations and delete all the edges labeled with any of them. 
In each case, the \emph{decreased ratio} are used as $\Delta$ in KGER.


It may seem that the random distortion operation we use for RQ2 can already reduce \emph{authentic} knowledge in a KG. The main difference is that there the number of facts/entities/relations remains the same, while here it decreases.

\subsection{What if KG is for cold-start users? (RQ4)}
One of the claimed advantages of KG-based RSs is their ability to improve recommendation accuracy for \emph{cold-start users}\footnote{A system may have insufficient historical interaction data of new users to predict their preference.} by providing extra information. 
We verify this by re-evaluating RQ 2/3 for cold-start users.
In particular, we randomly select a subset of users as ``new'' users, for whom only a small number of interactions are included in the training set compared to the other users.

\section{Implementation and Analysis}\label{sec:exp}
In this section, we implement the evaluation framework designed in Section \ref{sec:design}, \framework, by a series of experiments served to investigate the proposed RQs.

\subsection{Implementation Settings}
\begin{table*}[h]
\centering
\caption{Statistics of datasets.}
\label{table:dataset}
\resizebox{1\linewidth}{!}{\begin{tabular}{c|cccccccc}
\Xhline{2pt}
           & \# User & \# Item & \# Interaction & \# Entity & \# Relation & \# Fact & Sparsity of Dataset  \\ \Xhline{1pt} 
ML-1M    & 6038  & 3498 & 573637       & 77798  & 49       & 378151 & 97.284\% \\
Amazon-Books & 20474 & 4553 & 253726       & 12553   & 19       & 46522 & 99.727\% \\
BX & 17860 & 14910 & 69873 & 77903 & 25 & 151500 & 99.974\% \\
Last.FM &1872 &3846 &21173 &9366 &60 &15518 & 99.706\% \\
\Xhline{2pt}
\end{tabular}}
\end{table*}

\paragraph{Environment} The experimental environment is critical to ensure the validity and reproducibility of the results. To standardize the data format and evaluation process, we use RecBole\footnote{\href{https://recbole.io/}{https://recbole.io/}}~\cite{recbole,recbole[1.1.1],recbole[2.0]}, a Python-based RS library built on top of PyTorch. RecBole provides a unified framework for evaluating the recommendation accuracy of different recommendation models, processing datasets, and generating KGs. Our experiments are conducted on ubuntu20.04 using Intel Xeon (R) Gold 6226R CPU and NVIDIA GeForce RTX 3090 Ti GPU.
All experiments are run 5 times with average results reported.

\paragraph{Datasets}
We use the following four commonly used real-world datasets in our experiments:
\begin{itemize}[leftmargin=*]
    \item \textbf{MovieLens-1M (ML-1M)}\footnote{\href{https://grouplens.org/datasets/movielens/}{https://grouplens.org/datasets/movielens/}}~\cite{harper2015movielens} is a widely used dataset in movie recommendation released by GroupLens. It includes about 1 million ratings (ranging from 1 to 5) for movies.
    \item \textbf{Amazon-Books}\footnote{\href{https://cseweb.ucsd.edu/~jmcauley/datasets/amazon_v2/}{https://cseweb.ucsd.edu/\~jmcauley/datasets/amazon\_v2/}}~\cite{ni2019justifying} is part of the Amazon Review dataset, including book reviews and metadata (such as title, description and price etc.).
    \item \textbf{Book-Crossing (BX)}\footnote{\href{https://github.com/hwwang55/RippleNet/tree/master/data/book}{https://github.com/hwwang55/RippleNet/tree/master/data/book}}~\cite{bookcrossing} collects user ratings (ranging from 0 to 10) for different books from the Book-Crossing community.
    \item \textbf{Last.FM}\footnote{\href{https://grouplens.org/datasets/hetrec-2011/}{https://grouplens.org/datasets/hetrec-2011/}}~\cite{lastfm} is a widely used dataset in musician recommendation. It contains historical listening data for 2 thousand users in the Last.FM online music system. 
\end{itemize}
To ensure all the experiments can be completed with reasonable time and resources, we pre-process ML-1M and Amazon-Books following SOTA methods~\cite{li2022gromov,wang2018ripplenet}.
For ML-1M, we retained interactions with ratings no less than 4.
For Amazon-Books, we retained interactions with ratings no less than 4 and adopt 20-core setting, i.e., retaining users and items with more than 20 interactions.
For Last.FM and BX, we use the versions released in~\cite{wang2018ripplenet,wang2019KGCN} with no additional modifications.
W.l.o.g, the datasets are randomly split into training, validation, and test sets in the ratio of 0.8:0.1:0.1. For each positive item in the test set, we uniformly sample 50 unrated items as negative items.
The statistics of different datasets are summarized in Table~\ref{table:dataset}.

\paragraph{KG construction}
We follow the instruction of RecBole to construct KGs for ML-1M and Amazon-Books based on KB4Rec~\cite{Zhao-DI-2019} and use the KGs released in~\cite{wang2018ripplenet,wang2019KGCN} for BX and Last.FM.
The RQs to explore require us to make changes to KGs e.g., deleting entities as in RQ2.
As Recbole deletes the items in the user-item interaction data whose corresponding entities do not show up in the KG, it is challenging to change the KG without affecting the user-item interaction data. 
We introduce an additional self-to-self fact (\textit{<entity,self\_to\_self,entity>}, self-connected edges) for each entity that should not be deleted. Importantly, these added facts serve as placeholders and do not carry any inherent meaning. This modification ensures that the user-item interaction data remains unaffected while allowing us to make changes to KGs as needed.

\revision{\paragraph{Models} We select 10 SOTA KG-based RSs.
Specifically, CFKG~\cite{ai2018CFKG}, CKE~\cite{zhang2016CKE}, KTUP~\cite{cao2019KTUP}, KGRec~\cite{yang2023knowledge} and DiffKG~\cite{jiang2024diffkg} are embedding-based models.
KGCN~\cite{wang2019KGCN}, KGIN~\cite{wang2021KGIN}, RippleNet~\cite{wang2018ripplenet}, KGNN-LS~\cite{wang2019KGNNLS} and KGAT~\cite{wang2019kgat} are propagation-based models. 
The KG-based RSs we choose are mainly the relatively more influential and representative ones recognized by the domain of RSs, which also ensures a reliable benchmark in the analysis. Table~\ref{tab:citation} lists their citations (from Google Scholar due by 2024.11.20).}

\begin{table}[H]
\caption{\revision{Citations of KG-based RSs in consideration.}}
\label{tab:citation}
\centering
\resizebox{1\linewidth}{!}{\begin{tabular}{c|cccccccccc}
\Xhline{2pt}
           & CFKG & CKE & KGCN & KGIN & RippleNet & KGNNLS & KTUP & KGAT & KGRec & DiffKG \\ \Xhline{1pt}
Citations    & 468 & 1889 & 1010 & 437 & 1298 & 622 & 634 & 2080 & 64 & 22 \\
Publication Year & 2018 & 2016 & 2019 & 2021 & 2018 & 2019 & 2019 & 2019 & 2023 & 2024 \\
\Xhline{2pt}
\end{tabular}}
\end{table}

\paragraph{Metrics} To evaluate the efficiency of KG utilization in recommendation, we use our proposed KGER metric (Definition~\ref{Definition:kger}). 
For evaluating recommendation accuracy, multiple metrics exist such as MRR (mean reciprocal rank), HR (hit ratio), NDCG (normalized discounted cumulative gain), precision, and recall. 

\begin{itemize}
    \item \textbf{MRR.} MRR measures on average how well a RS captures user preference, specifically by measuring whether the first preferred item is ranked top, defined as:
   \begin{equation*}
        MRR=\frac{1}{|\mathcal{U}|}\sum\nolimits_{u\in \mathcal{U}} \frac{1}{rank_u^*}
    \end{equation*}
where $rank_u^*$ means the rank of the first preferred item of user $u$ in his recommendation list. 
    \item \textbf{Hit@10.}  Hit rate is a way of calculating how often you recommend at least 1 
    preferred item to users. If the recommendation list contains at least 1 item in the test set, 
    we call it a hit.
    \begin{equation*}
        Hit@K=\frac{1}{|\mathcal{U}|}\sum_{u\in \mathcal{U} } \delta \left(\hat{R}(u)\cap R(u)\ne \emptyset \right)
    \end{equation*}
    where $\delta$ is an indicator function. $\delta(x)=1$ if $x$ is true and 0 otherwise. 
    $\hat{R}(u)$ is the recommendation list and $R(u)$ is the positive items in the test set 
    for user $u$.
    \item \textbf{NDCG@10.}  NDCG (Normalized Discounted Cumulative Gain) is a measure of 
    ranking quality. It give higher scores to the hits at top ranks.
    \begin{equation*}
    \begin{matrix}
        NDCG@K=\frac{1}{|\mathcal{U}|}
        \sum_{u\in\mathcal{U}}
        \left(\frac{1}{\sum_{i=1}^{min(|R(u)|,K)}\frac{1}{log_2(i+1)}}
        \sum_{i=1}^{K}\delta (\hat{R}_i(u)\in R(u))\frac{1}{log_2(i+1)}\right)
    \end{matrix}
    \end{equation*}   
    where $\hat{R}_i(u)$ is the i-th item recommended to user $u$.
    \item \textbf{Precision@10.}  Precision is the fraction of correctly recommended items out of 
    all the recommended items for users.
    \begin{equation*}
        Precision@K=\frac{1}{|\mathcal{U}|}\sum_{u\in \mathcal{U}} 
        \frac{|\hat{R}(u)\cap R(u)|}{|\hat{R}(u)|}
    \end{equation*}
    \item \textbf{Recall@10.}  Recall is the fraction of correctly recommended item out of all 
    the preferred items for users.
    \begin{equation*}
        Recall@K=\frac{1}{|\mathcal{U}|}\sum_{u\in \mathcal{U}} 
        \frac{|\hat{R}(u)\cap R(u)|}{|R(u)|}
    \end{equation*}
\end{itemize}

\paragraph{Hyperparameter tuning} We apply a grid search on different models and datasets to find the best settings using HyperOPT~\cite{bergstra2013making}.
We follow the instruction of RecBole to determine the hyperparameter search range and settings\footnote{\href{https://recbole.io/hyperparameters/knowledge.html}{https://recbole.io/hyperparameters/knowledge.html}}.
Following the guidance of~\cite{wang2019kgat}, we apply early stopping if MRR does not increase for 50 successive epochs.
The results of hyperparameter tuning are presented in Table~\ref{tab:hyper1} and Table~\ref{tab:hyper2}.
Due to time and computational resource constraints, we do not re-tune hyperparameters when modifying KGs in subsequent experiments. However, as demonstrated by the additional experiments in Appendix~\ref{sec:hyperparameter experiment}, re-tuning the hyperparameters for the modified knowledge graphs has minimal impact on the results. Therefore, the main conclusions of our study remain valid regardless of whether hyperparameter tuning is performed after each alteration of KGs.

\begin{table}[htbp]
  \centering
  \caption{\revision{Hyperparameter tuning settings on ML-1M and Amazon-Books. The best hyperparameter settings are bolded.}}
    \resizebox{1\linewidth}{!}{\begin{tabular}{|l|l|l|}
    \hline
          & \multicolumn{1}{c|}{ML-1M} & \multicolumn{1}{c|}{Amazon-Books} \\
    \hline
    \multirow{3}[2]{*}{CFKG} & learning\_rate in [1e-2,\textbf{5e-3},1e-3,5e-4,1e-4] & learning\_rate in [\textbf{1e-2},5e-3,1e-3,5e-4,1e-4] \\
          & loss\_function in [`inner\_product', \textbf{`transe'}] & loss\_function in [`inner\_product', \textbf{`transe'}] \\
          & margin in [0.5,\textbf{1.0},2.0] & margin in [0.5,\textbf{1.0},2.0] \\
    \hline
    \multirow{3}[2]{*}{CKE} & learning\_rate in [5e-5,1e-4,5e-4,7e-4,\textbf{1e-3}] & learning\_rate in [5e-5,1e-4,5e-4,7e-4,\textbf{1e-3}] \\
          & kg\_embedding\_size in [16,32,\textbf{64},128] & kg\_embedding\_size in [16,32,\textbf{64},128] \\
          & reg\_weights in [\textbf{[0.1,0.1]},[0.01,0.01],[0.001,0.001]] & reg\_weights in [\textbf{[0.1,0.1]},[0.01,0.01],[0.001,0.001]] \\
    \hline
    \multirow{5}[2]{*}{KGCN} & learning\_rate in [0.002,\textbf{0.001},0.0005] & learning\_rate in [\textbf{0.002},0.001,0.0005] \\
          & n\_iter in [\textbf{1},2] & n\_iter in [\textbf{1},2] \\
          & aggregator in [`sum',\textbf{`concat'},`neighbor'] & aggregator in [`sum',\textbf{`concat'},`neighbor'] \\
          & l2\_weight in [1e-3,\textbf{1e-5},1e-7] & l2\_weight in [1e-3,\textbf{1e-5},1e-7] \\
          & neighbor\_sample\_size in [\textbf{4}] & neighbor\_sample\_size in [\textbf{4}] \\
    \hline
    \multirow{6}[2]{*}{KGIN} & learning\_rate in [\textbf{1e-4},1e-3,5e-3] & learning\_rate in [1e-4,1e-3,\textbf{5e-3}] \\
          & node\_dropout\_rate in [0.1,0.3,\textbf{0.5}] & node\_dropout\_rate in [0.1,\textbf{0.3},0.5] \\
          & mess\_dropout\_rate in [0.0,\textbf{0.1}] & mess\_dropout\_rate in [0.0,\textbf{0.1}] \\
          & context\_hops in [2,\textbf{3}] & context\_hops in [\textbf{2},3] \\
          & n\_factors in [\textbf{4},8] & n\_factors in [\textbf{4},8] \\
          & ind in [`cosine',\textbf{`distance'}] & ind in [`cosine',\textbf{`distance'}] \\
    \hline
    \multirow{3}[2]{*}{RippleNet} & learning\_rate in [\textbf{0.001}, 0.005, 0.01, 0.05] & learning\_rate in [\textbf{0.001}, 0.005, 0.01, 0.05] \\
          & n\_memory in [4, 8,\textbf{16},32] & n\_memory in [4, 8,\textbf{16},32] \\
          & training\_neg\_sample\_num in [1, 2, 5, \textbf{10}] & training\_neg\_sample\_num in [1, 2, \textbf{5}, 10] \\
    \hline
    \multirow{6}[2]{*}{KGNNLS} & learning\_rate in [0.002,\textbf{0.001},0.0005] & learning\_rate in [\textbf{0.002},0.001,0.0005] \\
          & n\_iter in [\textbf{1},2] & n\_iter in [\textbf{1},2] \\
          & aggregator in [\textbf{`sum'}] & aggregator in [\textbf{`sum'}] \\
          & l2\_weight in [1e-3,\textbf{1e-5}] & l2\_weight in [\textbf{1e-3},1e-5] \\
          & neighbor\_sample\_size in [\textbf{4}] & neighbor\_sample\_size in [\textbf{4}] \\
          & ls\_weight in [1,0.5,0.1,\textbf{0.01},0.001] & ls\_weight in [1,0.5,0.1,\textbf{0.01},0.001] \\
    \hline
    \multirow{5}[2]{*}{KTUP} & learning\_rate in [1e-2,5e-3,\textbf{1e-3},5e-4,1e-4] & learning\_rate in [1e-2,\textbf{5e-3},1e-3,5e-4,1e-4] \\
          & L1\_flag in [True, \textbf{False}] & L1\_flag in [True, \textbf{False}] \\
          & use\_st\_gumbel in [\textbf{True}, False] & use\_st\_gumbel in [True, \textbf{False}] \\
          & train\_rec\_step in [8,\textbf{10}] & train\_rec\_step in [\textbf{8},10] \\
          & train\_kg\_step in [0,1,2,3,4,\textbf{5}] & train\_kg\_step in [0,1,2,3,4,\textbf{5}] \\
    \hline
    \multirow{4}[2]{*}{KGAT} & learning\_rate in [1e-2,5e-3,1e-3,\textbf{5e-4},1e-4] & learning\_rate in [1e-2,5e-3,\textbf{1e-3},5e-4,1e-4] \\
          & layers in [`[64,32,16]',`[64,64,64]',\textbf{`[128,64,32]'}] & layers in [`[64,32,16]',`[64,64,64]',\textbf{`[128,64,32]'}] \\
          & reg\_weight in [1e-4,\textbf{5e-5},1e-5,5e-6,1e-6] & reg\_weight in [1e-4,5e-5,1e-5,\textbf{5e-6},1e-6] \\
          & mess\_dropout in [\textbf{0.1},0.2,0.3,0.4,0.5]] & mess\_dropout in [\textbf{0.1},0.2,0.3,0.4,0.5]] \\
    \hline
    \multirow{8}[2]{*}{KGRec} & decay\_weight in [5e-5, 1e-5, \textbf{5e-6}, 1e-6] & decay\_weight in [5e-5, 1e-5, 5e-6, \textbf{1e-6}] \\
          & learning\_rate in [5e-4, \textbf{1e-4}] & learning\_rate in [\textbf{5e-4}, 1e-4] \\
          & context\_hops in [1, 2, \textbf{3}] & context\_hops in [1, \textbf{2}, 3] \\
          & node\_dropout\_rate in [0.1, \textbf{0.3}] & node\_dropout\_rate in [\textbf{0.1}, 0.3] \\
          & mess\_dropout\_rate in [0.0, \textbf{0.1}] & mess\_dropout\_rate in [0.0, \textbf{0.1}] \\
          & mae\_msize in [64, 128, \textbf{256}, 512] & mae\_msize in [\textbf{64}, 128, 256, 512] \\
          & cl\_coef in [0.0005, \textbf{0.001}, 0.01] & cl\_coef in [0.0005, 0.001, \textbf{0.01}] \\
          & tau in [0.1, \textbf{0.2}] & tau in [\textbf{0.1}, 0.2] \\
          & cl\_drop in [0.1, \textbf{0.3}, 0.5] & cl\_drop in [0.1, 0.3, \textbf{0.5}] \\
    \hline
    \multirow{5}[2]{*}{DiffKG} & reg\_weight in [1e-3,1e-4,\textbf{1e-5}] & reg\_weight in [1e-3,1e-4,\textbf{1e-5}] \\
          & mess\_dropout\_rate in [\textbf{0.1},0.2] & mess\_dropout\_rate in [0.1,\textbf{0.2}] \\
          & res\_lambda in [0.0,\textbf{0.5}] & res\_lambda in [0.0,\textbf{0.5}] \\
          & e\_loss in [0.01,0.1,\textbf{0.5}] & e\_loss in [0.01,\textbf{0.1},0.5] \\
          & temperature in [0.1,\textbf{1}] & temperature in [0.1,\textbf{1}] \\
    \hline
    \end{tabular}}%
  \label{tab:hyper1}%
\end{table}%
\begin{table}[h]
  \centering
  \caption{\revision{Hyperparameter tuning settings on BX and Last.FM. The best hyperparameter settings are bolded.}}
  \resizebox{1\linewidth}{!}{\begin{tabular}{|l|l|l|}
    \hline
          & \multicolumn{1}{c|}{BX} & \multicolumn{1}{c|}{Last.FM} \\
    \hline
    \multirow{3}[2]{*}{CFKG} & learning\_rate in [1e-2,5e-3,\textbf{1e-3},5e-4,1e-4] & learning\_rate in [\textbf{1e-2},5e-3,1e-3,5e-4,1e-4] \\
          & loss\_function in [`inner\_product', \textbf{`transe'}] & loss\_function in [`inner\_product', \textbf{`transe'}] \\
          & margin in [0.5,1.0,\textbf{2.0}] & margin in [0.5,\textbf{1.0},2.0] \\
    \hline
    \multirow{3}[2]{*}{CKE} & learning\_rate in [5e-5,1e-4,5e-4,7e-4,\textbf{1e-3}] & learning\_rate in [5e-5,1e-4,5e-4,7e-4,\textbf{1e-3}] \\
          & kg\_embedding\_size in [\textbf{16},32,64,128] & kg\_embedding\_size in [16,32,64,\textbf{128}] \\
          & reg\_weights in [\textbf{[0.1,0.1]},[0.01,0.01],[0.001,0.001]] & reg\_weights in [\textbf{[0.1,0.1]},[0.01,0.01],[0.001,0.001]] \\
    \hline
    \multirow{5}[2]{*}{KGCN} & learning\_rate in [0.002,\textbf{0.001},0.0005] & learning\_rate in [0.002,\textbf{0.001},0.0005] \\
          & n\_iter in [\textbf{1},2] & n\_iter in [\textbf{1},2] \\
          & aggregator in [`sum',`concat',\textbf{`neighbor'}] & aggregator in [`sum',\textbf{`concat'},`neighbor'] \\
          & l2\_weight in [\textbf{1e-3},1e-5,1e-7] & l2\_weight in [1e-3,\textbf{1e-5},1e-7] \\
          & neighbor\_sample\_size in [\textbf{4}] & neighbor\_sample\_size in [\textbf{4}] \\
    \hline
    \multirow{6}[2]{*}{KGIN} & learning\_rate in [1e-4,1e-3,\textbf{5e-3}] & learning\_rate in [1e-4,1e-3,\textbf{5e-3}] \\
          & node\_dropout\_rate in [0.1,0.3,\textbf{0.5}] & node\_dropout\_rate in [0.1,0.3,\textbf{0.5}] \\
          & mess\_dropout\_rate in [\textbf{0.0},0.1] & mess\_dropout\_rate in [\textbf{0.0},0.1] \\
          & context\_hops in [\textbf{2},3] & context\_hops in [2,\textbf{3}] \\
          & n\_factors in [\textbf{4},8] & n\_factors in [\textbf{4},8] \\
          & ind in [`cosine',\textbf{`distance'}] & ind in [`cosine',\textbf{`distance'}] \\
    \hline
    \multirow{3}[2]{*}{RippleNet} & learning\_rate in [0.001, \textbf{0.005}, 0.01, 0.05] & learning\_rate in [\textbf{0.001}, 0.005, 0.01, 0.05] \\
          & n\_memory in [4, \textbf{8},16,32] & n\_memory in [4, \textbf{8},16,32] \\
          & training\_neg\_sample\_num in [1, 2, 5, \textbf{10}] & training\_neg\_sample\_num in [1, 2, 5, \textbf{10}] \\
    \hline
    \multirow{6}[2]{*}{KGNNLS} & learning\_rate in [\textbf{0.002},0.001,0.0005] & learning\_rate in [\textbf{0.002},0.001,0.0005] \\
          & n\_iter in [1,\textbf{2}] & n\_iter in [1,\textbf{2}] \\
          & aggregator in [\textbf{`sum'}] & aggregator in [\textbf{`sum'}] \\
          & l2\_weight in [1e-3,\textbf{1e-5}] & l2\_weight in [\textbf{1e-3},1e-5] \\
          & neighbor\_sample\_size in [\textbf{4}] & neighbor\_sample\_size in [\textbf{4}] \\
          & ls\_weight in [1,\textbf{0.5},0.1,0.01,0.001] & ls\_weight in [1,0.5,0.1,\textbf{0.01},0.001] \\
    \hline
    \multirow{5}[2]{*}{KTUP} & learning\_rate in [1e-2,\textbf{5e-3},1e-3,5e-4,1e-4] & learning\_rate in [\textbf{1e-2},5e-3,1e-3,5e-4,1e-4] \\
          & L1\_flag in [True, \textbf{False}] & L1\_flag in [True, \textbf{False}] \\
          & use\_st\_gumbel in [\textbf{True}, False] & use\_st\_gumbel in [\textbf{True}, False] \\
          & train\_rec\_step in [\textbf{8},10] & train\_rec\_step in [8,\textbf{10}] \\
          & train\_kg\_step in [\textbf{0},1,2,3,4,5] & train\_kg\_step in [0,\textbf{1},2,3,4,5] \\
    \hline
    \multirow{4}[2]{*}{KGAT} & learning\_rate in [1e-2,5e-3,\textbf{1e-3},5e-4,1e-4] & learning\_rate in [1e-2,\textbf{5e-3},1e-3,5e-4,1e-4] \\
          & layers in [`[64,32,16]',`[64,64,64]',\textbf{`[128,64,32]'}] & layers in [`[64,32,16]',`[64,64,64]',\textbf{`[128,64,32]'}] \\
          & reg\_weight in [1e-4,5e-5,1e-5,5e-6,\textbf{1e-6}] & reg\_weight in [\textbf{1e-4},5e-5,1e-5,5e-6,1e-6] \\
          & mess\_dropout in [0.1,0.2,\textbf{0.3},0.4,0.5]] & mess\_dropout in [\textbf{0.1},0.2,0.3,0.4,0.5]] \\
    \hline
    \multirow{8}[2]{*}{KGRec} & decay\_weight in [\textbf{5e-5}, 1e-5, 5e-6, 1e-6] & decay\_weight in [\textbf{5e-5}, 1e-5, 5e-6, 1e-6] \\
          & learning\_rate in [\textbf{5e-4}, 1e-4] & learning\_rate in [\textbf{5e-4}, 1e-4] \\
          & context\_hops in [1, \textbf{2}, 3] & context\_hops in [1, \textbf{2}, 3] \\
          & node\_dropout\_rate in [\textbf{0.1}, 0.3] & node\_dropout\_rate in [0.1, \textbf{0.3}] \\
          & mess\_dropout\_rate in [\textbf{0.0}, 0.1] & mess\_dropout\_rate in [\textbf{0.0}, 0.1] \\
          & mae\_msize in [64, 128, 256, \textbf{512}] & mae\_msize in [64, 128, 256, \textbf{512}] \\
          & cl\_coef in [0.0005, \textbf{0.001}, 0.01] & cl\_coef in [\textbf{0.0005}, 0.001, 0.01] \\
          & tau in [\textbf{0.1}, 0.2] & tau in [0.1, \textbf{0.2}] \\
          & cl\_drop in [0.1, \textbf{0.3}, 0.5] & cl\_drop in [\textbf{0.1}, 0.3, 0.5] \\
    \hline
    \multirow{5}[2]{*}{DiffKG} & reg\_weight in [1e-3,\textbf{1e-4},1e-5] & reg\_weight in [1e-3,1e-4,\textbf{1e-5}] \\
          & mess\_dropout\_rate in [\textbf{0.1},0.2] & mess\_dropout\_rate in [\textbf{0.1},0.2] \\
          & res\_lambda in [\textbf{0.0},0.5] & res\_lambda in [\textbf{0.0},0.5] \\
          & e\_loss in [0.01,0.1,\textbf{0.5}] & e\_loss in [0.01,\textbf{0.1},0.5] \\
          & temperature in [0.1,\textbf{1}] & temperature in [0.1,\textbf{1}] \\
    \hline
    \end{tabular}}%
  \label{tab:hyper2}%
\end{table}%

\subsection{Investigation on No Knowledge}\label{noknowledge}
\begin{table*}
\setlength\tabcolsep{4pt}
    \centering
    \caption{\revision{The MRR results of no knowledge experiment with the original/Interaction/Self KGs. KGER values are presented in brackets.}}
    \resizebox{1\linewidth}{!}{\begin{tabular}{c|ccccc|ccccc|ccccc|ccccc} 
        \Xhline{0.7mm}
         &  \multicolumn{5}{c|}{ML-1M}&  \multicolumn{5}{c|}{Amazon-Books}&  \multicolumn{5}{c|}{BX}&  \multicolumn{5}{c}{Last.FM}\\ 
         \Xcline{2-21}{0.35mm}
         & Original &  \multicolumn{2}{c}{Interaction} &\multicolumn{2}{c|}{Self} & Original &  \multicolumn{2}{c}{Interaction} &\multicolumn{2}{c|}{Self}& Original &  \multicolumn{2}{c}{Interaction} &\multicolumn{2}{c|}{Self} & Original &  \multicolumn{2}{c}{Interaction} &\multicolumn{2}{c}{Self}\\ 
         \Xhline{0.35mm}
         CFKG       &0.731  &\textbf{0.734} & \hspace{-7pt}\small{\underline{(-0.004)}}  &0.730 & \hspace{-7pt}\small{(\ 0.001)}  &0.541 &\textbf{0.581} & \hspace{-7pt}\small{\underline{(-0.074)}}  &0.516 & \hspace{-7pt}\small{(\ 0.046)}   &0.376    &\textbf{0.383} & \hspace{-7pt}\small{\underline{(-0.019)}}  &0.378 & \hspace{-7pt}\small{\underline{(-0.005)}}   &0.558 &\textbf{0.790} & \hspace{-7pt}\small{\underline{(-0.416)}}  &0.538 & \hspace{-7pt}\small{(\ 0.036)}   \\ 
         CKE        &0.729  &\textbf{0.735} & \hspace{-7pt}\small{\underline{(-0.008)}}  &0.727 & \hspace{-7pt}\small{(\ 0.003)}  &0.570 &\textbf{0.575} & \hspace{-7pt}\small{\underline{(-0.009)}}  &0.565 & \hspace{-7pt}\small{(\ 0.009)}   &0.342    &\textbf{0.353} & \hspace{-7pt}\small{\underline{(-0.032)}}  &0.342 & \hspace{-7pt}\small{(\ 0.000)}   &\textbf{0.565} &0.558 & \hspace{-7pt}\small{(\ 0.012)}  &0.561 & \hspace{-7pt}\small{(\ 0.007)}   \\ 
         KGCN       &0.705  &0.702 & \hspace{-7pt}\small{(\ 0.004)}  &\textbf{0.706} & \hspace{-7pt}\small{\underline{(-0.001)}}  &0.533 &0.537 & \hspace{-7pt}\small{\underline{(-0.008)}}  &\textbf{0.538} & \hspace{-7pt}\small{\underline{(-0.009)}}   &\textbf{0.340}    &0.322 & \hspace{-7pt}\small{(\ 0.053)}  &0.327 & \hspace{-7pt}\small{(\ 0.038)}   &\textbf{0.525} &0.506 & \hspace{-7pt}\small{(\ 0.036)}  &0.495 & \hspace{-7pt}\small{(\ 0.057)}   \\ 
         KGIN       &\textbf{0.750}  &0.739 & \hspace{-7pt}\small{(\ 0.015)}  &0.652 & \hspace{-7pt}\small{(\ 0.131)}  &0.548 &\textbf{0.551} & \hspace{-7pt}\small{\underline{(-0.005)}}  &0.530 & \hspace{-7pt}\small{(\ 0.033)}   &\textbf{0.349}    &0.282 & \hspace{-7pt}\small{(\ 0.192)}  &0.301 & \hspace{-7pt}\small{(\ 0.138)}   &\textbf{0.577} &0.474 & \hspace{-7pt}\small{(\ 0.179)}  &0.552 & \hspace{-7pt}\small{(\ 0.043)}   \\ 
         RippleNet  &\textbf{0.707}  &0.669 & \hspace{-7pt}\small{(\ 0.054)}  &0.662 & \hspace{-7pt}\small{(\ 0.064)}  &0.509 &\textbf{0.515} & \hspace{-7pt}\small{\underline{(-0.012)}}  &0.498 & \hspace{-7pt}\small{(\ 0.022)}   &0.362    &\textbf{0.373}& \hspace{-7pt}\small{\underline{(-0.030)}}  &0.369 & \hspace{-7pt}\small{\underline{(-0.019)}}   &0.509 &0.490 & \hspace{-7pt}\small{(\ 0.037)}  &\textbf{0.509} & \hspace{-7pt}\small{(\ \underline{0.000})}   \\ 
         KGNNLS     &0.704  &0.701 & \hspace{-7pt}\small{(\ 0.004)}  &\textbf{0.706} & \hspace{-7pt}\small{\underline{(-0.003)}}  &\textbf{0.533} &0.530 & \hspace{-7pt}\small{(\ 0.006)}  &0.528 & \hspace{-7pt}\small{(\ 0.009)}   &\textbf{0.349}    &0.344 & \hspace{-7pt}\small{(\ 0.014)}  &0.333 & \hspace{-7pt}\small{(\ 0.046)}   &0.500 &\textbf{0.508} & \hspace{-7pt}\small{\underline{(-0.016)}}  &0.494 & \hspace{-7pt}\small{(\ 0.012)}   \\  
         KTUP       &\textbf{0.711}  &0.689 & \hspace{-7pt}\small{(\ 0.031)}  &0.704 & \hspace{-7pt}\small{(\ 0.010)}  &0.392 &\textbf{0.442} & \hspace{-7pt}\small{\underline{(-0.128)}}  &0.398 & \hspace{-7pt}\small{\underline{(-0.015)}}   &\textbf{0.364}    &0.360 & \hspace{-7pt}\small{(\ 0.011)}  &0.357 & \hspace{-7pt}\small{(\ 0.019)}   &0.414 &\textbf{0.436} & \hspace{-7pt}\small{\underline{(-0.053)}}  &0.432 & \hspace{-7pt}\small{\underline{(-0.043)}}   \\ 
         KGAT       &0.734  &\textbf{0.738} & \hspace{-7pt}\small{\underline{(-0.005)}}  &0.736 & \hspace{-7pt}\small{\underline{(-0.003)}}  &0.591 &\textbf{0.606} & \hspace{-7pt}\small{\underline{(-0.025)}}  &0.588 & \hspace{-7pt}\small{(\ 0.005)}   &0.349    &\textbf{0.418} & \hspace{-7pt}\small{\underline{(-0.198)}}  &0.320 & \hspace{-7pt}\small{(\ 0.083)}   &0.566 &\textbf{0.688} & \hspace{-7pt}\small{\underline{(-0.216)}}  &0.574 & \hspace{-7pt}\small{\underline{(-0.014)}}   \\ 
         KGRec     & \textbf{0.672} & 0.533 & \hspace{-7pt}\small{(0.207)} & 0.363 & \hspace{-7pt}\small{(0.461)} & 0.522 & \textbf{0.558} & \hspace{-7pt}\small{\underline{(-0.069)}} & 0.504 & \hspace{-7pt}\small{(0.035)} & \textbf{0.341} & 0.294 & \hspace{-7pt}\small{(0.139)} & 0.257 & \hspace{-7pt}\small{(0.247)} & \textbf{0.538} & 0.537 & \hspace{-7pt}\small{(0.001)} & 0.417 & \hspace{-7pt}\small{(0.224)} \\ 
         DiffKG     &0.709&0.705&\hspace{-7pt}\small{(0.006)}&\textbf{0.712}&\hspace{-7pt}\small{\underline{(-0.004)}}&\textbf{0.591}&0.586&\hspace{-7pt}\small{(0.008)}&0.585&\hspace{-7pt}\small{(0.010)}&\textbf{0.380}&0.361&\hspace{-7pt}\small{(0.049)}&0.347&\hspace{-7pt}\small{(0.087)}&\textbf{0.598}&0.578&\hspace{-7pt}\small{(0.032)}&0.585&\hspace{-7pt}\small{(0.021)}\\ 
         \Xhline{0.7mm}
    \end{tabular}}
    \label{tab:noknowledge_result}
\end{table*}

The experiments in this subsection serves to answer RQ1: what if there is no KG? 
We remove the knowledge in a KG and investigate its effect on a KG-based RS. 
For Interaction KG, the original KG of a dataset is downgraded to the user-item interaction graph only, and for Self KG, only self-to-self facts are retained (Section~\ref{sec:design}).


The experimental results are presented in Table~\ref{tab:noknowledge_result}, where ``Original'' denotes the original KG of the dataset, ``Interaction'' denotes Interaction KG and ``Self'' denotes Self KG.
KGER values are also presented in brackets in Table~\ref{tab:noknowledge_result}, with positive (negative) values denoting positive (negative) effect of KGs on recommendation accuracy (their absolute values $|KGER|$ denote the amount of that effect). 
The highest MRR values are highlighted in bold while the negative KGER values (denoting negative effect of KGs) are underlined. 
There are multiple counter-intuitive observations.
\revision{For ML-1M, compared with when the original KG is used, 6 out of 10 RSs perform similarly or even better when the original KG is downgraded to Interaction/Self KG (with negative KGER values obtained).}
\revision{For Amazon-Books, almost all RSs (except for KGNNLS and DiffKG) perform better with the original KG downgraded to Interaction KG, with a mean KGER of $-0.032$.}
Similarly, for BX and Last.FM datasets, RSs do not necessarily perform better with the original KG, and may even perform worse than with Interaction/Self KG.
Also, there does not exist a RS that always achieves best performance across the four datasets when using the original KG.
CFKG always performs better with Interaction KG, with an average KGER of $-0.171$.
Moreover, it varies for different RSs under which setting the performance is the best, namely the role of KGs is highly dependent on which dataset and RS are used.
Finally, though both removing original KGs, the results obtained with Interaction KG and Self KG are not the same, implying graph characteristics like entity and fact number may influence.
Overall, these observations challenge the conclusions made by previous studies~\cite{wang2019kgat,ai2018CFKG,cao2019KTUP} that the use of KGs can help improve recommendations. 
And we obtain the core answer to RQ1:

\begin{tcolorbox}
\textbf{Answer to RQ1:} With a KG downgraded to the user-item interaction graph only  (or removed), the recommendation accuracy of a KG-based RS does not necessarily decrease. 
\end{tcolorbox}

The results for RQ1 force us to rethink the role of KGs in recommendation.
For a more fine-grained study, below we investigate the impact of knowledge \textit{authenticity} and \textit{amount} on recommendation performance under both normal and cold-start settings.

\subsection{Investigation on False Knowledge}\label{random}

\begin{figure*}[h]
\centering
    \begin{subfigure}{\textwidth}
    \includegraphics[width=\linewidth]{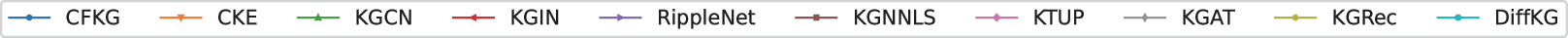}
    \end{subfigure}
    \centering
    \begin{subfigure}{0.23\textwidth}
    \includegraphics[width=\linewidth]{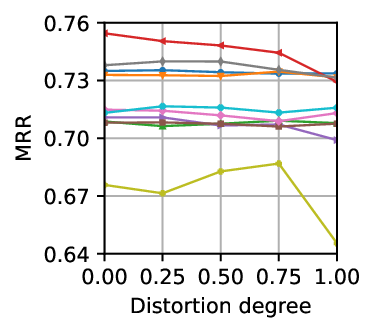}
    \end{subfigure}
    \begin{subfigure}{0.23\textwidth}
    \includegraphics[width=\linewidth]{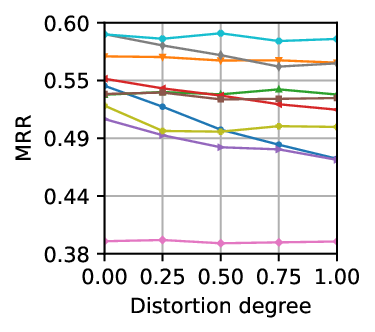}
    \end{subfigure}
    \begin{subfigure}{0.23\textwidth}
    \includegraphics[width=\linewidth]{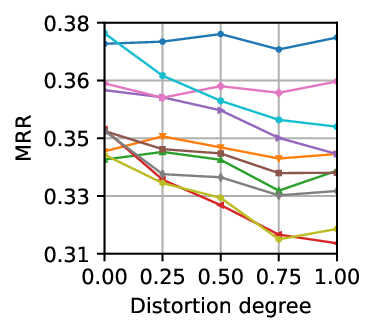}
    \end{subfigure}
    \begin{subfigure}{0.23\textwidth}
    \includegraphics[width=\linewidth]{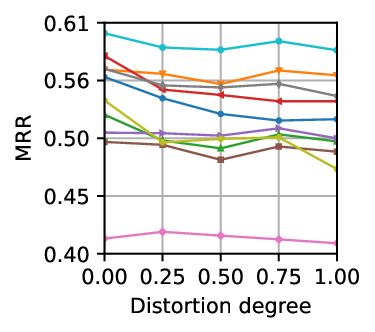}
    \end{subfigure}
    \begin{subfigure}{0.23\textwidth}
    \includegraphics[width=\linewidth]{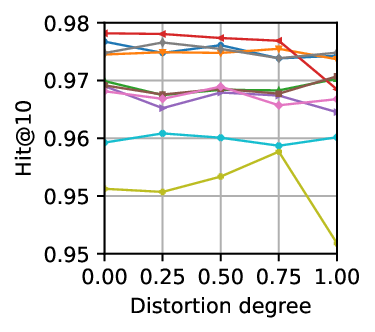}
    \end{subfigure}
    \begin{subfigure}{0.23\textwidth}
    \includegraphics[width=\linewidth]{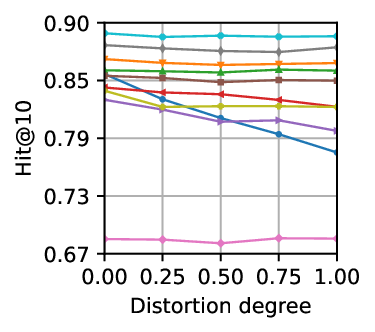}
    \end{subfigure}
    \begin{subfigure}{0.23\textwidth}
    \includegraphics[width=\linewidth]{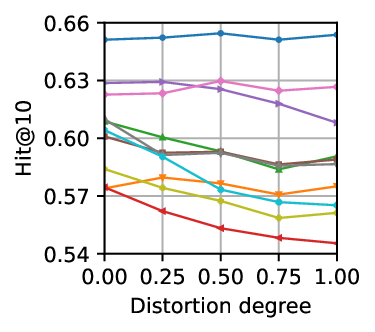}
    \end{subfigure}
    \begin{subfigure}{0.23\textwidth}
    \includegraphics[width=\linewidth]{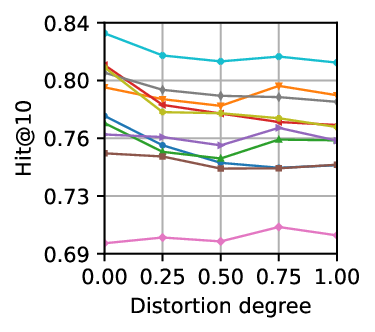}
    \end{subfigure}\\
    \begin{subfigure}{0.23\textwidth}
    \includegraphics[width=\linewidth]{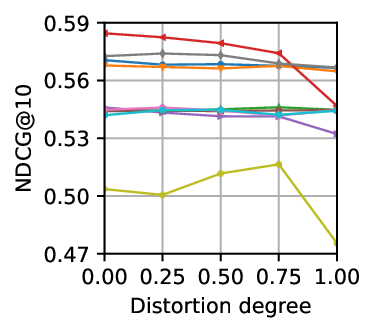}
    \end{subfigure}
    \begin{subfigure}{0.23\textwidth}
    \includegraphics[width=\linewidth]{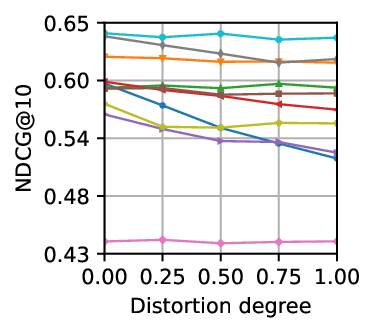}
    \end{subfigure}
    \begin{subfigure}{0.23\textwidth}
    \includegraphics[width=\linewidth]{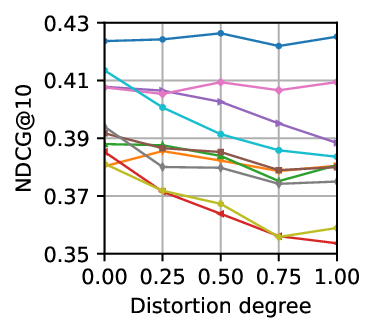}
    \end{subfigure}
    \begin{subfigure}{0.23\textwidth}
    \includegraphics[width=\linewidth]{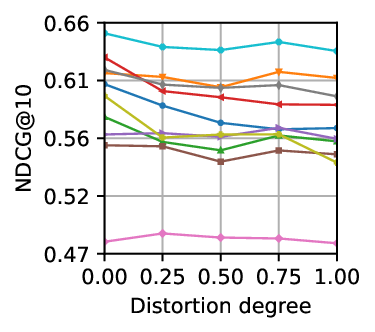}
    \end{subfigure}\\
    \begin{subfigure}{0.23\textwidth}
    \includegraphics[width=\linewidth]{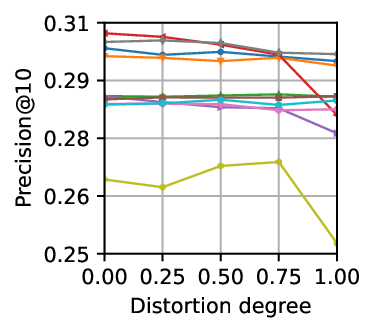}
    \end{subfigure}
    \begin{subfigure}{0.23\textwidth}
    \includegraphics[width=\linewidth]{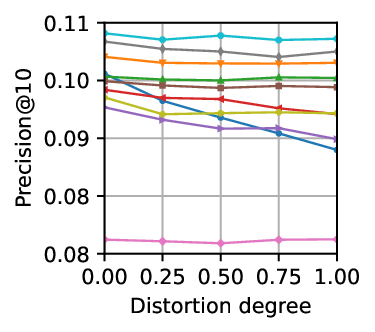}
    \end{subfigure}
    \begin{subfigure}{0.23\textwidth}
    \includegraphics[width=\linewidth]{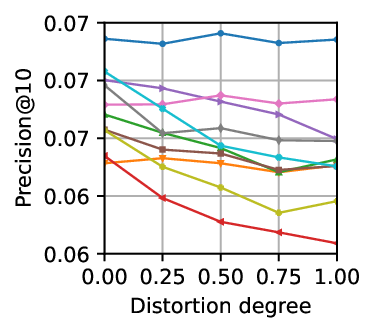}
    \end{subfigure}
    \begin{subfigure}{0.23\textwidth}
    \includegraphics[width=\linewidth]{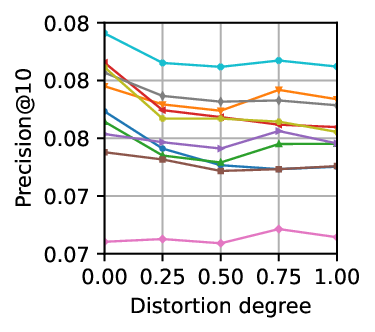}
    \end{subfigure}\\
    \begin{subfigure}{0.23\textwidth}
    \includegraphics[width=\linewidth]{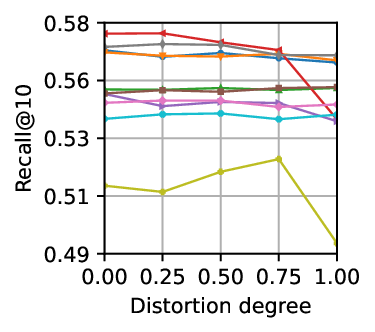}
    \caption{ML-1M}
    \end{subfigure}
    \begin{subfigure}{0.23\textwidth}
    \includegraphics[width=\linewidth]{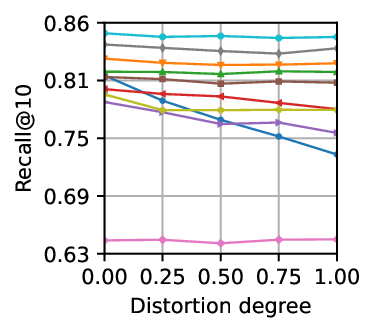}
    \caption{Amazon-Books}
    \end{subfigure}
    \begin{subfigure}{0.23\textwidth}
    \includegraphics[width=\linewidth]{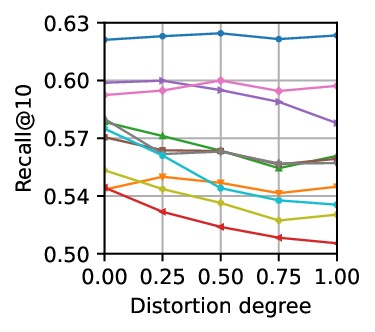}
    \caption{BX}
    \end{subfigure}
    \begin{subfigure}{0.23\textwidth}
    \includegraphics[width=\linewidth]{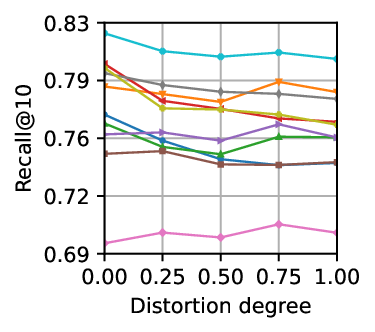}
    \caption{Last.FM}
    \end{subfigure}\\
    \caption{\revision{The results of false knowledge experiment. Horizontal axis denotes the distortion degree.}}
    \label{fig:random-kg}
\end{figure*}

The experiment in this subsection serves to answer RQ2: what if the knowledge is false?
We investigate whether the authenticity of knowledge is important to a KG-based RS.
Random distortion (Section \ref{sec:setup_RQ2}) is used to make changes to KG authenticity.

The experimental results are presented in Figure~\ref{fig:random-kg}.
W.l.o.g., we analyze the results using MRR due to their similarity under different metrics.
Counter-intuitively, across all the four datasets, only KGIN exhibits a consistent performance drop with the distortion degree increases.
\revision{MRR of all the other models fluctuates in a range of $87.01\%$ to $101.62\%$.}
For example, for ML-1M, MRR of KGAT improves at the beginning and decreases later.
Sometimes, the highest MRR is not observed in the original KG but in a distorted one, e.g., KTUP in Last.FM, KGAT in ML-1M, and CFKG in BX.
Among all the models, when all facts get distorted ($100\%$ distortion degree), KTUP is influenced the least with MRR decreasing by only $0.3\%$ on average. 
And MRR of CFKG decreases by $13.0\%$ and $6.8\%$ in Amazon-Books and Last.FM respectively but does not change significantly in ML-1M and BX.
We conjecture that CFKG is more efficient in exploiting KGs with fewer number of entities and facts (refer to Table~\ref{table:dataset}).
These observations challenge the common sense that more authentic knowledge should improve recommendations.

We also investigated KGER values of different models in different datasets.
The results are presented in Table~\ref{table:merged_kger}, including all the columns labeled ``Normal'' and ``F''\footnote{The results of KGER in different experiments (RQ2,RQ3,RQ4) are presented together in Table~\ref{table:merged_kger}.}.
The negative KGER values are highlighted with underline.
Distortion degree is set as $0.5$\footnote{Results for other distortion degrees are in Appendix~\ref{appendix:comparison}.}. 
Positive KGER values mean MRR decreased with random distortion, while negative values mean the other way, and the very small absolute values (e.g., $0.002$) imply that the KG makes little difference to the RS. 
We can observe multiple approximately non-positive KGER values, indicating that false knowledge does not necessarily degrades recommendation accuracy.
For most of the models (except for RippleNet and KGIN\footnote{We conjecture that RippleNet and KGIN can more effectively utilize the KG of ML-1M due to their specially designed knowledge propagation process.}), $|KGER|$ tends to be lower in the \emph{densest} dataset ML-1M compared to in the other datasets (who have more similar sparsity levels).
The red bars in Figure~\ref{fig:histogram} present the mean $|KGER|$ in different datasets, where in ML-1M it is the least.
Overall, the core answer to RQ2 is:
\begin{tcolorbox}
\textbf{Answer to RQ2:} 
    With more knowledge in a KG randomly distorted, the recommendation accuracy of a KG-based RS does not necessarily decrease. 
\end{tcolorbox}

\subsection{Investigation on Decreasing Knowledge}\label{decrease}

\begin{figure*}[h]
    \centering
    \begin{subfigure}{\textwidth}
    \includegraphics[width=\linewidth]{figure/legend.eps}
    \end{subfigure}
    \centering
    \begin{subfigure}{0.23\textwidth}
    \includegraphics[width=\linewidth]{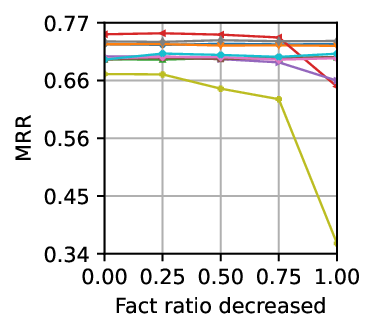}
    \end{subfigure}
    \begin{subfigure}{0.23\textwidth}
    \includegraphics[width=\linewidth]{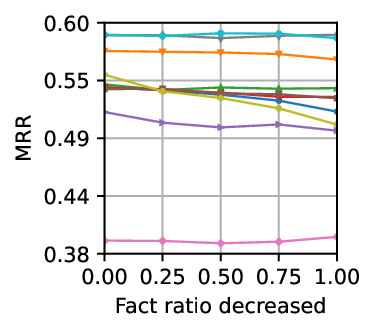}
    \end{subfigure}
    \begin{subfigure}{0.23\textwidth}
    \includegraphics[width=\linewidth]{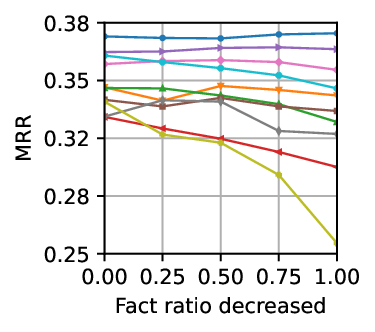}
    \end{subfigure}
    \begin{subfigure}{0.23\textwidth}
    \includegraphics[width=\linewidth]{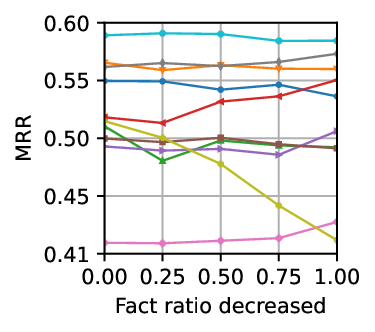}
    \end{subfigure}
    \begin{subfigure}{0.23\textwidth}
    \includegraphics[width=\linewidth]{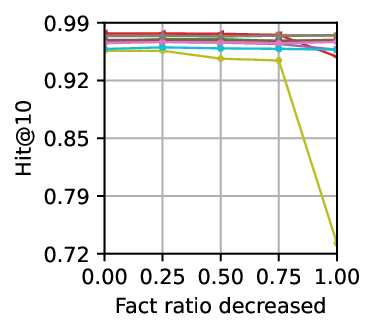}
    \end{subfigure}
    \begin{subfigure}{0.23\textwidth}
    \includegraphics[width=\linewidth]{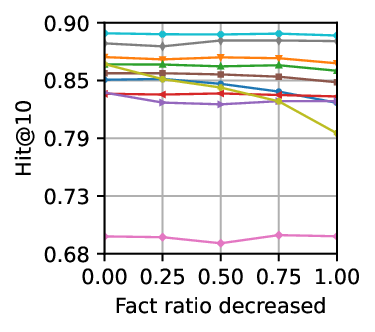}
    \end{subfigure}
    \begin{subfigure}{0.23\textwidth}
    \includegraphics[width=\linewidth]{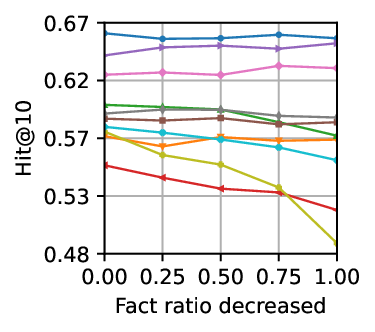}
    \end{subfigure}
    \begin{subfigure}{0.23\textwidth}
    \includegraphics[width=\linewidth]{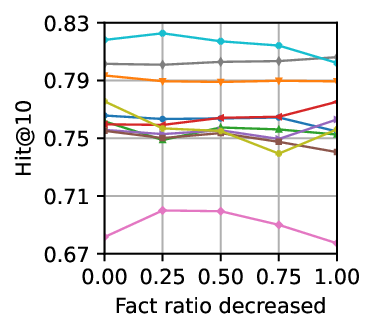}
    \end{subfigure}\\
    \begin{subfigure}{0.23\textwidth}
    \includegraphics[width=\linewidth]{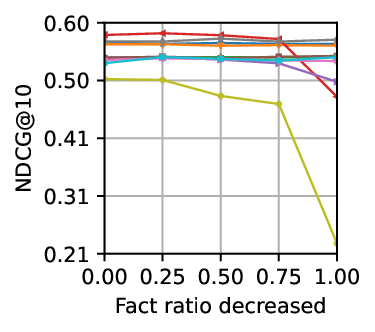}
    \end{subfigure}
    \begin{subfigure}{0.23\textwidth}
    \includegraphics[width=\linewidth]{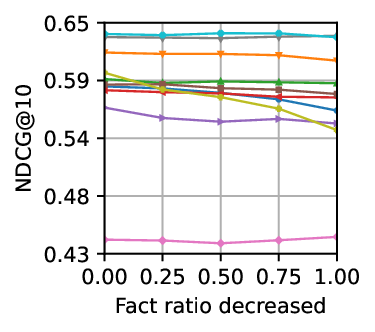}
    \end{subfigure}
    \begin{subfigure}{0.23\textwidth}
    \includegraphics[width=\linewidth]{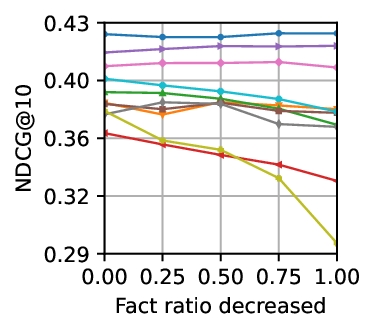}
    \end{subfigure}
    \begin{subfigure}{0.23\textwidth}
    \includegraphics[width=\linewidth]{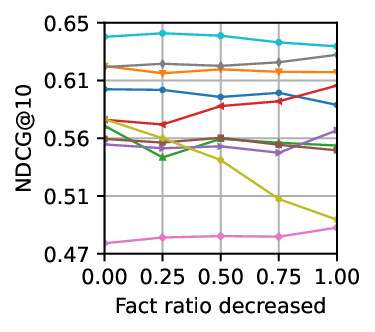}
    \end{subfigure}\\
    \begin{subfigure}{0.23\textwidth}
    \includegraphics[width=\linewidth]{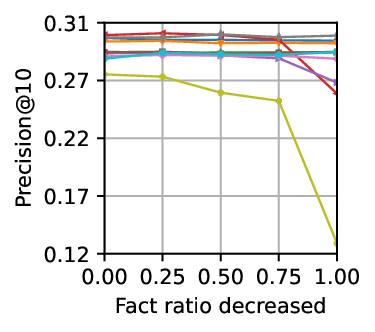}
    \end{subfigure}
    \begin{subfigure}{0.23\textwidth}
    \includegraphics[width=\linewidth]{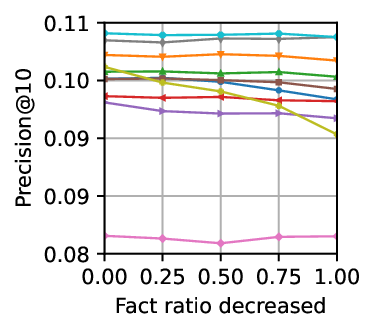}
    \end{subfigure}
    \begin{subfigure}{0.23\textwidth}
    \includegraphics[width=\linewidth]{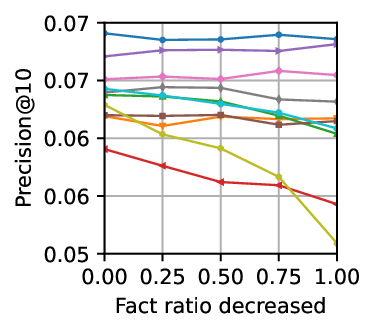}
    \end{subfigure}
    \begin{subfigure}{0.23\textwidth}
    \includegraphics[width=\linewidth]{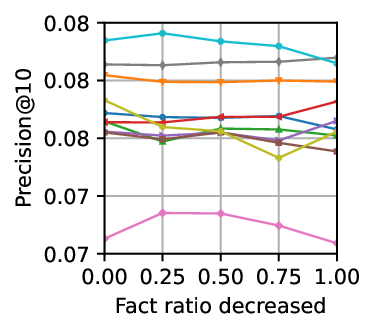}
    \end{subfigure}\\
    \begin{subfigure}{0.23\textwidth}
    \includegraphics[width=\linewidth]{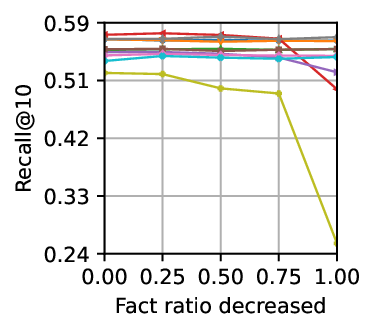}
    \caption{ML-1M}
    \end{subfigure}
    \begin{subfigure}{0.23\textwidth}
    \includegraphics[width=\linewidth]{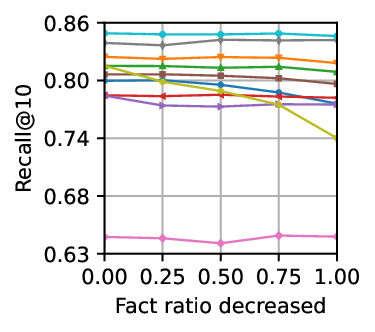}
    \caption{Amazon-Books}
    \end{subfigure}
    \begin{subfigure}{0.23\textwidth}
    \includegraphics[width=\linewidth]{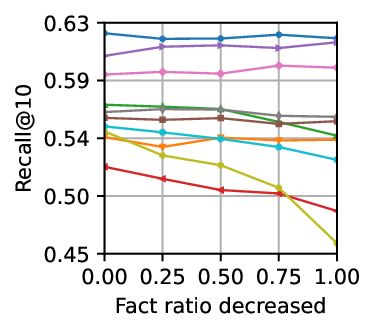}
    \caption{BX}
    \end{subfigure}
    \begin{subfigure}{0.23\textwidth}
    \includegraphics[width=\linewidth]{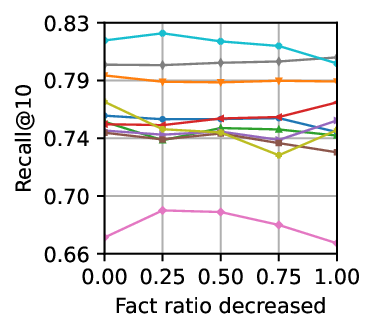}
    \caption{Last.FM}
    \end{subfigure}\\
    \caption{\revision{The results of decreasing facts. Horizontal axis denotes decreased fact ratio.}}
    \label{deletioin fact}
\end{figure*}

\begin{figure*}[h]
    \centering
    \begin{subfigure}{\textwidth}
    \includegraphics[width=\linewidth]{figure/legend.eps}
    \end{subfigure}
    \centering
    \begin{subfigure}{0.23\textwidth}
    \includegraphics[width=\linewidth]{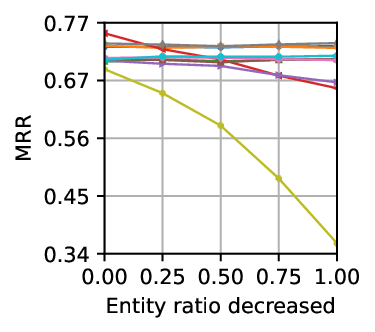}
    \end{subfigure}
    \begin{subfigure}{0.23\textwidth}
    \includegraphics[width=\linewidth]{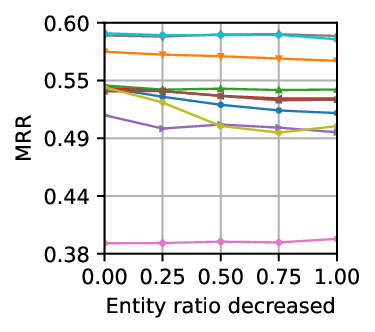}
    \end{subfigure}
    \begin{subfigure}{0.23\textwidth}
    \includegraphics[width=\linewidth]{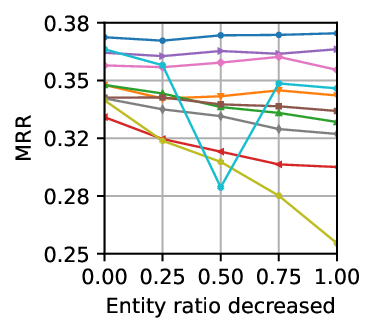}
    \end{subfigure}
    \begin{subfigure}{0.23\textwidth}
    \includegraphics[width=\linewidth]{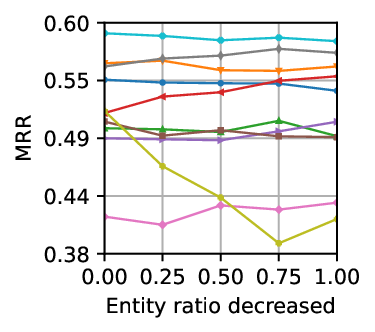}
    \end{subfigure}
    \begin{subfigure}{0.23\textwidth}
    \includegraphics[width=\linewidth]{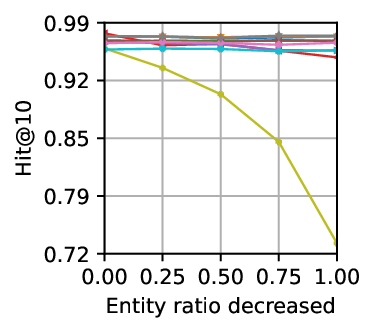}
    \end{subfigure}
    \begin{subfigure}{0.23\textwidth}
    \includegraphics[width=\linewidth]{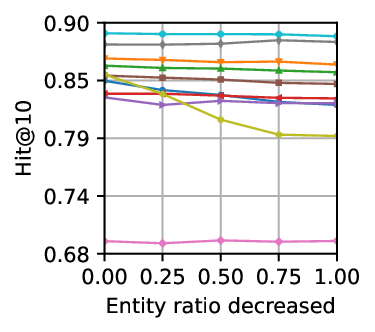}
    \end{subfigure}
    \begin{subfigure}{0.23\textwidth}
    \includegraphics[width=\linewidth]{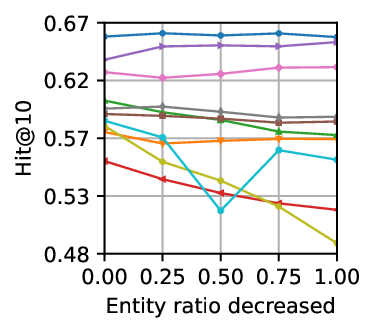}
    \end{subfigure}
    \begin{subfigure}{0.23\textwidth}
    \includegraphics[width=\linewidth]{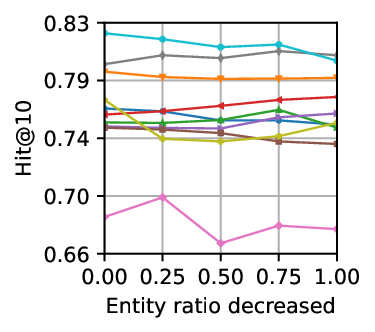}
    \end{subfigure}\\
    \begin{subfigure}{0.23\textwidth}
    \includegraphics[width=\linewidth]{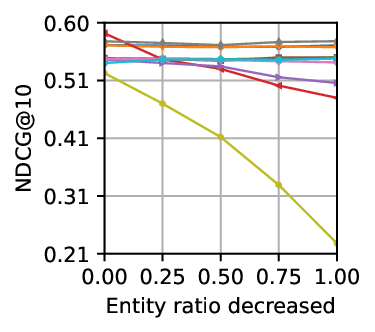}
    \end{subfigure}
    \begin{subfigure}{0.23\textwidth}
    \includegraphics[width=\linewidth]{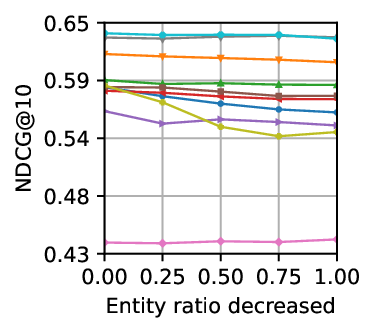}
    \end{subfigure}
    \begin{subfigure}{0.23\textwidth}
    \includegraphics[width=\linewidth]{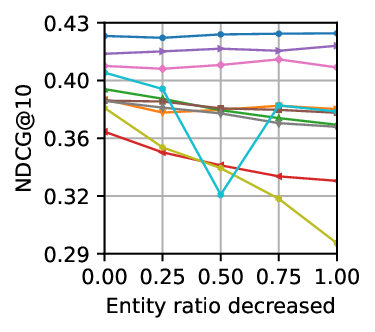}
    \end{subfigure}
    \begin{subfigure}{0.23\textwidth}
    \includegraphics[width=\linewidth]{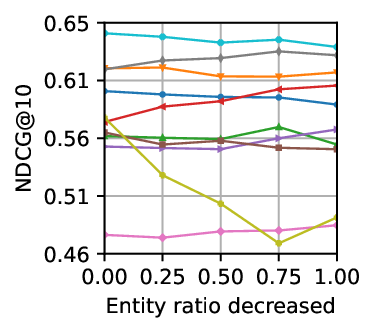}
    \end{subfigure}\\
    \begin{subfigure}{0.23\textwidth}
    \includegraphics[width=\linewidth]{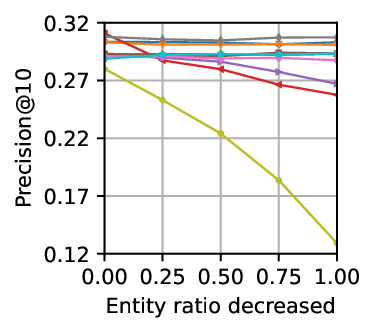}
    \end{subfigure}
    \begin{subfigure}{0.23\textwidth}
    \includegraphics[width=\linewidth]{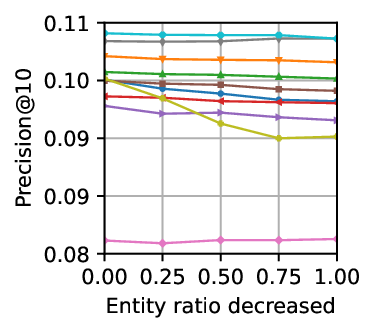}
    \end{subfigure}
    \begin{subfigure}{0.23\textwidth}
    \includegraphics[width=\linewidth]{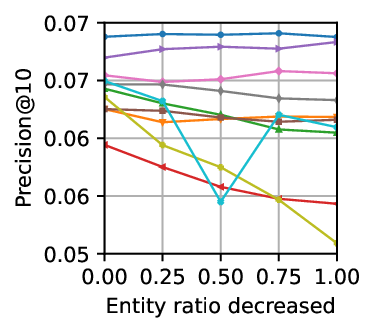}
    \end{subfigure}
    \begin{subfigure}{0.23\textwidth}
    \includegraphics[width=\linewidth]{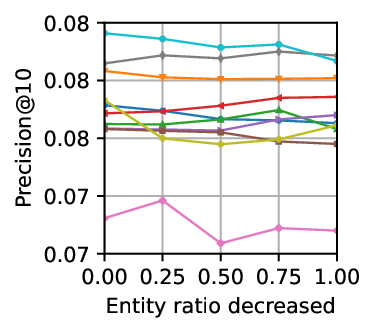}
    \end{subfigure}\\
    \begin{subfigure}{0.23\textwidth}
    \includegraphics[width=\linewidth]{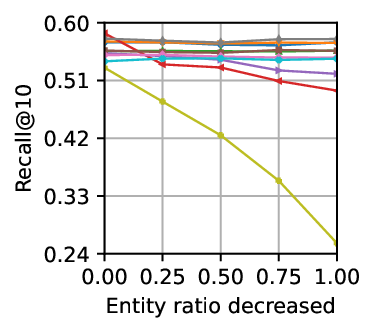}
    \caption{ML-1M}
    \end{subfigure}
    \begin{subfigure}{0.23\textwidth}
    \includegraphics[width=\linewidth]{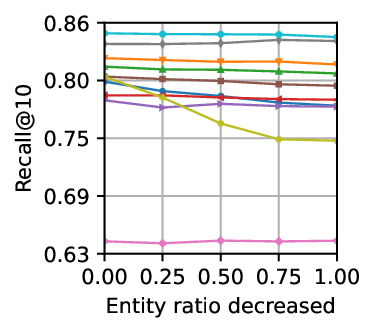}
    \caption{Amazon-Books}
    \end{subfigure}
    \begin{subfigure}{0.23\textwidth}
    \includegraphics[width=\linewidth]{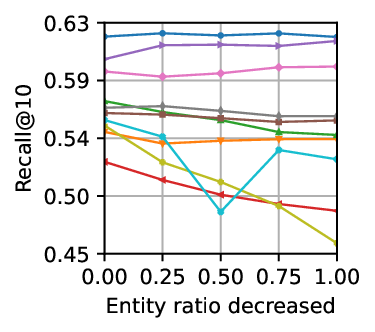}
    \caption{BX}
    \end{subfigure}
    \begin{subfigure}{0.23\textwidth}
    \includegraphics[width=\linewidth]{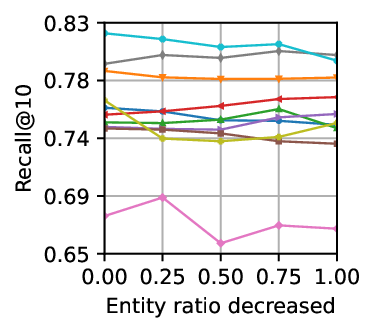}
    \caption{Last.FM}
    \end{subfigure}\\
    \caption{\revision{The results of decreasing entities. Horizontal axis denotes decreased entity ratio.}}
    \label{deletioin entity}
\end{figure*}

\begin{figure*}[h]
    \centering
    \begin{subfigure}{\textwidth}
    \includegraphics[width=\linewidth]{figure/legend.eps}
    \end{subfigure}
    \centering
    \begin{subfigure}{0.23\textwidth}
    \includegraphics[width=\linewidth]{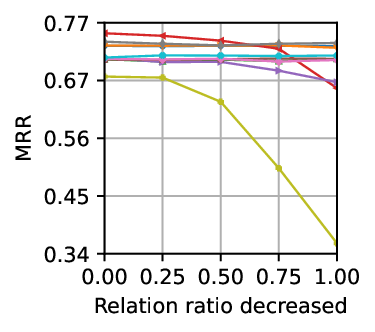}
    \end{subfigure}
    \begin{subfigure}{0.23\textwidth}
    \includegraphics[width=\linewidth]{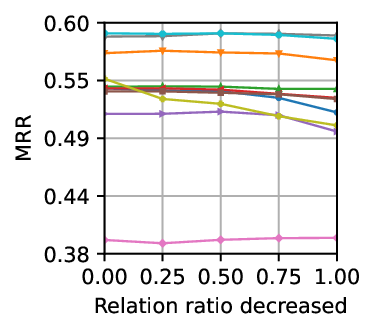}
    \end{subfigure}
    \begin{subfigure}{0.23\textwidth}
    \includegraphics[width=\linewidth]{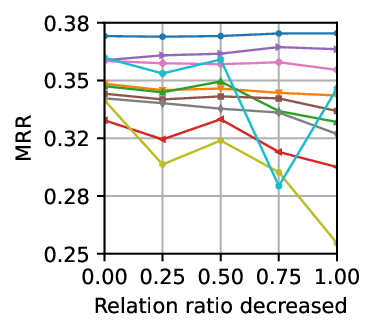}
    \end{subfigure}
    \begin{subfigure}{0.23\textwidth}
    \includegraphics[width=\linewidth]{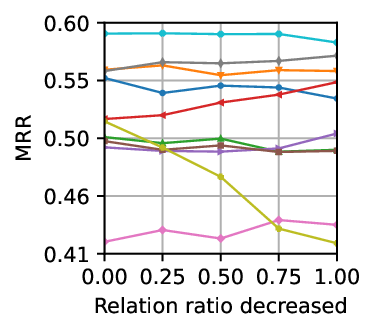}
    \end{subfigure}
    \begin{subfigure}{0.23\textwidth}
    \includegraphics[width=\linewidth]{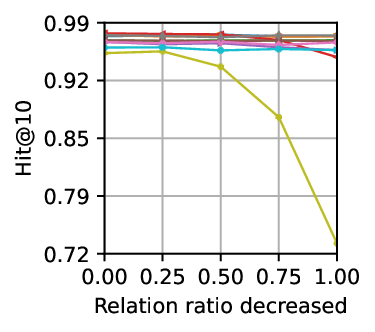}
    \end{subfigure}
    \begin{subfigure}{0.23\textwidth}
    \includegraphics[width=\linewidth]{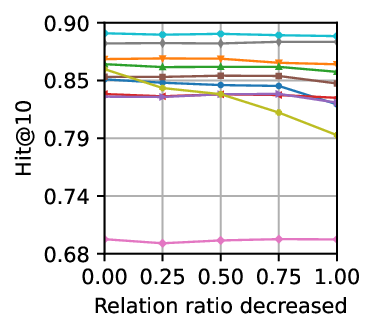}
    \end{subfigure}
    \begin{subfigure}{0.23\textwidth}
    \includegraphics[width=\linewidth]{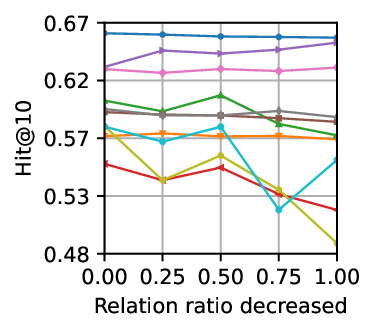}
    \end{subfigure}
    \begin{subfigure}{0.23\textwidth}
    \includegraphics[width=\linewidth]{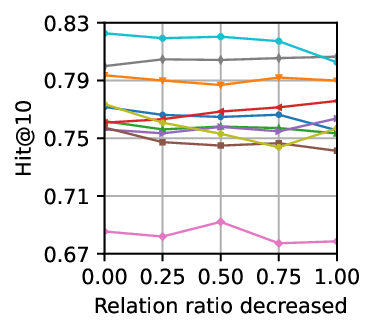}
    \end{subfigure}\\
    \begin{subfigure}{0.23\textwidth}
    \includegraphics[width=\linewidth]{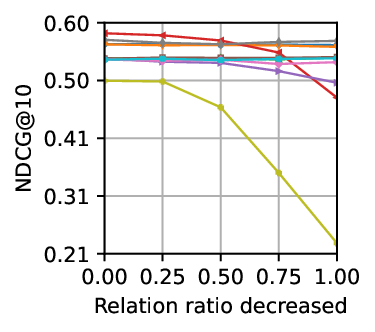}
    \end{subfigure}
    \begin{subfigure}{0.23\textwidth}
    \includegraphics[width=\linewidth]{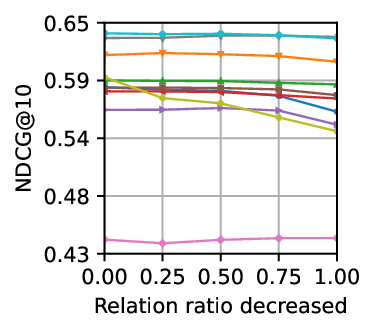}
    \end{subfigure}
    \begin{subfigure}{0.23\textwidth}
    \includegraphics[width=\linewidth]{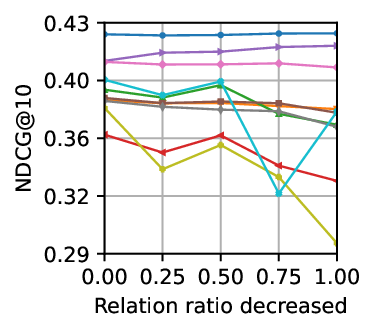}
    \end{subfigure}
    \begin{subfigure}{0.23\textwidth}
    \includegraphics[width=\linewidth]{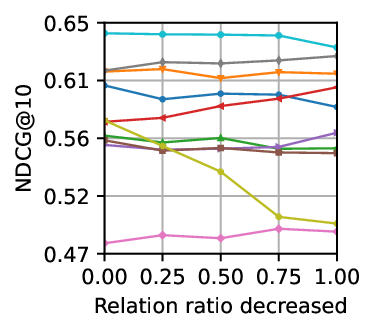}
    \end{subfigure}\\
    \begin{subfigure}{0.23\textwidth}
    \includegraphics[width=\linewidth]{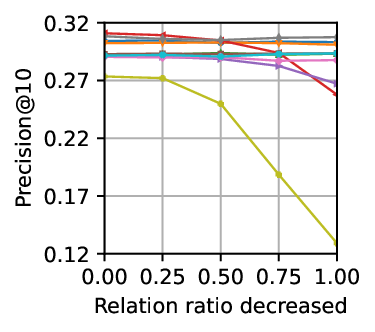}
    \end{subfigure}
    \begin{subfigure}{0.23\textwidth}
    \includegraphics[width=\linewidth]{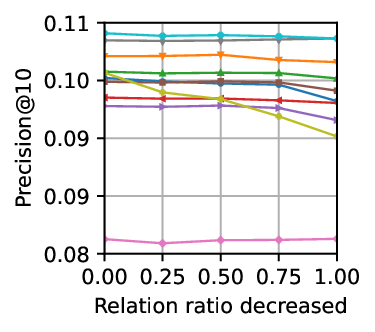}
    \end{subfigure}
    \begin{subfigure}{0.23\textwidth}
    \includegraphics[width=\linewidth]{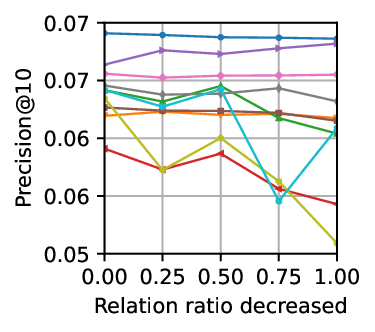}
    \end{subfigure}
    \begin{subfigure}{0.23\textwidth}
    \includegraphics[width=\linewidth]{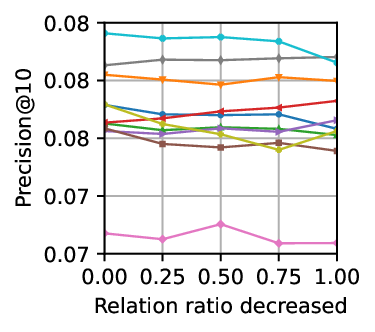}
    \end{subfigure}\\
    \begin{subfigure}{0.23\textwidth}
    \includegraphics[width=\linewidth]{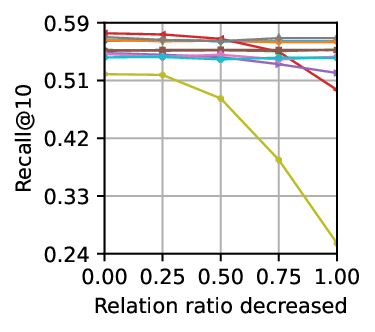}
    \caption{ML-1M}
    \end{subfigure}
    \begin{subfigure}{0.23\textwidth}
    \includegraphics[width=\linewidth]{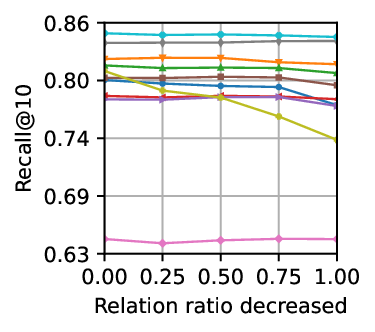}
    \caption{Amazon-Books}
    \end{subfigure}
    \begin{subfigure}{0.23\textwidth}
    \includegraphics[width=\linewidth]{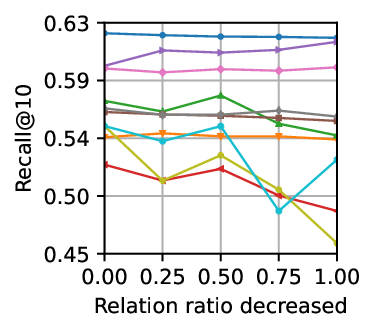}
    \caption{BX}
    \end{subfigure}
    \begin{subfigure}{0.23\textwidth}
    \includegraphics[width=\linewidth]{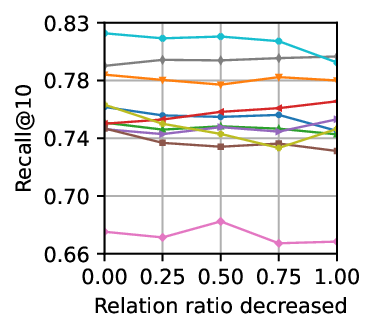}
    \caption{Last.FM}
    \end{subfigure}\\
    \caption{\revision{The results of decreasing relations. Horizontal axis denotes decreased relation ratio.}}
    \label{deletioin relation}
\end{figure*}

This experiment serves to answer RQ3: what if the knowledge decreased?
We investigate by randomly deleting a ratio of facts, entities or relations (refer to Section~\ref{sec:design} for the detailed operations).

Figure~\ref{deletioin fact}-\ref{deletioin relation} presents the results of decreasing knowledge of a KG, including that of decreasing facts, entities and relations. 
Observe that whichever of deleting facts, entities or relations, different KG-based RSs exhibit different trends of the changes in recommendation accuracy. 
Take MRR as an example.
\revision{Generally, MRR of most the models including CKE, CFKG, KGCN, KGNNLS, KGAT, and DiffKG, does not show consistent monotonicity with the decrease of fact/entity/relation, which is the case across different datasets.} 
\revision{For instance, their MRR typically fluctuates within a range of $94.33\%$ to $102.86\%$ with the number of facts decreased.}
KGIN/RippleNet seems to efficiently exploit KGs sometimes e.g., when its MRR decreases with more facts/entities/relations deleted in ML-1M, but not always (consider Last.FM, BX datasets).
Across all the four datasets, the MRR of KGRec decreases consistently with the decrease of facts.
It is worth noting that when the decreased ratio is below $0.5$, deleting entities generally has a stronger impact on MRR than deleting facts.
This may be because, compared with directly deleting facts, deleting entities would result in more of them being deleted when the decreased ratio is small.
For example, for the fact \textit{<Mike, interact, Interstellar>}, deleting either entity ``Mike'' or ``Interstellar'' would result in the deletion of it.
Further, we conduct additional experiments in Appendix C to investigate the impact of specific relation types in KGs on performance of RSs. The results indicate that there are large differences in the ability of different RSs to utilize the same type of relations.

We also investigate KGER of different models in different datasets.
The results are presented in Table~\ref{table:merged_kger}, including all the columns labeled ``Normal'' and ``D''.\footnote{The decreased ratio is $0.5$ with results of other ratios in Appendix \ref{appendix:comparison}.} 
The negative KGER values are highlighted with underline.
We can observe that multiple negative values exist, indicating that randomly decreasing knowledge does not necessarily decrease recommendation accuracy.
Among all the models, KGAT performs the worst in KG utilization, with KGER being $-0.026$ on average. 
Similarly as in false knowledge experiment, we can observe that for most models (except for RippleNet and KGIN), $|KGER|$ tends to be lower in the \emph{densest} dataset ML-1M than in the other datasets.
The green bars in Figure 3 present the mean $|KGER|$ in different datasets, where in ML-1M it is the least.
Overall, we provide the core answer to RQ3:

\begin{tcolorbox}
\textbf{Answer to RQ3:} 
    With more knowledge in a KG randomly decreased, the recommendation accuracy of a KG-based RS does not necessarily decrease.
\end{tcolorbox}

With the answers to RQ2,3, we can conclude that neither false nor less knowledge in a KG necessarily decreases recommendation accuracy of a KG-based RS.

\subsection{Investigation under the Cold-start Setting}\label{coldstart}

\begin{figure*}[h]
    \centering
    \begin{subfigure}{\textwidth}
    \includegraphics[width=\linewidth]{figure/legend.eps}
    \end{subfigure}
    \centering
    \begin{subfigure}{0.23\textwidth}
    \includegraphics[width=\linewidth]{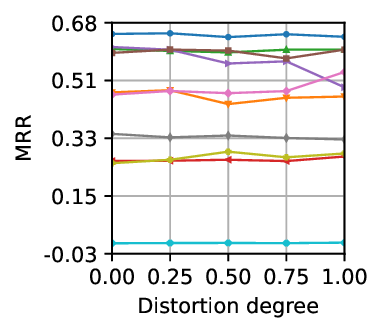}
    \end{subfigure}
    \begin{subfigure}{0.23\textwidth}
    \includegraphics[width=\linewidth]{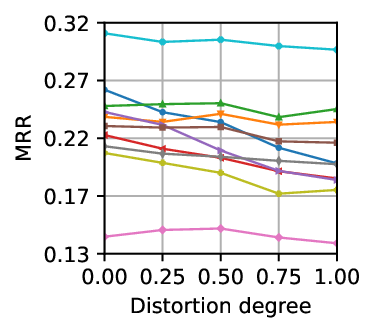}
    \end{subfigure}
    \begin{subfigure}{0.23\textwidth}
    \includegraphics[width=\linewidth]{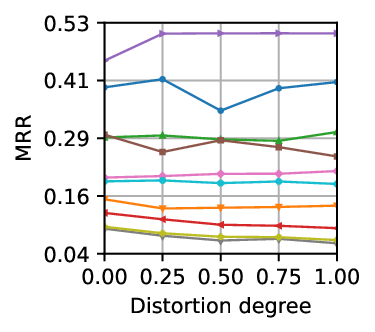}
    \end{subfigure}
    \begin{subfigure}{0.23\textwidth}
    \includegraphics[width=\linewidth]{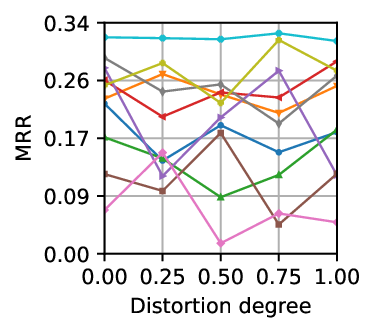}
    \end{subfigure}\\
    \begin{subfigure}{0.23\textwidth}
    \includegraphics[width=\linewidth]{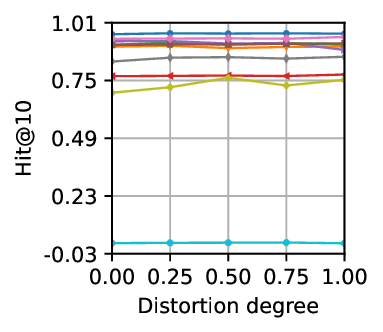}
    \end{subfigure}
    \begin{subfigure}{0.23\textwidth}
    \includegraphics[width=\linewidth]{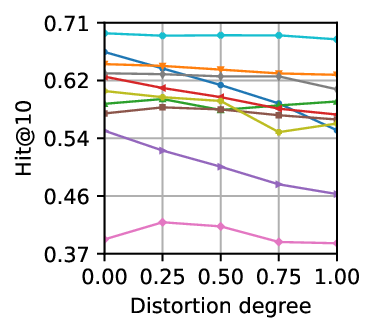}
    \end{subfigure}
    \begin{subfigure}{0.23\textwidth}
    \includegraphics[width=\linewidth]{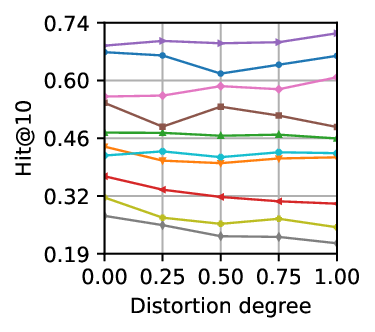}
    \end{subfigure}
    \begin{subfigure}{0.23\textwidth}
    \includegraphics[width=\linewidth]{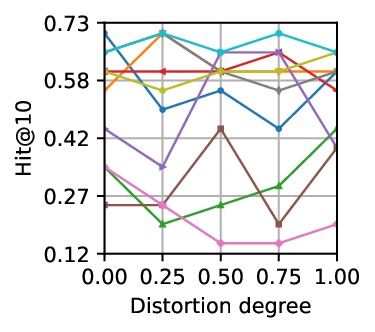}
    \end{subfigure}\\
    \begin{subfigure}{0.23\textwidth}
    \includegraphics[width=\linewidth]{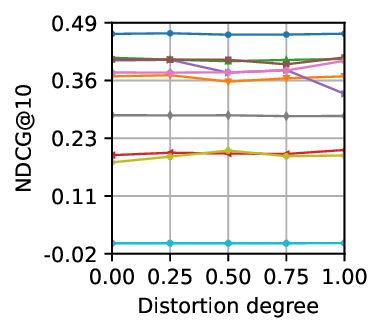}
    \end{subfigure}
    \begin{subfigure}{0.23\textwidth}
    \includegraphics[width=\linewidth]{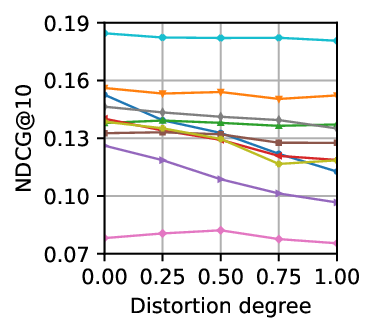}
    \end{subfigure}
    \begin{subfigure}{0.23\textwidth}
    \includegraphics[width=\linewidth]{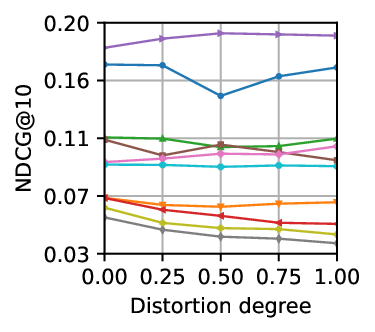}
    \end{subfigure}
    \begin{subfigure}{0.23\textwidth}
    \includegraphics[width=\linewidth]{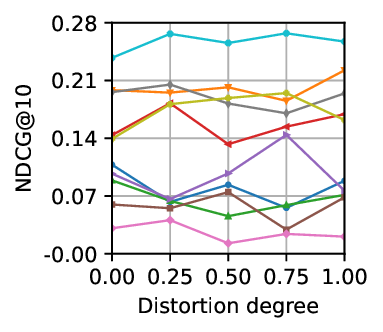}
    \end{subfigure}\\
    \begin{subfigure}{0.23\textwidth}
    \includegraphics[width=\linewidth]{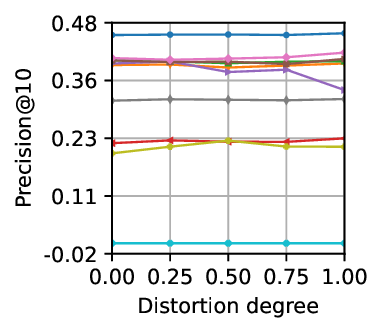}
    \end{subfigure}
    \begin{subfigure}{0.23\textwidth}
    \includegraphics[width=\linewidth]{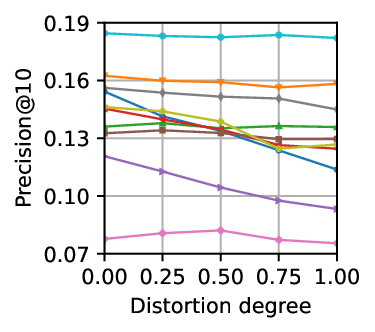}
    \end{subfigure}
    \begin{subfigure}{0.23\textwidth}
    \includegraphics[width=\linewidth]{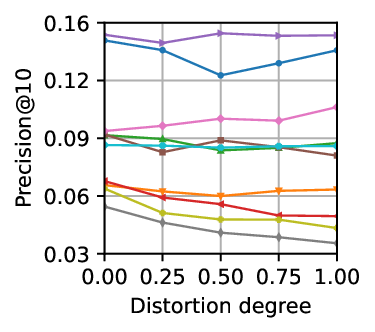}
    \end{subfigure}
    \begin{subfigure}{0.23\textwidth}
    \includegraphics[width=\linewidth]{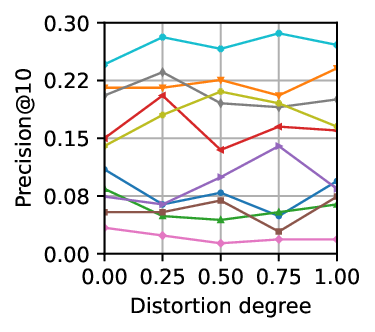}
    \end{subfigure}\\
    \begin{subfigure}{0.23\textwidth}
    \includegraphics[width=\linewidth]{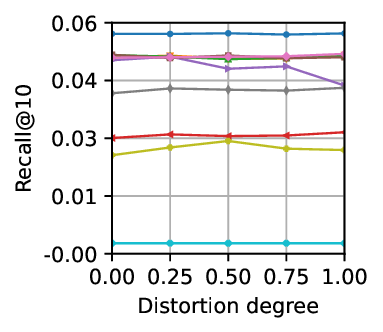}
    \caption{ML-1M}
    \end{subfigure}
    \begin{subfigure}{0.23\textwidth}
    \includegraphics[width=\linewidth]{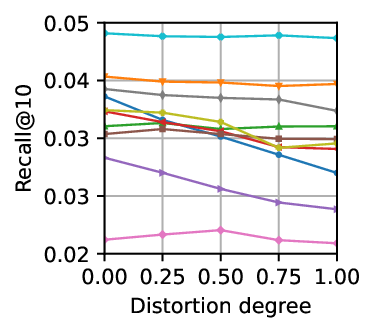}
    \caption{Amazon-Books}
    \end{subfigure}
    \begin{subfigure}{0.23\textwidth}
    \includegraphics[width=\linewidth]{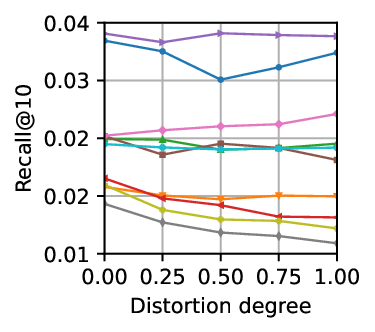}
    \caption{BX}
    \end{subfigure}
    \begin{subfigure}{0.23\textwidth}
    \includegraphics[width=\linewidth]{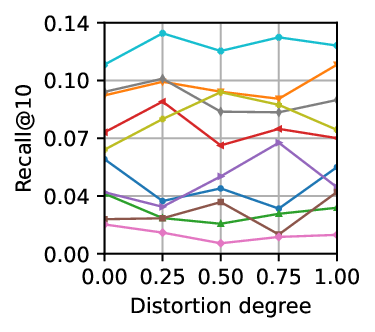}
    \caption{Last.FM}
    \end{subfigure}\\
    \caption{\revision{The results of cold-start experiment with false knowledge, $T$=1. Horizontal axis denotes distortion degree.}}
    \label{fig:cold-start false t1}
\end{figure*}

\begin{figure*}[h]
    \centering
    \begin{subfigure}{\textwidth}
    \includegraphics[width=\linewidth]{figure/legend.eps}
    \end{subfigure}
    \begin{subfigure}{0.23\textwidth}
    \includegraphics[width=\linewidth]{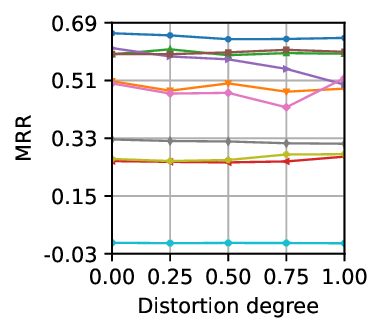}
    \end{subfigure}
    \begin{subfigure}{0.23\textwidth}
    \includegraphics[width=\linewidth]{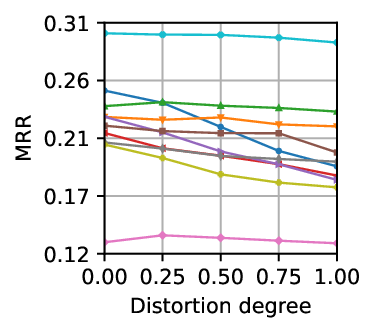}
    \end{subfigure}
    \begin{subfigure}{0.23\textwidth}
    \includegraphics[width=\linewidth]{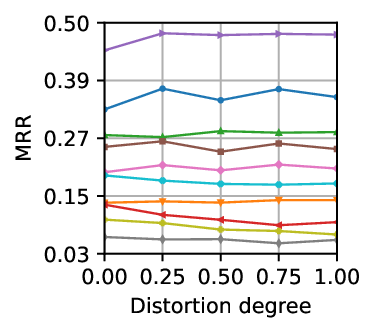}
    \end{subfigure}
    \begin{subfigure}{0.23\textwidth}
    \includegraphics[width=\linewidth]{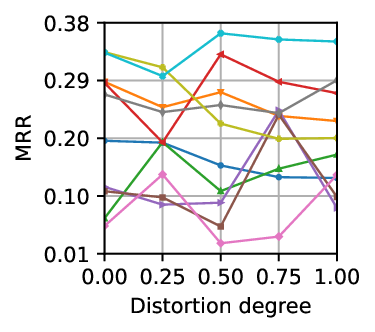}
    \end{subfigure}
    \begin{subfigure}{0.23\textwidth}
    \includegraphics[width=\linewidth]{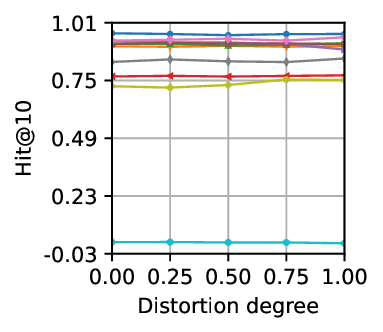}
    \end{subfigure}
    \begin{subfigure}{0.23\textwidth}
    \includegraphics[width=\linewidth]{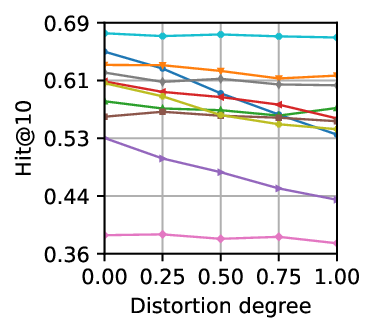}
    \end{subfigure}
    \begin{subfigure}{0.23\textwidth}
    \includegraphics[width=\linewidth]{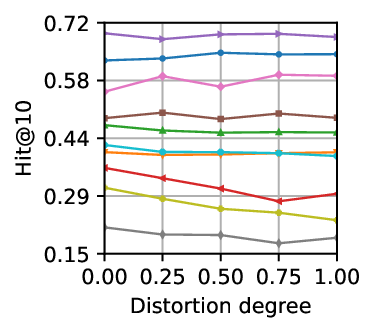}
    \end{subfigure}
    \begin{subfigure}{0.23\textwidth}
    \includegraphics[width=\linewidth]{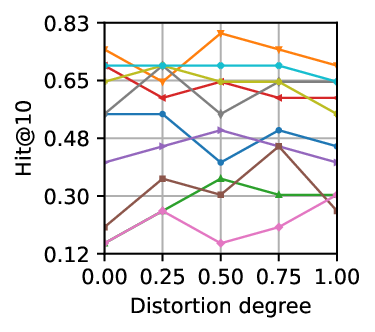}
    \end{subfigure}\\
    \begin{subfigure}{0.23\textwidth}
    \includegraphics[width=\linewidth]{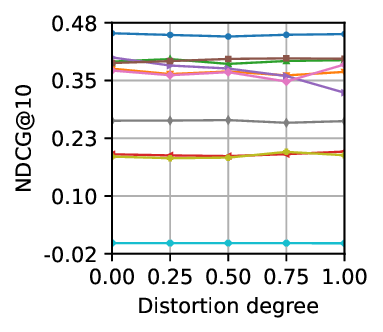}
    \end{subfigure}
    \begin{subfigure}{0.23\textwidth}
    \includegraphics[width=\linewidth]{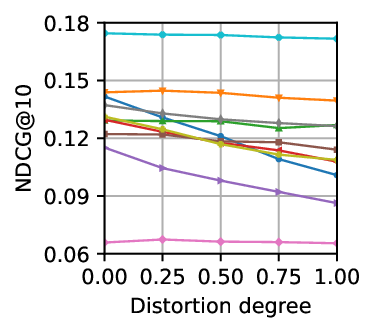}
    \end{subfigure}
    \begin{subfigure}{0.23\textwidth}
    \includegraphics[width=\linewidth]{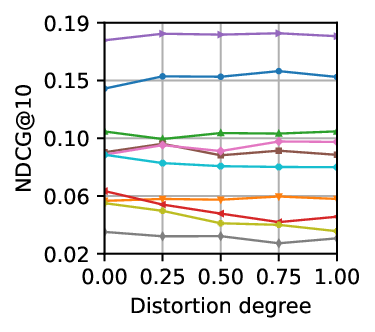}
    \end{subfigure}
    \begin{subfigure}{0.23\textwidth}
    \includegraphics[width=\linewidth]{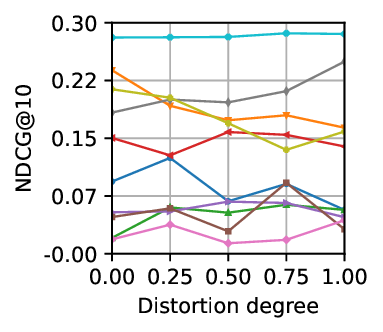}
    \end{subfigure}\\
    \begin{subfigure}{0.23\textwidth}
    \includegraphics[width=\linewidth]{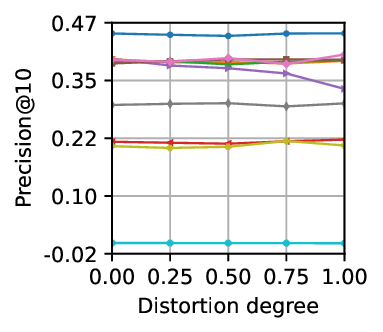}
    \end{subfigure}
    \begin{subfigure}{0.23\textwidth}
    \includegraphics[width=\linewidth]{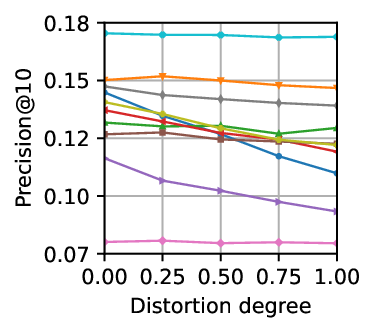}
    \end{subfigure}
    \begin{subfigure}{0.23\textwidth}
    \includegraphics[width=\linewidth]{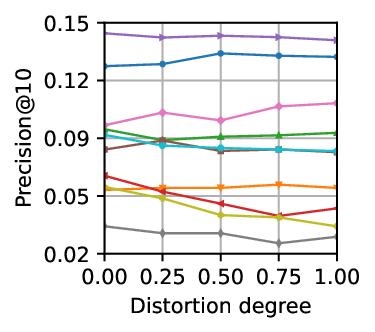}
    \end{subfigure}
    \begin{subfigure}{0.23\textwidth}
    \includegraphics[width=\linewidth]{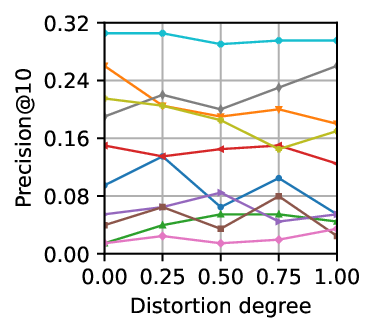}
    \end{subfigure}\\
    \begin{subfigure}{0.23\textwidth}
    \includegraphics[width=\linewidth]{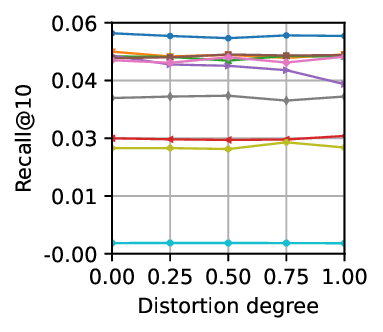}
    \caption{ML-1M}
    \end{subfigure}
    \begin{subfigure}{0.23\textwidth}
    \includegraphics[width=\linewidth]{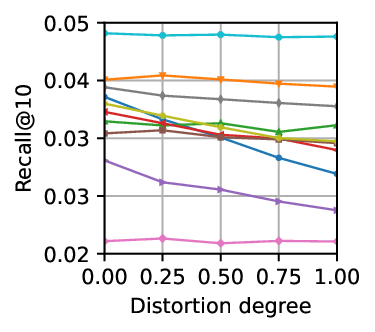}
    \caption{Amazon-Books}
    \end{subfigure}
    \begin{subfigure}{0.23\textwidth}
    \includegraphics[width=\linewidth]{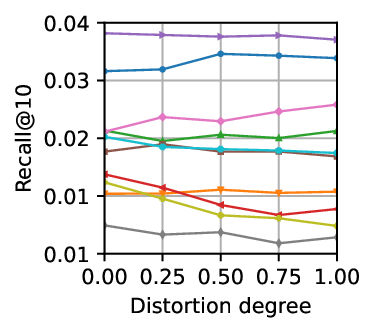}
    \caption{BX}
    \end{subfigure}
    \begin{subfigure}{0.23\textwidth}
    \includegraphics[width=\linewidth]{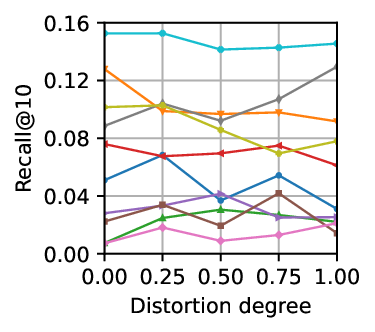}
    \caption{Last.FM}
    \end{subfigure}\\
    \caption{\revision{The results of cold-start experiment with false knowledge, $T$=3. Horizontal axis denotes distortion degree.}}
    \label{fig:cold-start false t3}
\end{figure*}

\begin{figure*}[h]
    \centering
    \begin{subfigure}{\textwidth}
    \includegraphics[width=\linewidth]{figure/legend.eps}
    \end{subfigure}
    \centering
    \begin{subfigure}{0.23\textwidth}
    \includegraphics[width=\linewidth]{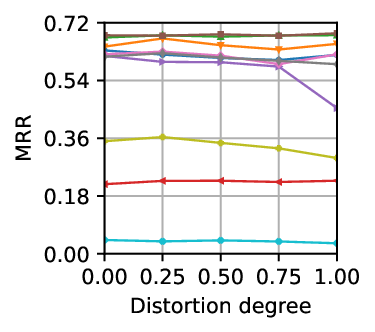}
    \end{subfigure}
    \begin{subfigure}{0.23\textwidth}
    \includegraphics[width=\linewidth]{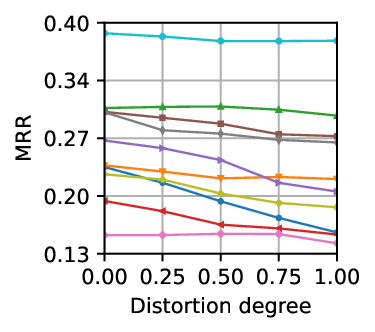}
    \end{subfigure}
    \begin{subfigure}{0.23\textwidth}
    \includegraphics[width=\linewidth]{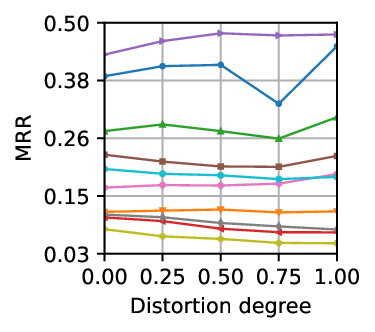}
    \end{subfigure}
    \begin{subfigure}{0.23\textwidth}
    \includegraphics[width=\linewidth]{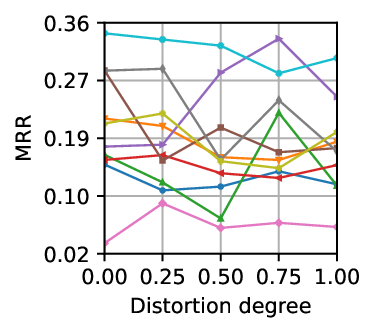}
    \end{subfigure}\\
    \begin{subfigure}{0.23\textwidth}
    \includegraphics[width=\linewidth]{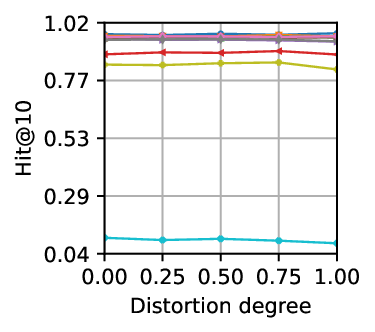}
    \end{subfigure}
    \begin{subfigure}{0.23\textwidth}
    \includegraphics[width=\linewidth]{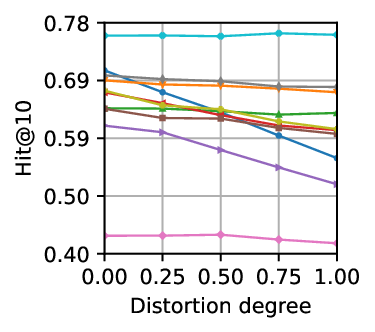}
    \end{subfigure}
    \begin{subfigure}{0.23\textwidth}
    \includegraphics[width=\linewidth]{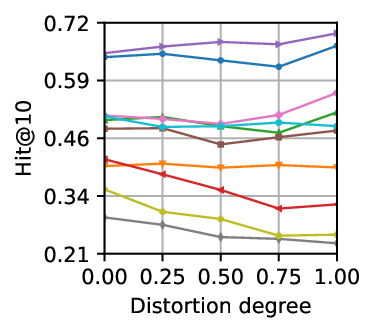}
    \end{subfigure}
    \begin{subfigure}{0.23\textwidth}
    \includegraphics[width=\linewidth]{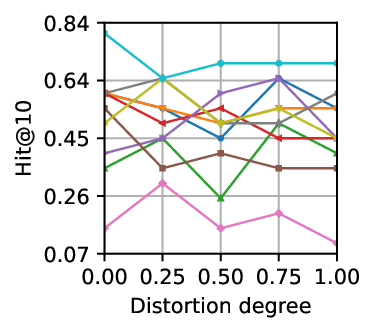}
    \end{subfigure}\\
    \begin{subfigure}{0.23\textwidth}
    \includegraphics[width=\linewidth]{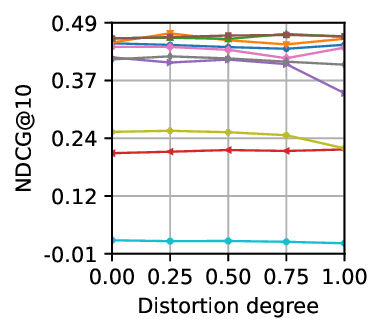}
    \end{subfigure}
    \begin{subfigure}{0.23\textwidth}
    \includegraphics[width=\linewidth]{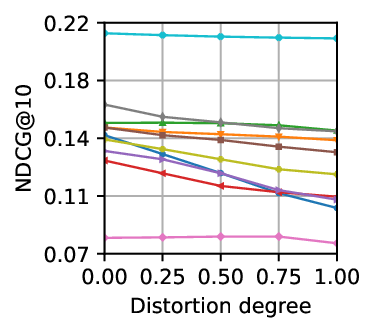}
    \end{subfigure}
    \begin{subfigure}{0.23\textwidth}
    \includegraphics[width=\linewidth]{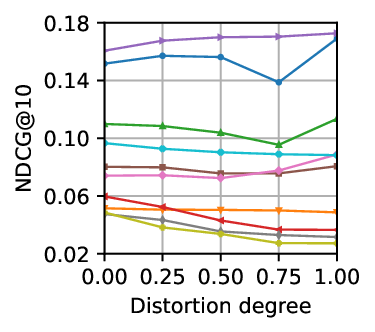}
    \end{subfigure}
    \begin{subfigure}{0.23\textwidth}
    \includegraphics[width=\linewidth]{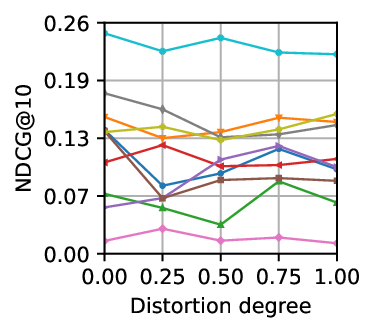}
    \end{subfigure}\\
    \begin{subfigure}{0.23\textwidth}
    \includegraphics[width=\linewidth]{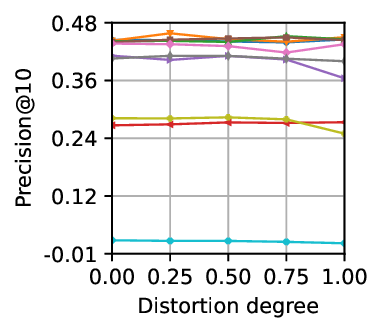}
    \end{subfigure}
    \begin{subfigure}{0.23\textwidth}
    \includegraphics[width=\linewidth]{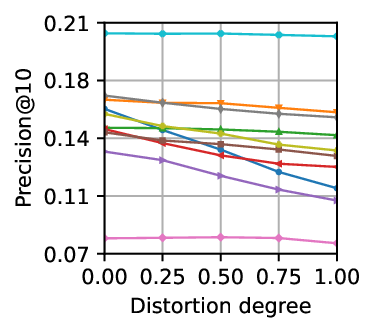}
    \end{subfigure}
    \begin{subfigure}{0.23\textwidth}
    \includegraphics[width=\linewidth]{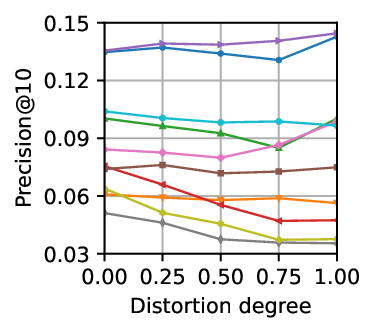}
    \end{subfigure}
    \begin{subfigure}{0.23\textwidth}
    \includegraphics[width=\linewidth]{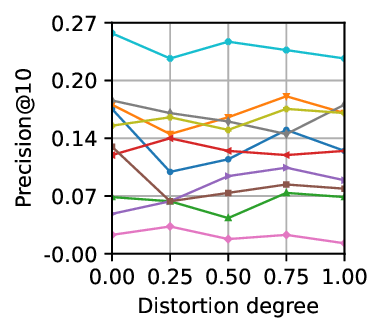}
    \end{subfigure}\\
    \begin{subfigure}{0.23\textwidth}
    \includegraphics[width=\linewidth]{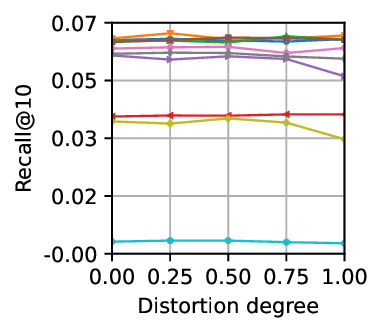}
    \caption{ML-1M}
    \end{subfigure}
    \begin{subfigure}{0.23\textwidth}
    \includegraphics[width=\linewidth]{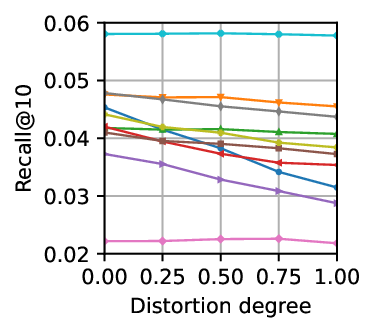}
    \caption{Amazon-Books}
    \end{subfigure}
    \begin{subfigure}{0.23\textwidth}
    \includegraphics[width=\linewidth]{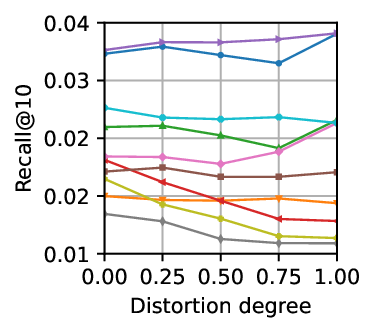}
    \caption{BX}
    \end{subfigure}
    \begin{subfigure}{0.23\textwidth}
    \includegraphics[width=\linewidth]{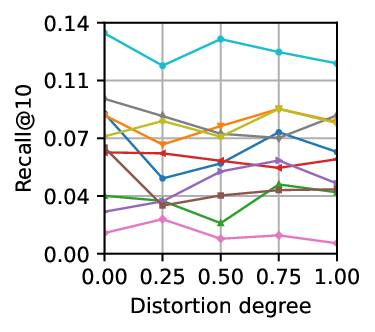}
    \caption{Last.FM}
    \end{subfigure}\\
    \caption{\revision{The results of cold-start experiment with false knowledge, $T$=5. Horizontal axis denotes distortion degree.}}
    \label{fig:cold-start false t5}
\end{figure*}

\begin{figure*}[h]
    \centering
    \begin{subfigure}{\textwidth}
    \includegraphics[width=\linewidth]{figure/legend.eps}
    \end{subfigure}
    \centering
    \begin{subfigure}{0.23\textwidth}
    \includegraphics[width=\linewidth]{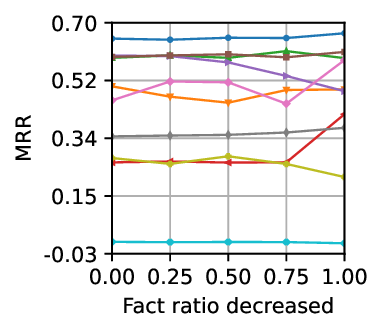}
    \end{subfigure}
    \begin{subfigure}{0.23\textwidth}
    \includegraphics[width=\linewidth]{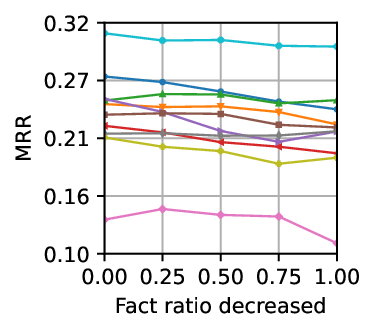}
    \end{subfigure}
    \begin{subfigure}{0.23\textwidth}
    \includegraphics[width=\linewidth]{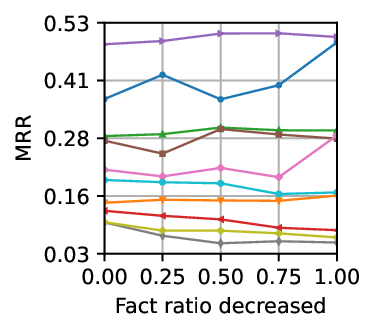}
    \end{subfigure}
    \begin{subfigure}{0.23\textwidth}
    \includegraphics[width=\linewidth]{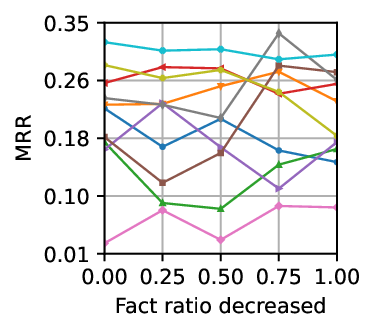}
    \end{subfigure}\\
    \begin{subfigure}{0.23\textwidth}
    \includegraphics[width=\linewidth]{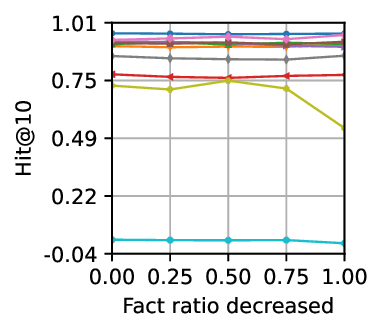}
    \end{subfigure}
    \begin{subfigure}{0.23\textwidth}
    \includegraphics[width=\linewidth]{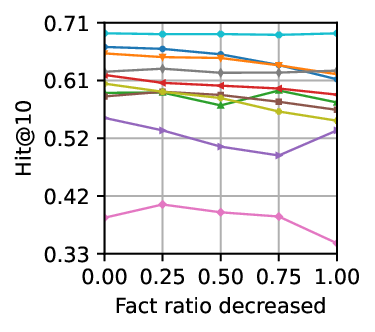}
    \end{subfigure}
    \begin{subfigure}{0.23\textwidth}
    \includegraphics[width=\linewidth]{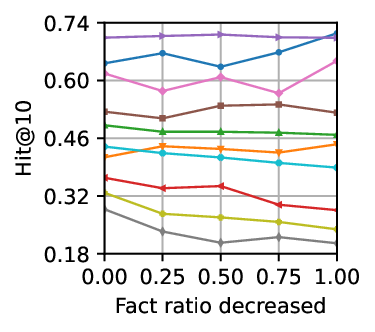}
    \end{subfigure}
    \begin{subfigure}{0.23\textwidth}
    \includegraphics[width=\linewidth]{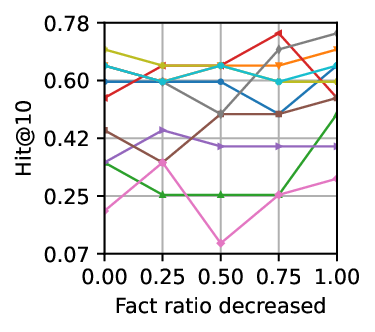}
    \end{subfigure}\\
    \begin{subfigure}{0.23\textwidth}
    \includegraphics[width=\linewidth]{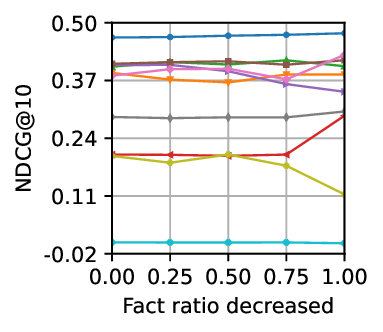}
    \end{subfigure}
    \begin{subfigure}{0.23\textwidth}
    \includegraphics[width=\linewidth]{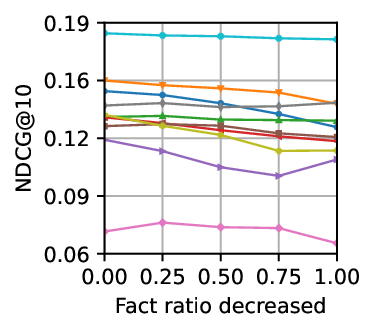}
    \end{subfigure}
    \begin{subfigure}{0.23\textwidth}
    \includegraphics[width=\linewidth]{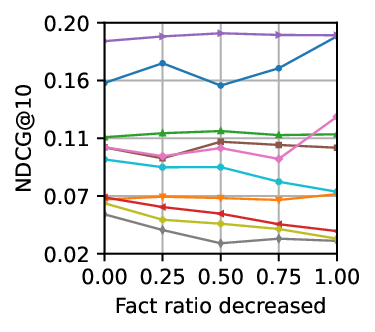}
    \end{subfigure}
    \begin{subfigure}{0.23\textwidth}
    \includegraphics[width=\linewidth]{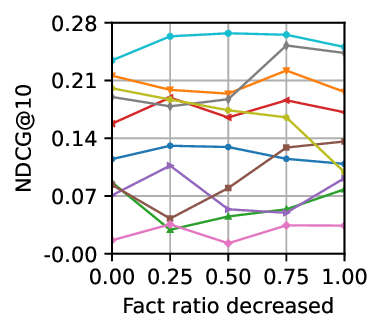}
    \end{subfigure}\\
    \begin{subfigure}{0.23\textwidth}
    \includegraphics[width=\linewidth]{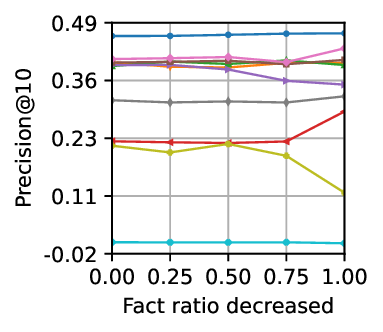}
    \end{subfigure}
    \begin{subfigure}{0.23\textwidth}
    \includegraphics[width=\linewidth]{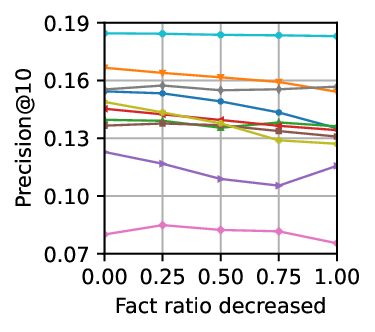}
    \end{subfigure}
    \begin{subfigure}{0.23\textwidth}
    \includegraphics[width=\linewidth]{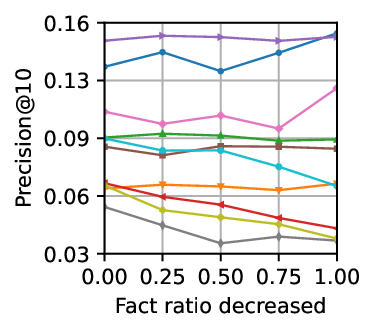}
    \end{subfigure}
    \begin{subfigure}{0.23\textwidth}
    \includegraphics[width=\linewidth]{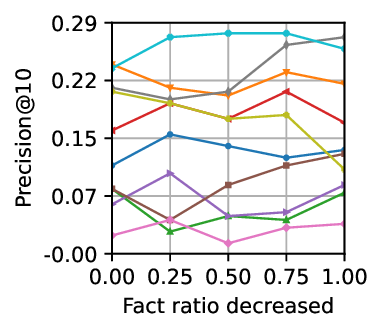}
    \end{subfigure}\\
    \begin{subfigure}{0.23\textwidth}
    \includegraphics[width=\linewidth]{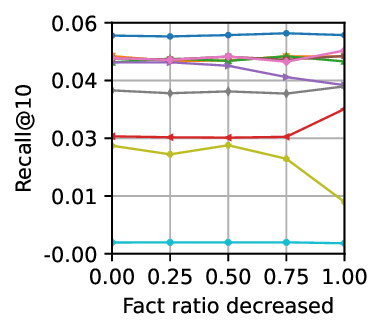}
    \caption{ML-1M}
    \end{subfigure}
    \begin{subfigure}{0.23\textwidth}
    \includegraphics[width=\linewidth]{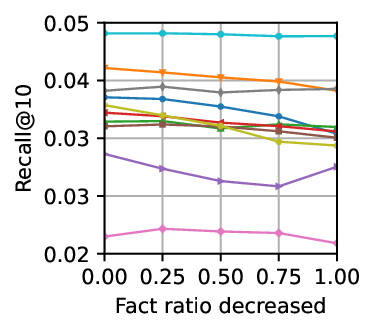}
    \caption{Amazon-Books}
    \end{subfigure}
    \begin{subfigure}{0.23\textwidth}
    \includegraphics[width=\linewidth]{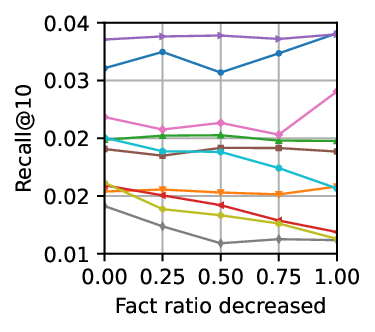}
    \caption{BX}
    \end{subfigure}
    \begin{subfigure}{0.23\textwidth}
    \includegraphics[width=\linewidth]{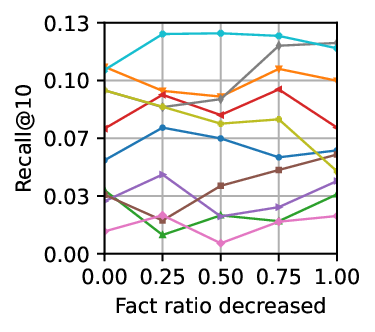}
    \caption{Last.FM}
    \end{subfigure}\\
    \caption{\revision{The results of cold-start experiment with decreased knowledge, $T$=1. Horizontal axis denotes decreased fact ratio.}}
    \label{fig:cold-start decrease t1}
    
\end{figure*}

\begin{figure*}[h]
    \centering
    \begin{subfigure}{\textwidth}
    \includegraphics[width=\linewidth]{figure/legend.eps}
    \end{subfigure}
    \begin{subfigure}{0.23\textwidth}
    \includegraphics[width=\linewidth]{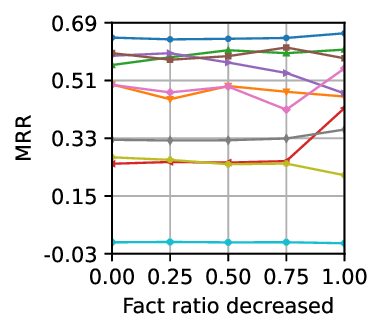}
    \end{subfigure}
    \begin{subfigure}{0.23\textwidth}
    \includegraphics[width=\linewidth]{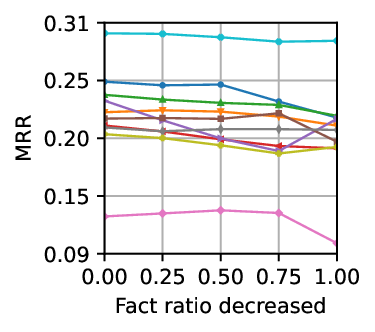}
    \end{subfigure}
    \begin{subfigure}{0.23\textwidth}
    \includegraphics[width=\linewidth]{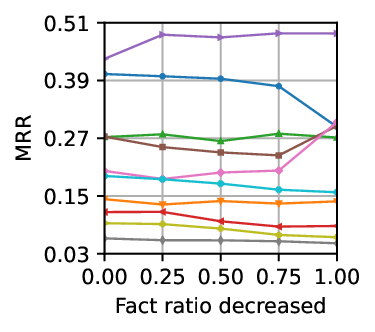}
    \end{subfigure}
    \begin{subfigure}{0.23\textwidth}
    \includegraphics[width=\linewidth]{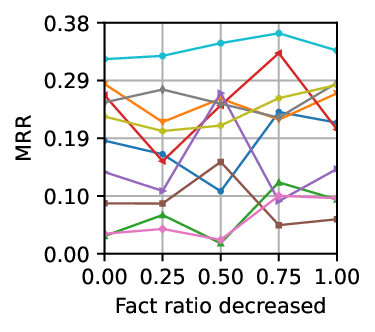}
    \end{subfigure}
    \begin{subfigure}{0.23\textwidth}
    \includegraphics[width=\linewidth]{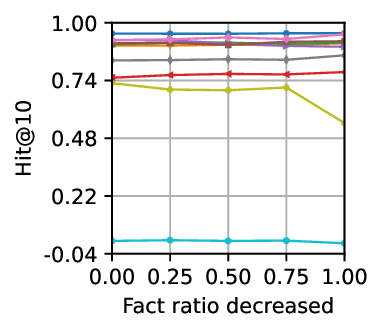}
    \end{subfigure}
    \begin{subfigure}{0.23\textwidth}
    \includegraphics[width=\linewidth]{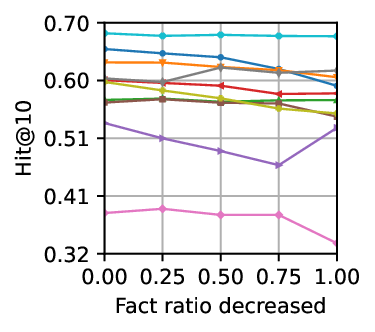}
    \end{subfigure}
    \begin{subfigure}{0.23\textwidth}
    \includegraphics[width=\linewidth]{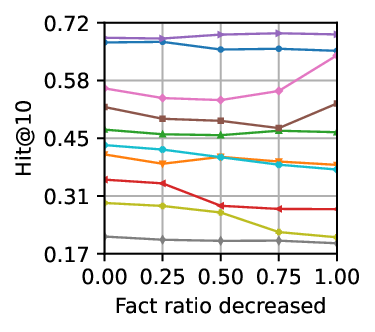}
    \end{subfigure}
    \begin{subfigure}{0.23\textwidth}
    \includegraphics[width=\linewidth]{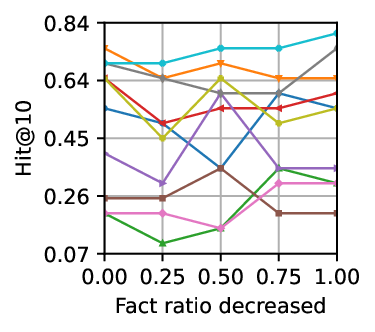}
    \end{subfigure}\\
    \begin{subfigure}{0.23\textwidth}
    \includegraphics[width=\linewidth]{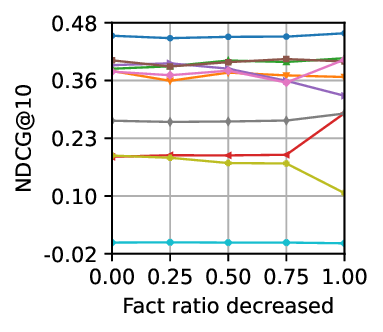}
    \end{subfigure}
    \begin{subfigure}{0.23\textwidth}
    \includegraphics[width=\linewidth]{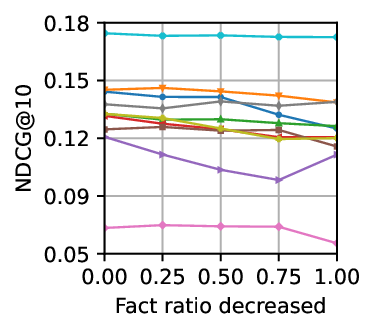}
    \end{subfigure}
    \begin{subfigure}{0.23\textwidth}
    \includegraphics[width=\linewidth]{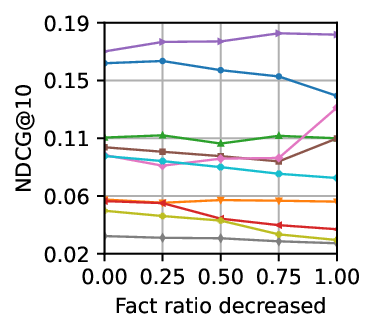}
    \end{subfigure}
    \begin{subfigure}{0.23\textwidth}
    \includegraphics[width=\linewidth]{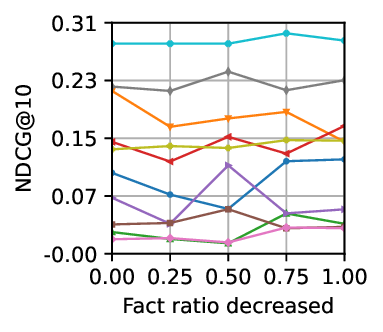}
    \end{subfigure}\\
    \begin{subfigure}{0.23\textwidth}
    \includegraphics[width=\linewidth]{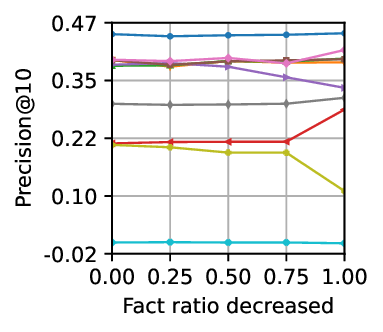}
    \end{subfigure}
    \begin{subfigure}{0.23\textwidth}
    \includegraphics[width=\linewidth]{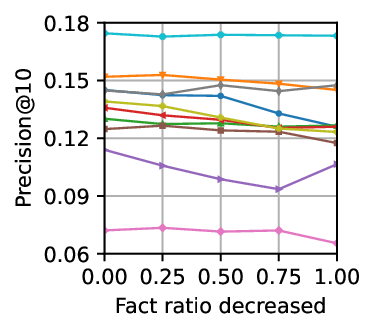}
    \end{subfigure}
    \begin{subfigure}{0.23\textwidth}
    \includegraphics[width=\linewidth]{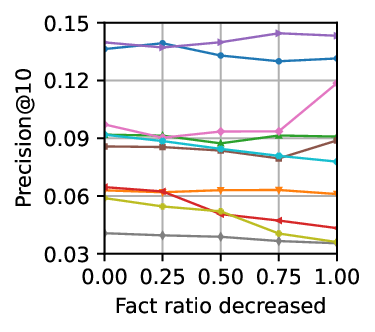}
    \end{subfigure}
    \begin{subfigure}{0.23\textwidth}
    \includegraphics[width=\linewidth]{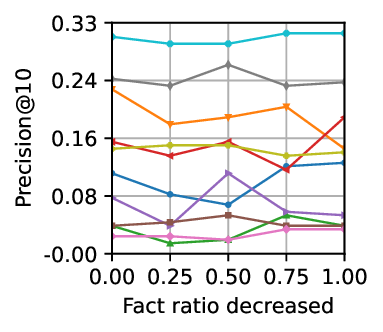}
    \end{subfigure}\\
    \begin{subfigure}{0.23\textwidth}
    \includegraphics[width=\linewidth]{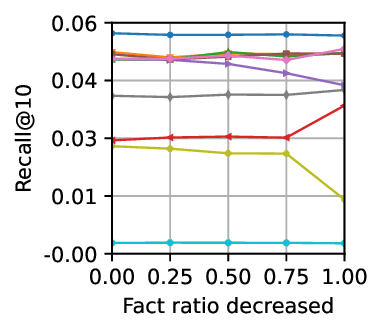}
    \caption{ML-1M}
    \end{subfigure}
    \begin{subfigure}{0.23\textwidth}
    \includegraphics[width=\linewidth]{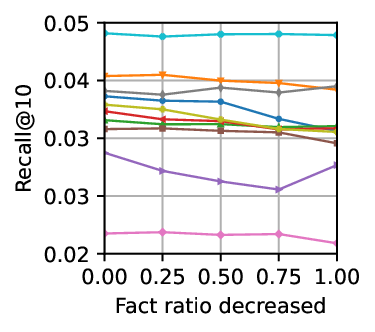}
    \caption{Amazon-Books}
    \end{subfigure}
    \begin{subfigure}{0.23\textwidth}
    \includegraphics[width=\linewidth]{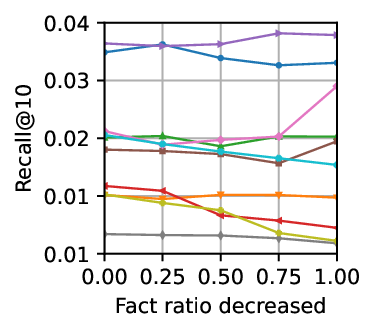}
    \caption{BX}
    \end{subfigure}
    \begin{subfigure}{0.23\textwidth}
    \includegraphics[width=\linewidth]{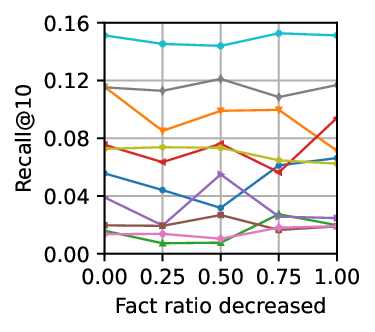}
    \caption{Last.FM}
    \end{subfigure}\\
    \caption{\revision{The results of cold-start experiment with decreased knowledge, $T$=3. Horizontal axis denotes decreased fact ratio.}}
    \label{fig:cold-start decrease t3}
    
\end{figure*}

\begin{figure*}[h]
    \centering
    \begin{subfigure}{\textwidth}
    \includegraphics[width=\linewidth]{figure/legend.eps}
    \end{subfigure}
    \centering
    \begin{subfigure}{0.23\textwidth}
    \includegraphics[width=\linewidth]{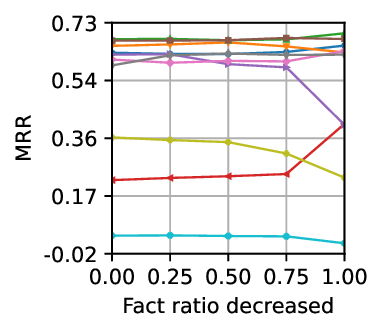}
    \end{subfigure}
    \begin{subfigure}{0.23\textwidth}
    \includegraphics[width=\linewidth]{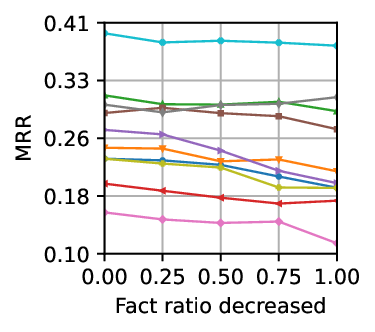}
    \end{subfigure}
    \begin{subfigure}{0.23\textwidth}
    \includegraphics[width=\linewidth]{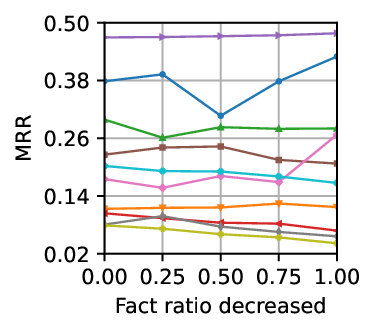}
    \end{subfigure}
    \begin{subfigure}{0.23\textwidth}
    \includegraphics[width=\linewidth]{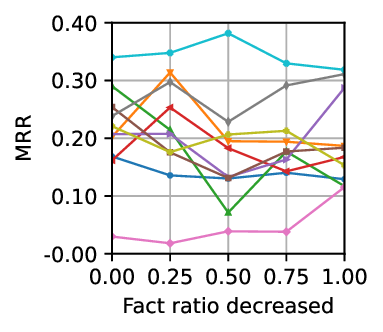}
    \end{subfigure}\\
    \begin{subfigure}{0.23\textwidth}
    \includegraphics[width=\linewidth]{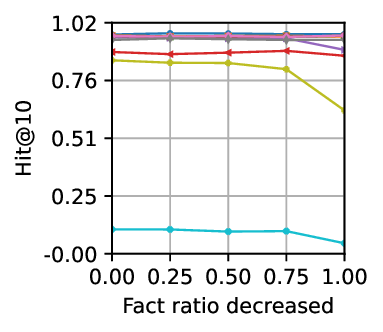}
    \end{subfigure}
    \begin{subfigure}{0.23\textwidth}
    \includegraphics[width=\linewidth]{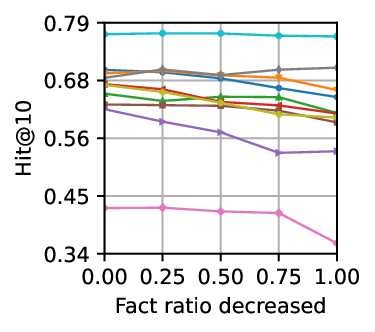}
    \end{subfigure}
    \begin{subfigure}{0.23\textwidth}
    \includegraphics[width=\linewidth]{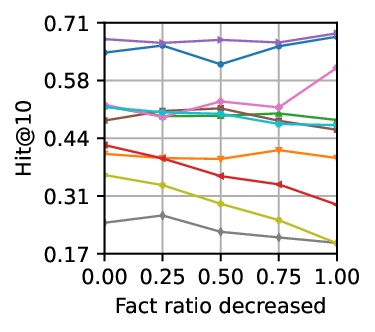}
    \end{subfigure}
    \begin{subfigure}{0.23\textwidth}
    \includegraphics[width=\linewidth]{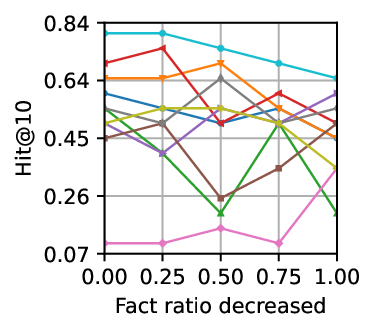}
    \end{subfigure}\\
    \begin{subfigure}{0.23\textwidth}
    \includegraphics[width=\linewidth]{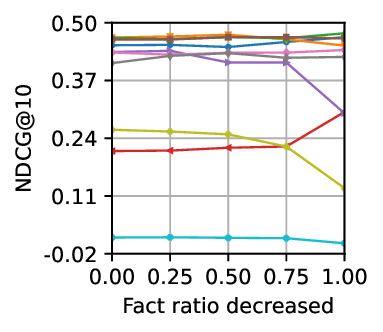}
    \end{subfigure}
    \begin{subfigure}{0.23\textwidth}
    \includegraphics[width=\linewidth]{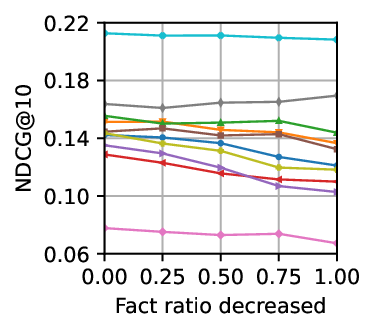}
    \end{subfigure}
    \begin{subfigure}{0.23\textwidth}
    \includegraphics[width=\linewidth]{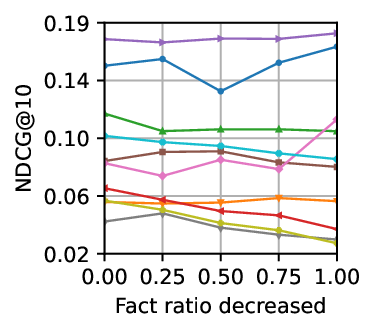}
    \end{subfigure}
    \begin{subfigure}{0.23\textwidth}
    \includegraphics[width=\linewidth]{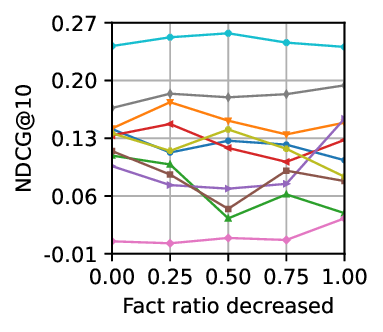}
    \end{subfigure}\\
    \begin{subfigure}{0.23\textwidth}
    \includegraphics[width=\linewidth]{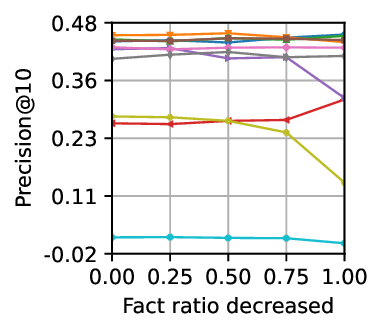}
    \end{subfigure}
    \begin{subfigure}{0.23\textwidth}
    \includegraphics[width=\linewidth]{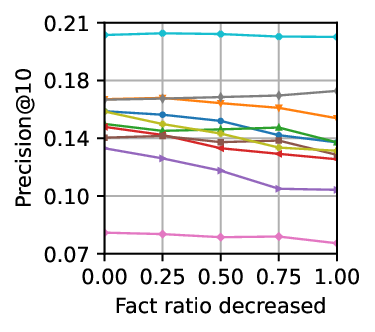}
    \end{subfigure}
    \begin{subfigure}{0.23\textwidth}
    \includegraphics[width=\linewidth]{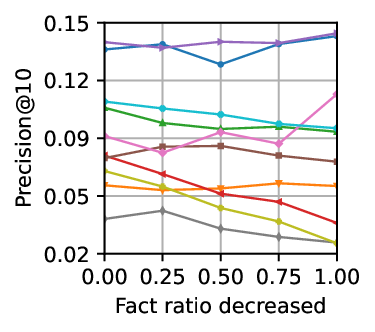}
    \end{subfigure}
    \begin{subfigure}{0.23\textwidth}
    \includegraphics[width=\linewidth]{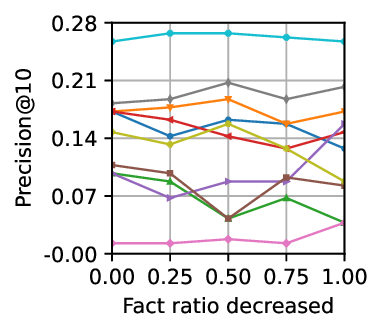}
    \end{subfigure}\\
    \begin{subfigure}{0.23\textwidth}
    \includegraphics[width=\linewidth]{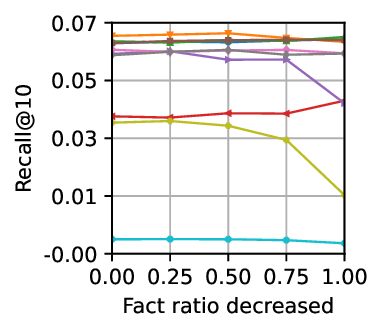}
    \caption{ML-1M}
    \end{subfigure}
    \begin{subfigure}{0.23\textwidth}
    \includegraphics[width=\linewidth]{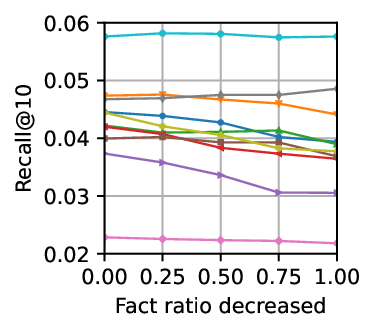}
    \caption{Amazon-Books}
    \end{subfigure}
    \begin{subfigure}{0.23\textwidth}
    \includegraphics[width=\linewidth]{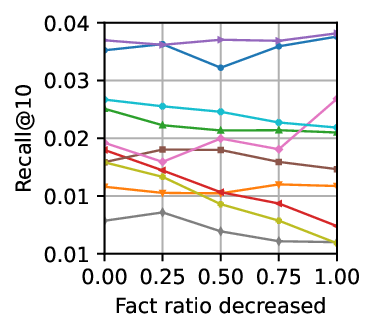}
    \caption{BX}
    \end{subfigure}
    \begin{subfigure}{0.23\textwidth}
    \includegraphics[width=\linewidth]{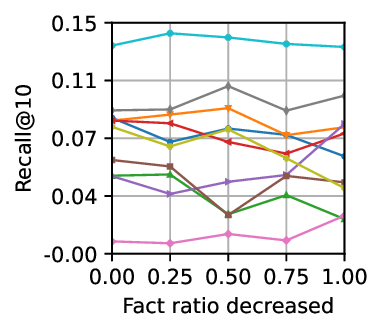}
    \caption{Last.FM}
    \end{subfigure}\\
    \caption{\revision{The results of cold-start experiment with decreased knowledge, $T$=5. Horizontal axis denotes decreased fact ratio.}}
    \label{fig:cold-start decrease t5}
    
\end{figure*}

An important reason to use KGs in RSs is that they may help improve recommendation accuracy for cold-start users with additional information.
We explore the role of KGs under the cold-start setting.

Following the experiment settings of~\cite{chen2013effective, bobadilla2012collaborative, zhang2020alleviating}, we consider users with interactions fewer than a threshold $T$ as cold-start users.
In particular, we first randomly select $10\%$ of users who have interacted with more than 25 items. Then, for each user, we choose $T: T=\{1, 3, 5\}$ of his/her interacted items to put in the training set and put the rest into the test set. In so doing, these sampled users can be considered as cold-start users. 
Then we use random distortion (as in false knowledge experiment) and random deletion (as in decreasing knowledge experiment) to study how MRR of cold-start users varies.

Regarding false knowledge, the results are presented in Figure~\ref{fig:cold-start false t1}-\ref{fig:cold-start false t5}. 
We choose $T =3$ to analyze the results as they show similar trend when $T$ ranges from 1 to 5.
Surprisingly, MRR of most of the models does not necessarily decrease with more knowledge randomly distorted.
Rather, it fluctuates and sometimes even increases with more facts distorted. 
For instance, the highest MRR values of KGIN (with ML-1M), CFKG (BX), KGAT (Last.FM) are all not obtained with the original KG but a distorted one.
And with BX, MRR of RippleNet even increases by $7.3\%$ when all its facts are distorted.
Among all the datasets: 1) in Last.FM, MRR values generally fluctuate the most sharply.
This may be due to the fact that Last.FM has the least number of interactions and accordingly RSs' performance with it is not stable enough for cold-start users.
2) Contrarily, in Amazon-Books, MRR values of quite a few models (RippleNet, KGIN, KGAT and CFKG) show a relatively more obvious downtrend (e.g., CFKG's decreases by $25\%$).
This implies that they do utilize the KG of Amazon-Books more efficiently than that of the other datasets for cold-start users.
3) \revision{In BX, however, MRR values of most models appear to be the least related to distortion degree, with an average change of only \revision{$1.7\%$}} (a similar trend is also observed for random knowledge deletion in BX in Figure~\ref{fig:cold-start decrease t3}). This implies that the KG of BX is exploited the least for cold-start users.


Regarding decreasing knowledge, the results are presented in Figure \ref{fig:cold-start decrease t1}-\ref{fig:cold-start decrease t5}. 
Across all the four datasets, the results (trend of the change in MRR with decreasing ratio) are very similar to those in Figure~\ref{fig:cold-start false t1}-\ref{fig:cold-start false t5} where knowledge is randomly distorted.
Generally, MRR values do not necessarily decrease with more fact decreased.
They fluctuate the most sharply in Last.FM and the least in BX, while showing a relatively more obvious downtrend in Amazon-Books.

We further present KGER values of different models in different datasets and settings (F/D) in Table~\ref{table:merged_kger} (including all the columns labeled cold-start).
The negative KGER values are highlighted with underline.
We can observe that there are multiple negative values, meaning that KGs do not necessarily improve recommendation accuracy for cold-start users, which is also observed for normal users (left side of Table~\ref{table:merged_kger}).
The overall range of KGER is much larger ($-1.89$ to $0.882$) than that for normal users, however.
And given a dataset, model and setting, often $|KGER|$ for cold-start users is larger than that for normal users.
These indicate that KGs do have more influence on cold-start users, which may not necessarily be positive though.
Overall, we provide the core answer to RQ4:

\begin{tcolorbox}
\textbf{Answer to RQ4:} 
With more knowledge randomly distorted or deleted, recommendation accuracy of a KG-based RS for cold-start users does not necessarily decrease.
\end{tcolorbox}

\begin{table*}
    \centering
    \caption{\revision{KGER under false knowledge (F) and decreased knowledge (D), for normal and cold-start users, F/D ratio is 0.5, T = 3.}}
    \resizebox{1\linewidth}{!}{
    \begin{tabular}{c|cc|cc|cc|cc|cc|cc|cc|cc} 
        \Xhline{0.7mm}
         &  \multicolumn{8}{c|}{Normal} & \multicolumn{8}{c}{Cold-Start} \\
         \Xcline{2-17}{0.35mm}
         &  \multicolumn{2}{c|}{ML-1M}&  \multicolumn{2}{c|}{Amazon-Books}&  \multicolumn{2}{c|}{BX}    & \multicolumn{2}{c|}{Last.FM}&  \multicolumn{2}{c|}{ML-1M}&  \multicolumn{2}{c|}{Amazon-Books}&  \multicolumn{2}{c|}{BX}    & \multicolumn{2}{c}{Last.FM}\\ 
         \Xcline{2-17}{0.35mm}
         &  F&  D&  F&  D&F&  D&  F&  D&  F&  D&  F&  D&  F&  D&  F&  D\\ 
         \Xhline{0.35mm}
         CFKG       &0.002 &\underline{-0.002} & {0.156} &0.026 &\underline{-0.016} &0.006 &0.118 &0.027&0.056 &0.012 &0.236 &0.022 &\underline{-0.116} &0.048 &0.410 & {0.882}   \\ 
         CKE        &0.002 &0.006 &0.014 &0.005 &\underline{-0.008} &\underline{-0.004} &0.046 &0.007&0.026 &0.020 &0.004 &\underline{-0.006} &\underline{-0.004} &0.054 &0.116 &0.172   \\ 
         KGCN       &0.003 &\underline{-0.007} &\underline{-0.002} &0.011 &0.000 &0.025 &0.114 & {0.047}&0.008 &\underline{-0.166} &\underline{-0.004} &0.062 &\underline{-0.060} &0.062 &\underline{-1.280} &0.744   \\ 
         KGIN       & {0.017} &0.002 &0.060 &0.018 & {0.136} & {0.077} & {0.122} &\underline{-0.051}&0.032 &\underline{-0.030} &0.172 &0.122 & {0.482} & {0.330} &\underline{-0.320} &0.128   \\ 
         RippleNet  &0.012 & {0.014} &0.106 & {0.056} &0.036 &\underline{-0.013} &0.010 &0.008&0.114 & {0.072} & {0.246} & {0.300} &\underline{-0.140} &\underline{-0.206} &0.428 &\underline{-1.890}  \\ 
         KGNNLS     &0.002 &\underline{-0.002} &0.020 &0.015 &0.040 &\underline{-0.006} &0.064 &\underline{-0.003}&\underline{-0.016} &0.032 &0.056 &0.002 &0.080 &0.242 & {1.012} &\underline{-1.588}  \\ 
         KTUP       &0.008 &\underline{-0.006} &0.010 &0.013 &0.004 &\underline{-0.013} &\underline{-0.012} &\underline{-0.008}&0.116 &0.022 &\underline{-0.056} &\underline{-0.090} &\underline{-0.040} &0.034 &0.988 &0.556   \\ 
         KGAT       &\underline{-0.005} &\underline{-0.006} &0.068 &0.009 &0.084 &\underline{-0.053} &0.058 &\underline{-0.003}& {0.166} &0.004 &0.106 &0.014 &0.142 &0.130 &0.128 &0.016   \\ 
         KGRec       &\underline{-0.020}&0.080&0.096&0.080&0.079&0.140&0.128&0.139&0.097  &0.159 &0.213 &0.102 &0.286 &0.241 &0.292 &0.131 \\ 
         DiffKG       &\underline{-0.008}&\underline{-0.024}&\underline{-0.004}&\underline{-0.005}&0.112 &0.040&0.049 &\underline{-0.004}&0.129 &0.051 &0.009 &0.025 &0.185 &0.167 &\underline{-0.185}&\underline{-0.163}\\ 
         \Xhline{0.7mm}
    \end{tabular}
    }
    \label{table:merged_kger}
\end{table*}

\subsection{Comparison between Authenticity and Amount of Knowledge}

\begin{figure}[h]
    \centering
    \includegraphics[width=0.5\linewidth]{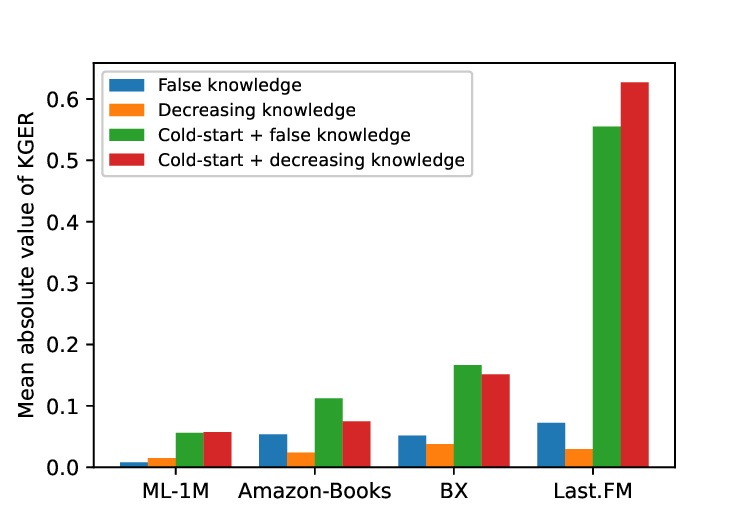}
    \caption{\revision{Comparison of mean absolute values of KGER across datasets and experiment settings.}}
    \label{fig:histogram}
\end{figure}

The previous experimental results show either randomly distorted or decreased knowledge of a KG has a limited effect on the recommendation accuracy of a KG-based RS.
Here, we analyze which of authenticity and amount of knowledge influences more, to answer RQ5.
Given a same ratio of facts, we compare KGER values under random distortion and that under random decrease.
To be consistent with the results in the previous sections, we still choose the distortion/decrease ratio to be $0.5$ and $T =3$ for cold-start setting. 

The results are presented in Figure \ref{fig:histogram}. 
\revision{Observe that for normal users, $|KGER|$ under random distortion is $0.079$ higher than that under random decrease.
However, for cold-start users, $|KGER|$ under random decrease is $0.021$ higher than that under random distortion.
We conjecture the reason may be that after random distortion, some facts may still be meaningful and helpful by accident and the learning model can capture them, while after random deletion, the affected facts turn useless.
For normal settings, RSs are able to find useful knowledge after random distortion as there are sufficient number of interactions, while for cold-start settings, RSs are more likely to be tricked by false knowledge.}
Overall, we provide the core answer to RQ5:

\begin{tcolorbox}
\textbf{Answer to RQ5:} 
For normal users, knowledge authenticity has a slightly stronger influence on KG utilization efficiency of KG-based RSs. For cold-start users, knowledge amount has a slightly stronger influence.
\end{tcolorbox}

\section{Discussion and Insights}

\revision{Among all the tested KG-based RSs, RippleNet and KGRec usually present a positive KGER across datasets and also relatively higher KGER for cold-start users.
Both of them use attention mechanism to explore potential recommendation paths and filter useless or incorrect facts in KGs, assigning higher weights to neighboring relations and entities that might interest users, thus generating more efficient and interpretable recommendation paths.
This may explain their high utilization efficiency of KGs, and it also enlightens us to find ways to distinguish knowledge in a KG and strengthen the parts which have a stronger connection to user preferences.}
CFKG combines KGs and user-item interactions to create a new graph and generate recommendation results based on it. This may result in low KGER if the KG is much smaller than the user-item interaction graph, which is often the case. 
We thus suggest to treat the KG and user-item interactions separately instead of simply combining them together.
KGCN treats the KG as an undirected graph, which is unreasonable and may lead to the loss of useful knowledge. For example, <\textit{Dunkirk, belongs\_to, Historical Film}> is completely different from <\textit{Historical Film, belongs\_to, Dunkirk}> which is incorrect. 
We suggest to treat the KG as a directed graph and make full use of edge directions.

Our proposed framework and KGER metric can be generalized to analyze how much other types of (structured) side information (e.g., social network) are exploited in supporting a system's decision performance.
Moreover, our findings may also be of great significance for privacy protection in the domain of RSs.
For example, we may randomly add fake content to a KG to ``hide'' the real knowledge.

\section{Conclusion}
In this paper, we propose a novel evaluation framework \framework to quantitatively assess whether (and how much) KGs improve the recommendation accuracy of KG-based RSs.
We propose a novel metric KGER to measure KG utilization efficiency in KG-based RSs.
And we design a series of experiments to systematically investigate the role of KGs, namely, how knowledge existence, authenticity, and amount influence recommendation accuracy for both normal and cold-start users.
The extensive experiments on four real-world datasets and 10 widely recognized KG-based RSs present that, counter-intuitively, whether for normal or cold-start users, the decrease of authentic knowledge in a KG does not necessarily decrease recommendation accuracy of the KG-based RSs.
And how KGs influence recommendation accuracy varies depending on datasets and KG-based RSs.
Relatively speaking, KGs tend to
have more influence (not necessarily positive) on recommendation accuracy for cold-start users, or sparser datasets.
Moreover, while for normal users, knowledge authenticity has a slightly stronger influence on KG utilization efficiency of KG-based RSs, for cold-start users, knowledge amount has a slightly stronger influence.

\begin{acks}
This work is supported by National Natural Science Foundation of China (NSFC) under grant 62106223 and Fundamental Research Funds for the Central Universities+226-2024-00048.
\end{acks}

\newpage

\bibliographystyle{ACM-Reference-Format}
\bibliography{reference}

\clearpage
\appendix
\renewcommand\thefigure{\Alph{section}\arabic{figure}}
\renewcommand\thetable{\Alph{section}\arabic{table}}
\setcounter{figure}{0}
\setcounter{table}{0}
\section*{Appendix}

\section{Additional Results of KGER}\label{appendix:comparison}
Additional experimental results of KGER are presented in Table \ref{table:normal comparison 0.25}-\ref{table:cold-start comparison 1.0 5}.
\begin{table}[H]
    \centering
    \caption{\revision{Under the normal settings, comparison of KGER under false knowledge and decreased knowledge, distortion (F) / decreased (D) ratio is 0.25.}}
    \begin{tabular}{c|cc|cc|cc|cc} 
        \Xhline{0.7mm}
         &  \multicolumn{2}{c|}{ML-1M}&  \multicolumn{2}{c|}{Amazon-Books}&  \multicolumn{2}{c|}{BX}&  \multicolumn{2}{c}{Last.FM}\\ 
         \Xcline{2-9}{0.35mm}
         &  F&  D&  F&  D&  F&  D&  F&  D\\ 
         \Xhline{0.35mm}
            CFKG &-0.002& 0.007& 0.149& 0.021& -0.007& 0.009& 0.136& 0.002\\
            CKE &0.001& -0.002& 0.005& 0.005& -0.055& 0.089& 0.027& 0.044\\
            KGCN &0.014& 0.004& -0.020& 0.038& -0.028& 0.003& 0.175& 0.227\\
            KGIN &0.021& -0.010& 0.068& 0.016& 0.181& 0.081& 0.209& 0.037\\
            RippleNet &0.000& 0.000& 0.122& 0.077& 0.025& -0.004& 0.004& 0.028\\
            KGNNLS &-0.001& -0.017& -0.010& -0.001& 0.065& 0.045& 0.021& 0.024\\
            KTUP &0.003& -0.015& -0.012& 0.003& 0.049& -0.019& -0.059& 0.004\\
            KGAT &-0.011& 0.005& 0.074& -0.001& 0.157& -0.115& 0.104& -0.024\\
            KGRec &0.026 &0.004 &0.187 &0.113 &0.104 &0.225 &0.280&0.109 \\
            DiffKG &-0.023&-0.063&0.029 &0.008 &0.140  &0.041 &0.085 &-0.012\\
         \Xhline{0.7mm}
    \end{tabular}
    \label{table:normal comparison 0.25}
\end{table}
\begin{table}[H]
    \centering
    \caption{\revision{Under the normal settings, comparison of KGER under false knowledge and decreased knowledge, distortion (F) / decreased (D) ratio is 0.75.}}
    \begin{tabular}{c|cc|cc|cc|cc} 
        \Xhline{0.7mm}
         &  \multicolumn{2}{c|}{ML-1M}&  \multicolumn{2}{c|}{Amazon-Books}&  \multicolumn{2}{c|}{BX}&  \multicolumn{2}{c}{Last.FM}\\ 
         \Xcline{2-9}{0.35mm}
         &  F&  D&  F&  D&  F&  D&  F&  D\\ 
         \Xhline{0.35mm}
            CFKG &0.002& 0.001& 0.140& 0.031& 0.006& -0.004& 0.093& 0.008\\
            CKE &-0.003& 0.004& 0.009& 0.007& 0.009& 0.006& 0.002& 0.012\\
            KGCN &-0.001& -0.002& -0.013& 0.010& 0.038& 0.036& 0.045& 0.042\\
            KGIN &0.018& 0.010& 0.060& 0.021& 0.126& 0.082& 0.094& -0.045\\
            RippleNet &0.007& 0.021& 0.077& 0.030& 0.055& -0.010& -0.010& 0.019\\
            KGNNLS &0.004& -0.005& 0.012& 0.012& 0.051& 0.015& 0.011& 0.013\\
            KTUP &0.011& 0.005& 0.004& 0.003& 0.011& -0.004& 0.002& -0.012\\
            KGAT &0.004& -0.001& 0.071& 0.001& 0.078& 0.034& 0.031& -0.010\\
            KGRec &-0.022&0.092&0.050&0.077&0.104&0.167&0.082&0.183\\
            DiffKG &0.000&-0.009&0.014  &-0.002&0.096 &0.042 &0.016 &0.010\\
         \Xhline{0.7mm}
    \end{tabular}
    \label{table:normal comparison 0.75}
\end{table}
\begin{table}[H]
    \centering
    \caption{\revision{Under the normal settings, comparison of KGER under false knowledge and decreased knowledge, distortion (F) / decreased (D) ratio is 1.0.}}
    \begin{tabular}{c|cc|cc|cc|cc} 
        \Xhline{0.7mm}
         &  \multicolumn{2}{c|}{ML-1M}&  \multicolumn{2}{c|}{Amazon-Books}&  \multicolumn{2}{c|}{BX}&  \multicolumn{2}{c}{Last.FM}\\ 
         \Xcline{2-9}{0.35mm}
         &  F&  D&  F&  D&  F&  D&  F&  D\\ 
         \Xhline{0.35mm}
            CFKG &0.002& 0.000& 0.130& 0.042& -0.005& -0.005& 0.068& 0.023\\
            CKE &0.002& 0.004& 0.011& 0.014& 0.002& 0.014& 0.009& 0.009\\
            KGCN &0.001& -0.006& -0.000& 0.006& 0.010& 0.057& 0.045& 0.035\\
            KGIN &0.033& 0.129& 0.054& 0.017& 0.102& 0.088& 0.070& -0.060\\
            RippleNet &0.016& 0.063& 0.077& 0.034& 0.056& -0.005& 0.010& -0.026\\
            KGNNLS &0.001& -0.006& 0.007& 0.016& 0.038& 0.019& 0.017& 0.016\\
            KTUP &0.002& -0.000& 0.001& -0.009& -0.002& 0.009& 0.010& -0.042\\
            KGAT &0.009& -0.001& 0.048& -0.001& 0.054& 0.030& 0.043& -0.020\\
            KGRec &0.044&0.462&0.039&0.085&0.068&0.242&0.114&0.194\\
            DiffKG &-0.004&-0.015&0.007&0.005&0.077&0.052&0.025&0.008\\
         \Xhline{0.7mm}
    \end{tabular}
    \label{table:normal comparison 1.0}
\end{table}
\begin{table}[H]
    \centering
    \caption{\revision{Under the cold-start settings, comparison of KGER under false knowledge and decreased knowledge, distortion (F) / decreased (D) ratio is 0.25, T=1.}}
    \begin{tabular}{c|cc|cc|cc|cc} 
        \Xhline{0.7mm}
         &  \multicolumn{2}{c|}{ML-1M}&  \multicolumn{2}{c|}{Amazon-Books}&  \multicolumn{2}{c|}{BX}&  \multicolumn{2}{c}{Last.FM}\\ 
         \Xcline{2-9}{0.35mm}
         &  F&  D&  F&  D&  F&  D&  F&  D\\ 
         \Xhline{0.35mm}
            CFKG &-0.013& 0.021& 0.282& 0.079& -0.174& -0.568& 1.503& 0.995\\
            CKE &-0.057& 0.262& 0.069& 0.047& 0.486& -0.178& -0.629& -0.022\\
            KGCN &0.040& -0.052& -0.025& -0.100& -0.058& -0.058& 0.715& 2.016\\
            KGIN &-0.015& -0.043& 0.200& 0.112& 0.414& 0.338& 0.834& -0.355\\
            RippleNet &0.055& 0.010& 0.174& 0.201& -0.505& -0.057& 2.308& -1.637\\
            KGNNLS &-0.068& -0.031& 0.022& -0.023& 0.495& 0.402& 0.828& 1.454\\
            KTUP &-0.097& -0.525& -0.157& -0.305& -0.061& 0.269& -5.074& -6.717\\
            KGAT &0.125& -0.031& 0.113& -0.008& 0.617& 1.123& 0.682& 0.161\\
            KGRec &-0.167&0.271&0.155&0.167&0.542&0.703&-0.523&0.263\\
            DiffKG &-0.516&0.435 &0.092 &0.088 &-0.041&0.100&0.016&0.154 \\
         \Xhline{0.7mm}
    \end{tabular}
    \label{table:cold-start comparison 0.25 1}
\end{table}
\begin{table}[H]
    \centering
    \caption{\revision{Under the cold-start settings, comparison of KGER under false knowledge and decreased knowledge, distortion (F) / decreased (D) ratio is 0.5, T=1.}}
    \begin{tabular}{c|cc|cc|cc|cc} 
        \Xhline{0.7mm}
         &  \multicolumn{2}{c|}{ML-1M}&  \multicolumn{2}{c|}{Amazon-Books}&  \multicolumn{2}{c|}{BX}&  \multicolumn{2}{c}{Last.FM}\\ 
         \Xcline{2-9}{0.35mm}
         &  F&  D&  F&  D&  F&  D&  F&  D\\ 
         \Xhline{0.35mm}
            CFKG &0.030& -0.009& 0.203& 0.106& 0.249& 0.004& 0.286& 0.136\\
            CKE &0.153& 0.208& -0.021& 0.018& 0.224& -0.069& -0.054& -0.235\\
            KGCN &0.035& -0.000& -0.019& -0.047& 0.028& -0.127& 1.014& 1.105\\
            KGIN &-0.031& 0.004& 0.166& 0.140& 0.383& 0.289& 0.139& -0.163\\
            RippleNet &0.166& 0.072& 0.259& 0.247& -0.256& -0.095& 0.529& -0.041\\
            KGNNLS &-0.024& -0.028& 0.007& -0.007& 0.080& -0.179& -1.012& 0.253\\
            KTUP &-0.019& -0.251& -0.095& -0.069& -0.077& -0.035& 1.465& -0.348\\
            KGAT &0.030& -0.028& 0.080& 0.018& 0.514& 0.879& 0.268& 0.241\\
            KGRec &-0.281&-0.041&0.155&0.123&0.416&0.354&0.208&0.049\\
            DiffKG &-0.335&0.089 &0.034 &0.041 &0.038  &0.074 &0.018 &0.064 \\
         \Xhline{0.7mm}
    \end{tabular}
    \label{table:cold-start comparison 0.5 1}
\end{table}
\begin{table}[H]
    \centering
    \caption{\revision{Under the cold-start settings, comparison of KGER under false knowledge and decreased knowledge, distortion (F) / decreased (D) ratio is 0.75, T=1.}}
    \begin{tabular}{c|cc|cc|cc|cc} 
        \Xhline{0.7mm}
         &  \multicolumn{2}{c|}{ML-1M}&  \multicolumn{2}{c|}{Amazon-Books}&  \multicolumn{2}{c|}{BX}&  \multicolumn{2}{c}{Last.FM}\\ 
         \Xcline{2-9}{0.35mm}
         &  F&  D&  F&  D&  F&  D&  F&  D\\ 
         \Xhline{0.35mm}
            CFKG &0.002& -0.004& 0.243& 0.118& 0.006& -0.108& 0.428& 0.363\\
            CKE &0.047& 0.031& 0.036& 0.042& 0.135& -0.037& 0.121& -0.278\\
            KGCN &0.005& -0.048& 0.049& 0.014& 0.033& -0.058& 0.425& 0.249\\
            KGIN &0.001& 0.000& 0.176& 0.119& 0.278& 0.382& 0.132& 0.079\\
            RippleNet &0.097& 0.144& 0.265& 0.220& -0.171& -0.065& 0.020& 0.459\\
            KGNNLS &0.038& 0.003& 0.072& 0.055& 0.119& -0.064& 0.827& -0.756\\
            KTUP &-0.031& 0.031& 0.006& -0.030& -0.052& 0.099& 0.098& -2.526\\
            KGAT &0.049& -0.048& 0.074& 0.011& 0.294& 0.529& 0.443& -0.524\\
            KGRec &-0.096&0.091&0.212&0.158&0.290&0.310&-0.355&0.181\\
            DiffKG &-0.090&0.100&0.045 &0.051 &0.038  &0.208 &-0.025&0.104 \\
         \Xhline{0.7mm}
    \end{tabular}
    \label{table:cold-start comparison 0.75 1}
\end{table}
\begin{table}[H]
    \centering
    \caption{\revision{Under the cold-start settings, comparison of KGER under false knowledge and decreased knowledge, distortion (F) / decreased (D) ratio is 1.0, T=1.}}
    \begin{tabular}{c|cc|cc|cc|cc} 
        \Xhline{0.7mm}
         &  \multicolumn{2}{c|}{ML-1M}&  \multicolumn{2}{c|}{Amazon-Books}&  \multicolumn{2}{c|}{BX}&  \multicolumn{2}{c}{Last.FM}\\ 
         \Xcline{2-9}{0.35mm}
         &  F&  D&  F&  D&  F&  D&  F&  D\\ 
         \Xhline{0.35mm}
            CFKG &0.013& -0.026& 0.231& 0.116& -0.028& -0.332& 0.188& 0.348\\
            CKE &0.027& 0.020& 0.017& 0.080& 0.084& -0.107& -0.082& -0.022\\
            KGCN &0.003& 0.001& 0.010& -0.001& -0.040& -0.042& -0.064& 0.057\\
            KGIN &-0.054& -0.589& 0.159& 0.118& 0.247& 0.328& -0.107& 0.005\\
            RippleNet &0.203& 0.189& 0.229& 0.125& -0.128& -0.032& 0.572& -0.066\\
            KGNNLS &-0.017& -0.027& 0.059& 0.052& 0.155& -0.015& -0.004& -0.515\\
            KTUP &-0.150& -0.280& 0.038& 0.169& -0.067& -0.341& 0.269& -1.820\\
            KGAT &0.050& -0.080& 0.068& -0.011& 0.317& 0.423& 0.090& -0.111\\
            KGRec &-0.118&0.220&0.144&0.091&0.278& 0.320&-0.086&0.358\\
            DiffKG &-0.310&0.610 &0.044&0.041&0.028&0.138&0.017&0.056\\
         \Xhline{0.7mm}
    \end{tabular}
    \label{table:cold-start comparison 1.0 1}
\end{table}
\begin{table}[H]
    \centering
    \caption{\revision{Under the cold-start settings, comparison of KGER under false knowledge and decreased knowledge, distortion (F) / decreased (D) ratio is 0.25, T=3.}}
    \begin{tabular}{c|cc|cc|cc|cc} 
        \Xhline{0.7mm}
         &  \multicolumn{2}{c|}{ML-1M}&  \multicolumn{2}{c|}{Amazon-Books}&  \multicolumn{2}{c|}{BX}&  \multicolumn{2}{c}{Last.FM}\\ 
         \Xcline{2-9}{0.35mm}
         &  F&  D&  F&  D&  F&  D&  F&  D\\ 
         \Xhline{0.35mm}
            CFKG &0.040& 0.034& 0.159& 0.053& -0.527& 0.047& 0.062& 0.472\\
            CKE &0.232& 0.371& 0.040& -0.032& -0.083& 0.312& 0.568& 0.890\\
            KGCN &-0.106& -0.173& -0.055& 0.074& 0.063& -0.086& -7.153& -4.312\\
            KGIN &0.037& -0.087& 0.231& 0.105& 0.646& -0.012& 1.340& 1.647\\
            RippleNet &0.171& -0.054& 0.222& 0.312& -0.315& -0.463& 0.961& 0.905\\
            KGNNLS &0.005& 0.139& 0.080& -0.010& -0.185& 0.321& 0.353& 0.016\\
            KTUP &0.251& 0.191& -0.178& -0.089& -0.300& 0.333& -5.927& -0.882\\
            KGAT &0.058& 0.025& 0.096& 0.069& 0.314& 0.245& 0.425& -0.337\\
            KGRec &0.097&0.117&0.213&0.072&0.286&0.081&0.292&0.418\\
            DiffKG &0.682 &-0.785&0.015 &0.007 &0.226 &0.144 &0.453 &0.067 \\
         \Xhline{0.7mm}
    \end{tabular}
    \label{table:cold-start comparison 0.25 3}
\end{table}
\begin{table}[H]
    \centering
    \caption{\revision{Under the cold-start settings, comparison of KGER under false knowledge and decreased knowledge, distortion (F) / decreased (D) ratio is 0.75, T=3.}}
    \begin{tabular}{c|cc|cc|cc|cc} 
        \Xhline{0.7mm}
         &  \multicolumn{2}{c|}{ML-1M}&  \multicolumn{2}{c|}{Amazon-Books}&  \multicolumn{2}{c|}{BX}&  \multicolumn{2}{c}{Last.FM}\\ 
         \Xcline{2-9}{0.35mm}
         &  F&  D&  F&  D&  F&  D&  F&  D\\ 
         \Xhline{0.35mm}
            CFKG &0.036& 0.002& 0.262& 0.098& -0.172& 0.084& 0.404& -0.333\\
            CKE &0.085& 0.061& 0.035& 0.024& -0.054& 0.086& 0.254& 0.277\\
            KGCN &-0.009& -0.088& 0.008& 0.052& -0.024& -0.035& -1.553& -3.671\\
            KGIN &0.006& -0.044& 0.157& 0.121& 0.436& 0.347& -0.007& -0.349\\
            RippleNet &0.141& 0.122& 0.225& 0.268& -0.102& -0.163& -1.377& 0.465\\
            KGNNLS &-0.029& -0.039& 0.037& -0.030& -0.038& 0.191& -1.463& 0.557\\
            KTUP &0.197& 0.208& -0.013& -0.034& -0.107& -0.005& 0.402& -2.325\\
            KGAT &0.048& -0.017& 0.086& 0.010& 0.274& 0.131& 0.150& 0.132\\
            KGRec &-0.069&0.096&0.200&0.118&0.317&0.350&0.554&-0.176\\
            DiffKG &0.133 &-0.063&0.016 &0.035 &0.134 &0.200&-0.084&-0.176\\
         \Xhline{0.7mm}
    \end{tabular}
    \label{table:cold-start comparison 0.75 3}
\end{table}
\begin{table}[H]
    \centering
    \caption{\revision{Under the cold-start settings, comparison of KGER under false knowledge and decreased knowledge, distortion (F) / decreased (D) ratio is 1.0, T=3.}}
    \begin{tabular}{c|cc|cc|cc|cc} 
        \Xhline{0.7mm}
         &  \multicolumn{2}{c|}{ML-1M}&  \multicolumn{2}{c|}{Amazon-Books}&  \multicolumn{2}{c|}{BX}&  \multicolumn{2}{c}{Last.FM}\\ 
         \Xcline{2-9}{0.35mm}
         &  F&  D&  F&  D&  F&  D&  F&  D\\ 
         \Xhline{0.35mm}
            CFKG &0.022& -0.021& 0.246& 0.133& -0.078& 0.273& 0.310& -0.156\\
            CKE &0.045& 0.076& 0.034& 0.056& -0.040& 0.032& 0.219& 0.055\\
            KGCN &-0.003& -0.086& 0.019& 0.082& -0.022& 0.003& -1.497& -1.872\\
            KGIN &-0.053& -0.693& 0.161& 0.101& 0.278& 0.249& 0.058& 0.214\\
            RippleNet &0.186& 0.200& 0.224& 0.072& -0.073& -0.122& 0.285& -0.037\\
            KGNNLS &-0.011& 0.027& 0.098& 0.099& 0.018& -0.082& 0.081& 0.305\\
            KTUP &-0.033& -0.103& 0.007& 0.199& -0.039& -0.504& -1.470& -1.633\\
            KGAT &0.041& -0.099& 0.075& 0.011& 0.095& 0.166& -0.085& -0.128\\
            KGRec &-0.056&0.206&0.169&0.059&0.310&0.315&0.411&-0.227\\
            DiffKG &0.221&0.556&0.025&0.024&0.088&0.178&-0.053&-0.044\\
         \Xhline{0.7mm}
    \end{tabular}
    \label{table:cold-start comparison 1.0 3}
\end{table}
\begin{table}[H]
    \centering
    \caption{\revision{Under the cold-start settings, comparison of KGER under false knowledge and decreased knowledge, distortion (F) / decreased (D) ratio is 0.25, T=5.}}
    \begin{tabular}{c|cc|cc|cc|cc} 
        \Xhline{0.7mm}
         &  \multicolumn{2}{c|}{ML-1M}&  \multicolumn{2}{c|}{Amazon-Books}&  \multicolumn{2}{c|}{BX}&  \multicolumn{2}{c}{Last.FM}\\ 
         \Xcline{2-9}{0.35mm}
         &  F&  D&  F&  D&  F&  D&  F&  D\\ 
         \Xhline{0.35mm}
            CFKG &0.082& 0.024& 0.315& 0.038& -0.209& -0.151& 1.037& 0.802\\
            CKE &-0.158& -0.028& 0.123& 0.017& -0.092& -0.072& 0.205& -2.265\\
            KGCN &-0.040& -0.005& -0.018& 0.148& -0.195& 0.497& 1.008& 1.063\\
            KGIN &-0.188& -0.130& 0.246& 0.189& 0.289& 0.376& -0.187& -2.336\\
            RippleNet &0.121& -0.008& 0.135& 0.088& -0.252& -0.005& -0.066& -0.007\\
            KGNNLS &0.007& -0.000& 0.093& -0.095& 0.240& -0.257& 1.859& 1.260\\
            KTUP &-0.054& 0.068& -0.005& 0.231& -0.122& 0.403& -7.636& 1.866\\
            KGAT &-0.072& -0.221& 0.283& 0.135& 0.197& -0.826& -0.039& -1.011\\
            KGRec &-0.148&0.096&0.111&0.111&0.710&0.336&-0.287&0.831\\
            DiffKG &0.376 &-0.076&0.039 &0.122 &0.190&0.206 &0.108 &-0.092\\
         \Xhline{0.7mm}
    \end{tabular}
    \label{table:cold-start comparison 0.25 5}
\end{table}
\begin{table}[H]
    \centering
    \caption{\revision{Under the cold-start settings, comparison of KGER under false knowledge and decreased knowledge, distortion (F) / decreased (D) ratio is 0.5, T=5.}}
    \begin{tabular}{c|cc|cc|cc|cc} 
        \Xhline{0.7mm}
         &  \multicolumn{2}{c|}{ML-1M}&  \multicolumn{2}{c|}{Amazon-Books}&  \multicolumn{2}{c|}{BX}&  \multicolumn{2}{c}{Last.FM}\\ 
         \Xcline{2-9}{0.35mm}
         &  F&  D&  F&  D&  F&  D&  F&  D\\ 
         \Xhline{0.35mm}
            CFKG &0.074& 0.014& 0.344& 0.070& -0.118& 0.375& 0.443& 0.467\\
            CKE &-0.014& -0.034& 0.129& 0.145& -0.082& -0.048& 0.531& 0.073\\
            KGCN &-0.008& 0.010& -0.012& 0.078& -0.000& 0.104& 1.163& 1.534\\
            KGIN &-0.098& -0.115& 0.286& 0.187& 0.437& 0.363& 0.254& -0.268\\
            RippleNet &0.064& 0.099& 0.173& 0.203& -0.199& -0.011& -1.257& 0.734\\
            KGNNLS &-0.009& -0.003& 0.093& 0.001& 0.206& -0.148& 0.589& 0.986\\
            KTUP &0.014& 0.014& -0.021& 0.173& -0.050& -0.073& -1.467& -0.733\\
            KGAT &0.008& -0.130& 0.168& 0.003& 0.308& 0.103& 0.921& 0.084\\
            KGRec &0.029&0.085&0.200&0.101&0.485&0.441&0.535&0.133\\
            DiffKG &0.057 &0.059  &0.047 &0.050&0.126 &0.112 &0.107 &-0.248\\
         \Xhline{0.7mm}
    \end{tabular}
    \label{table:cold-start comparison 0.5 5}
\end{table}
\begin{table}[H]
    \centering
    \caption{\revision{Under the cold-start settings, comparison of KGER under false knowledge and decreased knowledge, distortion (F) / decreased (D) ratio is 0.75, T=5.}}
    \begin{tabular}{c|cc|cc|cc|cc} 
        \Xhline{0.7mm}
         &  \multicolumn{2}{c|}{ML-1M}&  \multicolumn{2}{c|}{Amazon-Books}&  \multicolumn{2}{c|}{BX}&  \multicolumn{2}{c}{Last.FM}\\ 
         \Xcline{2-9}{0.35mm}
         &  F&  D&  F&  D&  F&  D&  F&  D\\ 
         \Xhline{0.35mm}
            CFKG &0.063& -0.005& 0.340& 0.135& 0.190& 0.000& 0.090& 0.230\\
            CKE &0.018& 0.004& 0.077& 0.084& 0.012& -0.126& 0.380& 0.055\\
            KGCN &-0.013& 0.001& 0.009& 0.036& 0.072& 0.084& -0.511& 0.529\\
            KGIN &-0.042& -0.120& 0.221& 0.177& 0.382& 0.266& 0.230& 0.162\\
            RippleNet &0.072& 0.088& 0.249& 0.267& -0.119& -0.012& -1.224& 0.290\\
            KGNNLS &0.003& -0.017& 0.118& 0.019& 0.141& 0.063& 0.563& 0.412\\
            KTUP &0.065& 0.013& -0.011& 0.101& -0.064& 0.046& -1.304& -0.444\\
            KGAT &0.027& -0.077& 0.145& -0.005& 0.289& 0.241& 0.204& -0.304\\
            KGRec &0.084&0.194&0.199&0.219&0.456&0.401& 0.423&0.049\\
            DiffKG &0.135 &0.075  &0.031 &0.042 &0.135 &0.142 &0.231 &0.041  \\
         \Xhline{0.7mm}
    \end{tabular}
    \label{table:cold-start comparison 0.75 5}
\end{table}
\begin{table}[H]
    \centering
    \caption{\revision{Under the cold-start settings, comparison of KGER under false knowledge and decreased knowledge, distortion (F) / decreased (D) ratio is 1.0, T=5.}}
    \begin{tabular}{c|cc|cc|cc|cc} 
        \Xhline{0.7mm}
         &  \multicolumn{2}{c|}{ML-1M}&  \multicolumn{2}{c|}{Amazon-Books}&  \multicolumn{2}{c|}{BX}&  \multicolumn{2}{c}{Last.FM}\\ 
         \Xcline{2-9}{0.35mm}
         &  F&  D&  F&  D&  F&  D&  F&  D\\ 
         \Xhline{0.35mm}
            CFKG &0.022& -0.036& 0.327& 0.166& -0.156& -0.135& 0.198& 0.241\\
            CKE &-0.013& 0.033& 0.068& 0.126& -0.009& -0.031& 0.159& 0.078\\
            KGCN &-0.011& -0.028& 0.029& 0.066& -0.101& 0.061& 0.283& 0.607\\
            KGIN &-0.050& -0.830& 0.203& 0.113& 0.288& 0.334& 0.049& -0.038\\
            RippleNet &0.263& 0.363& 0.226& 0.261& -0.094& -0.018& -0.424& -0.396\\
            KGNNLS &-0.009& -0.008& 0.096& 0.073& 0.010& 0.081& 0.401& 0.283\\
            KTUP &0.000& -0.045& 0.062& 0.254& -0.165& -0.515& -0.778& -3.435\\
            KGAT &0.038& -0.059& 0.119& -0.033& 0.273& 0.289& 0.417& -0.312\\
            KGRec &0.148&0.364&0.172&0.167&0.352&0.448&0.060&0.313\\
            DiffKG &0.226&0.622&0.022&0.041&0.079&0.171&0.107&0.064\\
         \Xhline{0.7mm}
    \end{tabular}
    \label{table:cold-start comparison 1.0 5}
\end{table}

\section{Additional Experiments: Impact of Hyperparameter Re-tuning on the Performance of KG Variants}\label{sec:hyperparameter experiment}

In this section, we present additional experiments to assess the impact of hyperparameter tuning when KGs are altered or reduced. Specifically, we aim to determine whether re-tuning the hyperparameters for each variant of the KG significantly affects the outcomes of the experiments presented in the main paper.

We conducted our analysis using the BX and Last.FM datasets, and the following KG-based RSs: CFKG, CKE, KGAT, and KTUP. 
We repeat the false knowledge experiments (Section~\ref{random}) and the decreasing facts experiments (Section~\ref{decrease}) in this section.
For each experiment, we compared the results obtained with and without re-hyperparameter tuning after modifying the KGs.

\begin{figure*}[h]
    \centering
    \begin{subfigure}{\textwidth}
    \includegraphics[width=\linewidth]{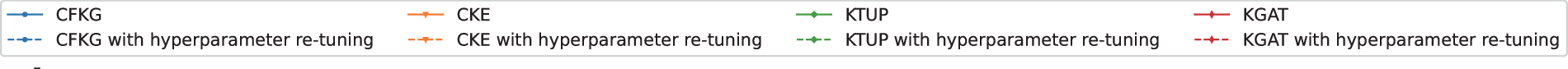}
    \end{subfigure}
    \centering
    \begin{subfigure}{0.19\textwidth}
    \includegraphics[width=\linewidth]{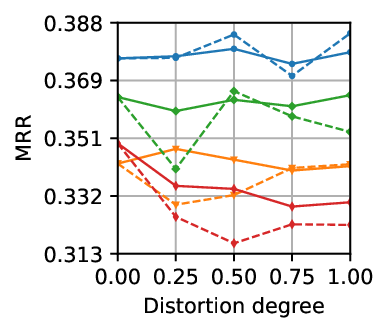}
    \end{subfigure}
    \begin{subfigure}{0.19\textwidth}
    \includegraphics[width=\linewidth]{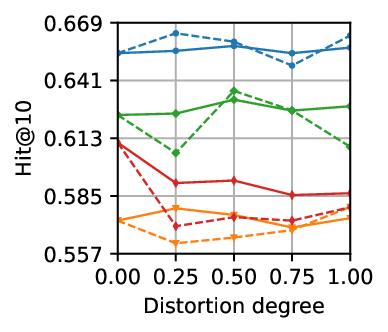}
    \end{subfigure}
    \begin{subfigure}{0.19\textwidth}
    \includegraphics[width=\linewidth]{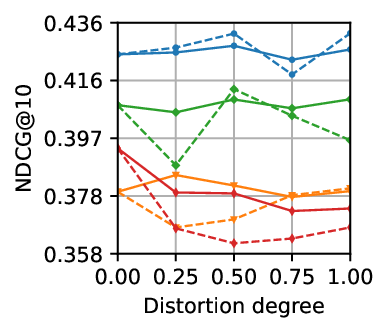}
    \end{subfigure}
    \begin{subfigure}{0.19\textwidth}
    \includegraphics[width=\linewidth]{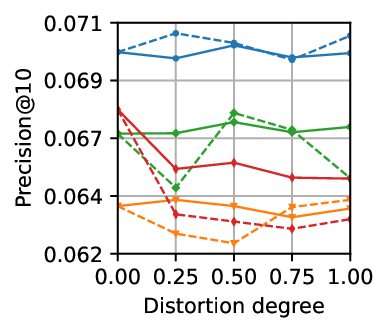}
    \end{subfigure}
    \begin{subfigure}{0.19\textwidth}
    \includegraphics[width=\linewidth]{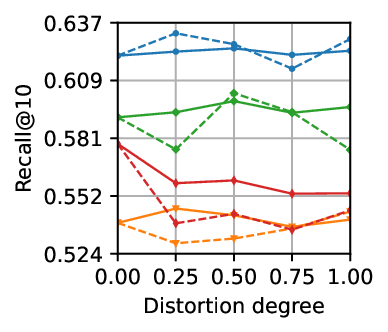}
    \end{subfigure}
    
    \caption{The results of hyperparameter re-tuning experiment with false knowledge on BX dataset. Horizontal axis denotes distortion degree.}
    \label{fig:hyperparameter false BX}
    
\end{figure*}

\begin{figure*}[h]
    \centering
    \begin{subfigure}{0.19\textwidth}
    \includegraphics[width=\linewidth]{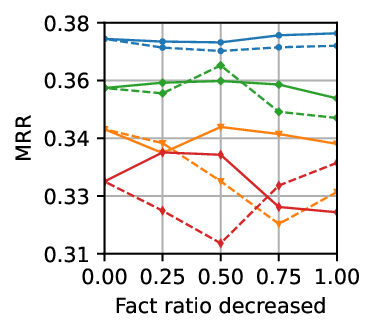}
    \end{subfigure}
    \begin{subfigure}{0.19\textwidth}
    \includegraphics[width=\linewidth]{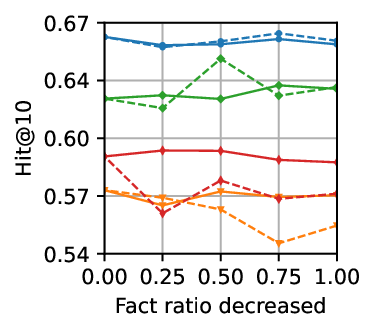}
    \end{subfigure}
    \begin{subfigure}{0.19\textwidth}
    \includegraphics[width=\linewidth]{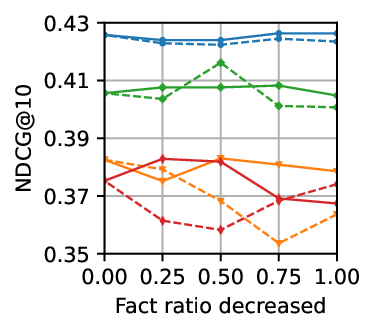}
    \end{subfigure}
    \begin{subfigure}{0.19\textwidth}
    \includegraphics[width=\linewidth]{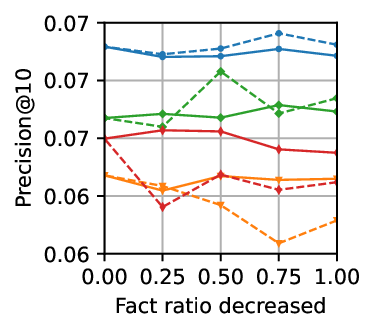}
    \end{subfigure}
    \begin{subfigure}{0.19\textwidth}
    \includegraphics[width=\linewidth]{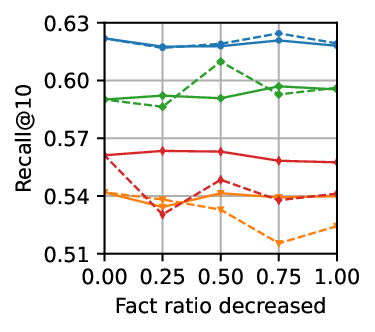}
    \end{subfigure}
    
    \caption{The results of hyperparameter re-tuning experiment with decreased facts on BX dataset. Horizontal axis denotes decreased fact ratio.}
    \label{fig:hyperparameter decrease BX}
    
\end{figure*}

\begin{figure*}[h]

    \centering
    \begin{subfigure}{0.19\textwidth}
    \includegraphics[width=\linewidth]{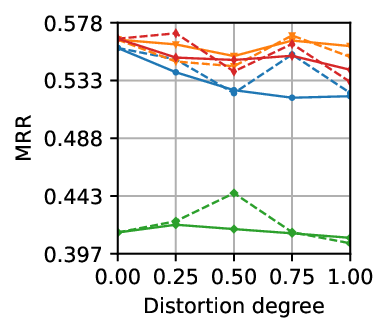}
    \end{subfigure}
    \begin{subfigure}{0.19\textwidth}
    \includegraphics[width=\linewidth]{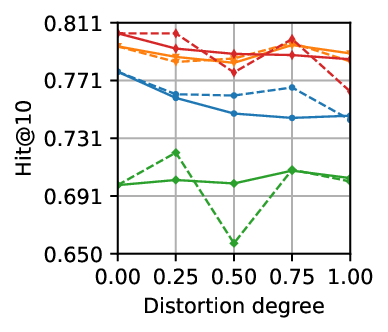}
    \end{subfigure}
    \begin{subfigure}{0.19\textwidth}
    \includegraphics[width=\linewidth]{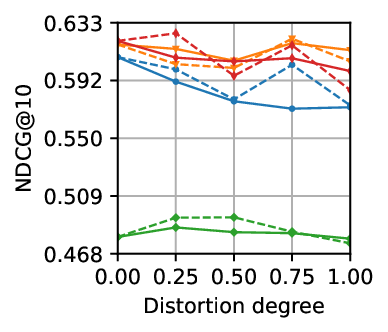}
    \end{subfigure}
    \begin{subfigure}{0.19\textwidth}
    \includegraphics[width=\linewidth]{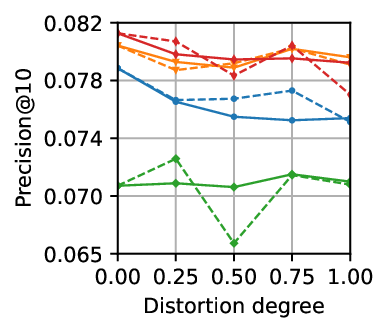}
    \end{subfigure}
    \begin{subfigure}{0.19\textwidth}
    \includegraphics[width=\linewidth]{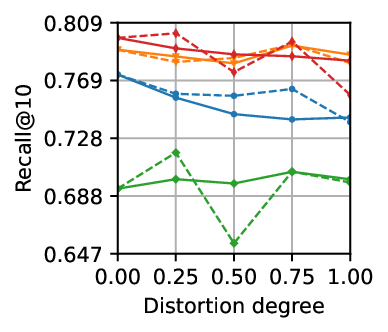}
    \end{subfigure}
    
    \caption{The results of hyperparameter re-tuning experiment with false knowledge on Last.FM dataset. Horizontal axis denotes distortion degree.}
    \label{fig:hyperparameter false lastfm}
    
\end{figure*}

\begin{figure*}[h]

    \centering
    \begin{subfigure}{0.19\textwidth}
    \includegraphics[width=\linewidth]{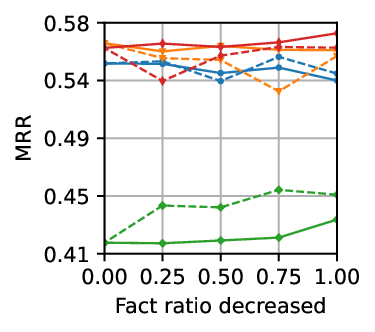}
    \end{subfigure}
    \begin{subfigure}{0.19\textwidth}
    \includegraphics[width=\linewidth]{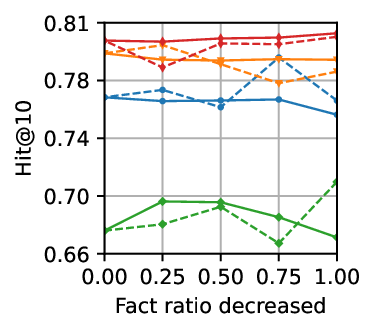}
    \end{subfigure}
    \begin{subfigure}{0.19\textwidth}
    \includegraphics[width=\linewidth]{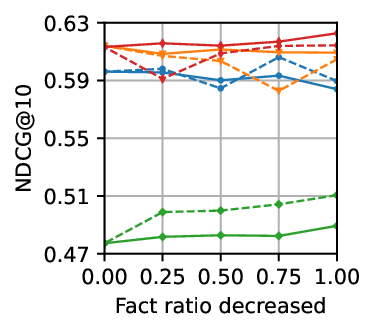}
    \end{subfigure}
    \begin{subfigure}{0.19\textwidth}
    \includegraphics[width=\linewidth]{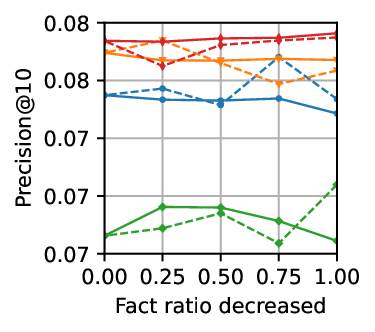}
    \end{subfigure}
    \begin{subfigure}{0.19\textwidth}
    \includegraphics[width=\linewidth]{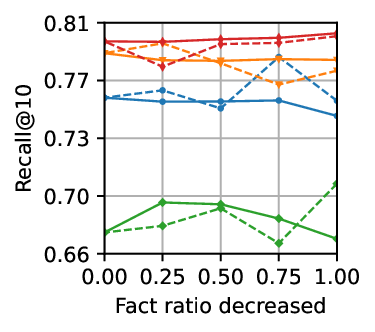}
    \end{subfigure}
    
    \caption{The results of hyperparameter re-tuning experiment with false knowledge on Last.FM dataset. Horizontal axis denotes decreased fact ratio.}
    \label{fig:hyperparameter decrease lastfm}
    
\end{figure*}

The experimental results are presented in Figure~\ref{fig:hyperparameter false BX} - Figure~\ref{fig:hyperparameter decrease lastfm}. The results indicate that, in the vast majority of cases, re-tuning the hyperparameters had minimal impact on the performance of the KG-based RSs. For CFKG, CKE, KGAT, and KTUP, the performance metrics were almost identical before and after hyperparameter tuning, with variations of 1.5\% on average.
The results show that the conclusions regarding the effects of false knowledge and decreasing facts on KGAT remain robust, regardless of whether hyperparameter tuning was applied.

\clearpage
\section{Additional Experiments: Impact of Specific Relations in KGs on the performance of RSs}\label{sec:specific relation removal}

\begin{table}[H]
  \belowrulesep=0pt
  \aboverulesep=0pt
    \centering
  \caption{Statistics on the number of most frequent relations in KGs}
    \begin{tabular}{|c|p{35pt}|l|c|}
    \toprule
    \multicolumn{1}{|c|}{Dataset} & Relation  Number & \multicolumn{1}{|c|}{Relation Type} & \# Relation \\
    \midrule
    \multirow{5}{*}{ML-1M} & 1     & film.actor.film & 92267 \\
\cmidrule{2-4}          & 2     & film.film.actor & 92254 \\
\cmidrule{2-4}          & 3     & film.film\_genre.films\_in\_this\_genre & 16659 \\
\cmidrule{2-4}          & 4     & film.film.genre & 16659 \\
\cmidrule{2-4}          & 5     & film.award\_nomination.film & 14353 \\
    \midrule
    \multirow{5}{*}{Amazon-Books} & 6     & media\_common.literary\_genre.books\_in\_this\_genre & 5969 \\
\cmidrule{2-4}          & 7     & book.book.genre & 5969 \\
\cmidrule{2-4}          & 8     & book.written\_work.author & 4397 \\
\cmidrule{2-4}          & 9     & book.author.works\_written & 4397 \\
\cmidrule{2-4}          & 10    & book.book\_subject.works & 4155 \\
    \midrule
    \multirow{5}{*}{BX} & 11    & book.author.works\_written & 66236 \\
\cmidrule{2-4}          & 12    & book.written\_work.author & 65118 \\
\cmidrule{2-4}          & 13    & book.written\_work.date\_of\_first\_publication & 13791 \\
\cmidrule{2-4}          & 14    & book.book.genre & 2568 \\
\cmidrule{2-4}          & 15    & comic\_books.publisher.comic\_book\_series\_published & 1208 \\
    \midrule
    \multirow{5}{*}{Last.FM} & 16    & film.actor.film & 4623 \\
\cmidrule{2-4}          & 17    & film.person\_or\_entity\_appearing\_in\_film.film & 4582 \\
\cmidrule{2-4}          & 18    & music.artist.origin & 2536 \\
\cmidrule{2-4}          & 19    & people.person.place\_of\_birth & 946 \\
\cmidrule{2-4}          & 20    & film.distributor.film & 500 \\
    \bottomrule
    \end{tabular}%
    \label{tab: relation statistic}
\end{table}

To further investigate the impact of specific relations within KGs on the performance of RSs, we conducted an additional experiment focusing on the most frequently occurring relations in the KGs across four datasets. Specifically, we selected the top-5 most common relations from each dataset, resulting in a total of 20 relations. These relations and their counts are listed in Table~\ref{tab: relation statistic}. For convenience, we numbered them as relation 1-20.

In this experiment, we systematically remove each relation from the KGs one at a time and measure the MRR of the RSs before and after removal. The experiments are conducted on four KG-based RSs: CFKG, CKE, KTUP, and KGAT. The results are presented in Table~\ref{tab:relation deletion result}, where numbers highlighted in red indicate an increase in MRR after the removal, and numbers highlighted in blue indicate a decrease.

Our findings indicate that none of the relations consistently improves performance across all four RSs. Furthermore, in 47.5\% of the cases (almost half of the cases), removing the relation actually leads to an increase in the RS's MRR, suggesting that not all relations within a KG are beneficial for enhancing RS performance.

These results reinforce one of the key conclusions of our study: KGs do not necessarily enhance the performance of RSs.

\begin{table}[H]
  \belowrulesep=0pt
  \aboverulesep=0pt
  \centering
  \caption{MRR of RSs after removing specific type of relations in KGs. Results highlighted in red indicate that the MRR is higher than the original result, and results highlighted in blue indicate that the MRR is lower than the original result.}
    \begin{tabular}{|c|c|r|r|r|r|}
    \toprule
    \multirow{2}{*}{Dataset} & \multirow{2}{*}{Relation Number} & \multicolumn{4}{c|}{Model} \\
\cmidrule{3-6}          &       & \multicolumn{1}{l|}{CFKG} & \multicolumn{1}{l|}{CKE} & \multicolumn{1}{l|}{KTUP} & \multicolumn{1}{l|}{KGAT} \\
    \midrule
    \multirow{6}{*}{ML-1M} & Original  & 0.731  & 0.728  & 0.705  & 0.731  \\
\cmidrule{2-6}          & 1     & \textcolor[rgb]{ .267,  .447,  .769}{0.729 } & \textcolor[rgb]{ .267,  .447,  .769}{0.727 } & \textcolor[rgb]{ 1,  0,  0}{0.706 } & \textcolor[rgb]{ 1,  0,  0}{0.738 } \\
\cmidrule{2-6}          & 2     & \textcolor[rgb]{ .267,  .447,  .769}{0.731 } & \textcolor[rgb]{ 1,  0,  0}{0.731 } & \textcolor[rgb]{ 1,  0,  0}{0.707 } & \textcolor[rgb]{ 1,  0,  0}{0.734 } \\
\cmidrule{2-6}          & 3     & \textcolor[rgb]{ .267,  .447,  .769}{0.730 } & \textcolor[rgb]{ 1,  0,  0}{0.729 } & \textcolor[rgb]{ 1,  0,  0}{0.706 } & \textcolor[rgb]{ 1,  0,  0}{0.738 } \\
\cmidrule{2-6}          & 4     & \textcolor[rgb]{ .267,  .447,  .769}{0.729 } & \textcolor[rgb]{ 1,  0,  0}{0.729 } & \textcolor[rgb]{ 1,  0,  0}{0.706 } & \textcolor[rgb]{ 1,  0,  0}{0.736 } \\
\cmidrule{2-6}          & 5     & \textcolor[rgb]{ .267,  .447,  .769}{0.729 } & \textcolor[rgb]{ 1,  0,  0}{0.729 } & \textcolor[rgb]{ 1,  0,  0}{0.710 } & \textcolor[rgb]{ 1,  0,  0}{0.734 } \\
    \midrule
    \multirow{6}{*}{Amazon-Books} & Original  & 0.538  & 0.574  & 0.394 & 0.591  \\
\cmidrule{2-6}          & 6     & \textcolor[rgb]{ 1,  0,  0}{0.538 } & \textcolor[rgb]{ 1,  0,  0}{0.575 } & \textcolor[rgb]{ 1,  0,  0}{0.395} & \textcolor[rgb]{ .267,  .447,  .769}{0.590 } \\
\cmidrule{2-6}          & 7     & \textcolor[rgb]{ 1,  0,  0}{0.539 } & \textcolor[rgb]{ .267,  .447,  .769}{0.572 } & \textcolor[rgb]{ 1,  0,  0}{0.394} & \textcolor[rgb]{ .267,  .447,  .769}{0.590 } \\
\cmidrule{2-6}          & 8     & \textcolor[rgb]{ 1,  0,  0}{0.539 } & \textcolor[rgb]{ .267,  .447,  .769}{0.572 } & \textcolor[rgb]{ 1,  0,  0}{0.395} & \textcolor[rgb]{ .267,  .447,  .769}{0.590 } \\
\cmidrule{2-6}          & 9     & \textcolor[rgb]{ .267,  .447,  .769}{0.535 } & \textcolor[rgb]{ .267,  .447,  .769}{0.571 } & \textcolor[rgb]{ 1,  0,  0}{0.395} & \textcolor[rgb]{ .267,  .447,  .769}{0.590 } \\
\cmidrule{2-6}          & 10    & \textcolor[rgb]{ 1,  0,  0}{0.538 } & \textcolor[rgb]{ .267,  .447,  .769}{0.573 } & \textcolor[rgb]{ 1,  0,  0}{0.396} & \textcolor[rgb]{ .267,  .447,  .769}{0.591 } \\
    \midrule
    \multirow{6}{*}{BX} & Original  & 0.377  & 0.348  & 0.361  & 0.337  \\
\cmidrule{2-6}          & 11    & \textcolor[rgb]{ .267,  .447,  .769}{0.375 } & \textcolor[rgb]{ .267,  .447,  .769}{0.344 } & \textcolor[rgb]{ .267,  .447,  .769}{0.360 } & \textcolor[rgb]{ .267,  .447,  .769}{0.328 } \\
\cmidrule{2-6}          & 12    & \textcolor[rgb]{ 1,  0,  0}{0.377 } & \textcolor[rgb]{ .267,  .447,  .769}{0.342 } & \textcolor[rgb]{ 1,  0,  0}{0.364 } & \textcolor[rgb]{ .267,  .447,  .769}{0.327 } \\
\cmidrule{2-6}          & 13    & \textcolor[rgb]{ .267,  .447,  .769}{0.377 } & \textcolor[rgb]{ .267,  .447,  .769}{0.341 } & \textcolor[rgb]{ 1,  0,  0}{0.363 } & \textcolor[rgb]{ .267,  .447,  .769}{0.324 } \\
\cmidrule{2-6}          & 14    & \textcolor[rgb]{ 1,  0,  0}{0.379 } & \textcolor[rgb]{ .267,  .447,  .769}{0.344 } & \textcolor[rgb]{ 1,  0,  0}{0.363 } & \textcolor[rgb]{ 1,  0,  0}{0.343 } \\
\cmidrule{2-6}          & 15    & \textcolor[rgb]{ 1,  0,  0}{0.378 } & \textcolor[rgb]{ .267,  .447,  .769}{0.345 } & \textcolor[rgb]{ 1,  0,  0}{0.362 } & \textcolor[rgb]{ 1,  0,  0}{0.344 } \\
    \midrule
    \multirow{6}{*}{Last.FM} & Original  & 0.548  & 0.566  & 0.417  & 0.566  \\
\cmidrule{2-6}          & 16    & \textcolor[rgb]{ .267,  .447,  .769}{0.544 } & \textcolor[rgb]{ .267,  .447,  .769}{0.558 } & \textcolor[rgb]{ .267,  .447,  .769}{0.412 } & \textcolor[rgb]{ .267,  .447,  .769}{0.560 } \\
\cmidrule{2-6}          & 17    & \textcolor[rgb]{ 1,  0,  0}{0.552 } & \textcolor[rgb]{ .267,  .447,  .769}{0.561 } & \textcolor[rgb]{ .267,  .447,  .769}{0.417 } & \textcolor[rgb]{ .267,  .447,  .769}{0.564 } \\
\cmidrule{2-6}          & 18    & \textcolor[rgb]{ 1,  0,  0}{0.551 } & \textcolor[rgb]{ .267,  .447,  .769}{0.559 } & \textcolor[rgb]{ .267,  .447,  .769}{0.412 } & \textcolor[rgb]{ 1,  0,  0}{0.570 } \\
\cmidrule{2-6}          & 19    & \textcolor[rgb]{ 1,  0,  0}{0.554 } & \textcolor[rgb]{ .267,  .447,  .769}{0.560 } & \textcolor[rgb]{ 1,  0,  0}{0.421 } & \textcolor[rgb]{ .267,  .447,  .769}{0.565 } \\
\cmidrule{2-6}          & 20    & \textcolor[rgb]{ 1,  0,  0}{0.551 } & \textcolor[rgb]{ .267,  .447,  .769}{0.565 } & \textcolor[rgb]{ .267,  .447,  .769}{0.417 } & \textcolor[rgb]{ 1,  0,  0}{0.570 } \\
    \bottomrule
    \end{tabular}%
  \label{tab:relation deletion result}%
\end{table}%

\clearpage
\revision{\section{Additional Metric: Knowledge Graph Utilization Score}\label{appendix:KGUS}
In this appendix, we introduce an alternative metric, the Knowledge Graph Utilization Score (KGUS), to assess the effectiveness of the knowledge graph utilization in RSs. This metric provides a different perspective compared to the original KGER by removing the division by the ratio. While KGER considers the efficiency of the knowledge graph's utilization relative to the number of facts removed or altered, KGUS focuses directly on the model's performance without scaling by a coefficient.

The formula for KGUS is as follows:
\begin{definition}\label{Definition:kgus}
\emph{KG Utilization Score (KGUS)}:
\[
    KGUS(\s, \g, \m, \Delta)=\frac{\m(\s, \g)-\m(\s, (\g {-} \Delta))}{ \m(\s, \g) }
\]
\end{definition}

The KGUS values under different settings are presented in Table~\ref{table:merged_kgus}. To be consistent with the results in the previous sections, we still choose the distortion/decrease ratio to be $0.5$ and $T =3$. Under this setting, the KGER values in Table~\ref{table:merged_kger} are twice the KGUS values presented in Table~\ref{table:merged_kgus}. Despite this difference in scale, the relative relationships between the values remain unchanged, and the sign of the values also remains consistent. The only notable difference is that the variance is four times greater for KGUS than for KGER. Therefore, whether we use KGER or KGUS as the metric, the conclusions drawn from the experiments remain unchanged.}

\begin{table*}
    \centering
    \caption{\revision{KGUS under false knowledge (F) and decreased knowledge (D), for normal and cold-start users, F/D ratio is 0.5, T = 3.}}
    \resizebox{1\linewidth}{!}{
    \begin{tabular}{c|cc|cc|cc|cc|cc|cc|cc|cc} 
        \Xhline{0.7mm}
         &  \multicolumn{8}{c|}{Normal} & \multicolumn{8}{c}{Cold-Start} \\
         \Xcline{2-17}{0.35mm}
         &  \multicolumn{2}{c|}{ML-1M}&  \multicolumn{2}{c|}{Amazon-Books}&  \multicolumn{2}{c|}{BX}    & \multicolumn{2}{c|}{Last.FM}&  \multicolumn{2}{c|}{ML-1M}&  \multicolumn{2}{c|}{Amazon-Books}&  \multicolumn{2}{c|}{BX}    & \multicolumn{2}{c}{Last.FM}\\ 
         \Xcline{2-17}{0.35mm}
         &  F&  D&  F&  D&F&  D&  F&  D&  F&  D&  F&  D&  F&  D&  F&  D\\ 
         \Xhline{0.35mm}
            CFKG&   0.001 & \underline{-0.001} & 0.078 & 0.013 & \underline{-0.008} & 0.003 & 0.059 & 0.013 & 0.028 & 0.006 & 0.118 & 0.011 & \underline{-0.058} & 0.024 & 0.205 & 0.441 \\ 
            CKE&   0.001 & 0.003 & 0.007 & 0.003 & \underline{-0.004} & \underline{-0.002} & 0.023 & 0.004 & 0.013 & 0.010 & 0.002 & \underline{-0.003} & \underline{-0.002} & 0.027 & 0.058 & 0.086 \\ 
            KGCN&   0.002 & \underline{-0.004} & \underline{-0.001} & 0.005 & 0.000 & 0.013 & 0.057 & 0.024 & 0.004 & \underline{-0.083} & \underline{-0.002} & 0.031 & \underline{-0.030} & 0.031 & \underline{-0.640} & 0.372 \\ 
            KGIN&   0.009 & 0.001 & 0.030 & 0.009 & 0.068 & 0.038 & 0.061 & \underline{-0.025} & 0.016 & \underline{-0.015} & 0.086 & 0.061 & 0.241 & 0.165 & \underline{-0.160} & 0.064 \\ 
            RippleNet&   0.006 & 0.007 & 0.053 & 0.028 & 0.018 & \underline{-0.006} & 0.005 & 0.004 & 0.057 & 0.036 & 0.123 & 0.150 & \underline{-0.070} & \underline{-0.103} & 0.214 & \underline{-0.945} \\ 
            KGNNLS&   0.001 & \underline{-0.001} & 0.010 & 0.007 & 0.020 & \underline{-0.003} & 0.032 & \underline{-0.002} & \underline{-0.008} & 0.016 & 0.028 & 0.001 & 0.040 & 0.121 & 0.506 & \underline{-0.794} \\ 
            KTUP&   0.004 & \underline{-0.003} & 0.005 & 0.006 & 0.002 & \underline{-0.006} & \underline{-0.006} & \underline{-0.004} & 0.058 & 0.011 & \underline{-0.028} & \underline{-0.045} & \underline{-0.020} & 0.017 & 0.494 & 0.278 \\ 
            KGAT&   \underline{-0.003} & \underline{-0.003} & 0.034 & 0.004 & 0.042 & \underline{-0.026} & 0.029 & \underline{-0.002} & 0.083 & 0.002 & 0.053 & 0.007 & 0.071 & 0.065 & 0.064 & 0.008 \\ 
            KGRec&   \underline{-0.010} & 0.040 & 0.048 & 0.040 & 0.040 & 0.070 & 0.064 & 0.070 & 0.049 & 0.080 & 0.106 & 0.051 & 0.143 & 0.120 & 0.146 & 0.066 \\ 
            DiffKG&   \underline{-0.004} & \underline{-0.012} & \underline{-0.002} & \underline{-0.003} & 0.056 & 0.020 & 0.025 & \underline{-0.002} & 0.065 & 0.025 & 0.004 & 0.013 & 0.092 & 0.084 & \underline{-0.092} & \underline{-0.082} \\ 
         \Xhline{0.7mm}
    \end{tabular}
    }
    \label{table:merged_kgus}
\end{table*}

\end{document}